\documentclass[preprint]{jfm}
\pdfoutput=1
\usepackage{ifpdf}
\usepackage{upmath}
\usepackage{setspace}
\usepackage[english]{babel}
\usepackage[latin1]{inputenc}
\usepackage{epsfig,float}
\usepackage{multirow,subfig}
\usepackage{booktabs,tabularx,graphicx}

\usepackage{efbox}

\usepackage{vmargin}
\usepackage{amsmath,amssymb,amsfonts}
\usepackage{float}
\usepackage{psfrag,wrapfig,tabularx}
\usepackage{inputenc,fontenc}
\usepackage{longtable,graphics}
\usepackage{picinpar}
\usepackage{supertabular,marvosym}
\usepackage{ulem} 
\usepackage{rotating} 
\usepackage{color}
\usepackage{natbib}
\usepackage{placeins}
\newcolumntype{C}{>{\centering\arraybackslash}X} %% this is for the figures table labeling Reda

\graphicspath{{./figs/}}

\usepackage {latexsym} %para simbolos
\efboxsetup{linecolor=black,linewidth=10pt}

\title[Square cylinder in the interface of two different-velocity streams]{
	Square cylinder in the interface of two different-velocity streams}
\author[R. El Mansy, W. Sarwar, J.M. Bergad\`a, F. Mellibovsky ]
{R.\ns E\ls L\ns M\ls A\ls N\ls S\ls Y$^{1}$, \ns W.\ns S\ls A\ls R\ls W\ls A\ls R$^2$,  \ns J.\ns M.\ns B\ls E\ls R\ls G\ls A\ls D\ls A $^{1}$, \ns F.\ns M\ls E\ls L\ls L\ls I\ls B\ls O\ls V\ls S\ls K\ls Y$^2$  }

\affiliation{$^1$Fluid Mechanics Department, Universitat Polit\`ecnica de Catalunya, 08034, Barcelona, Spain  \\[\affilskip]
	$^2$ Department of Physics, Aerospace Engineering Division, Universitat Polit\`ecnica de Catalunya, 08034, Barcelona, Spain }

\begin{document}
	
\newcommand{\D}{\displaystyle}
\maketitle
	
\begin{abstract}
		
  We investigate the incompressible flow past a square cylinder immersed in the wake of an upstream nearby splitter plate separating two streams of different velocity. The bottom stream Reynolds number, based on the square side, $Re_B=56$ is kept constant while the top-to-bottom Reynolds numbers ratio $R\equiv Re_T/Re_B$ is increased in the range $R\in[1,6.5]$, corresponding to a coupled variation of the bulk Reynolds number $Re\equiv(Re_T+Re_B)/2\in[56,210]$ and an {\it equivalent} nondimensional shear parameter $K\equiv2(R-1)/(R+1)\in[0,1.4667]$. The onset of vortex-shedding, at $R=2.1\pm0.1$ (corresponding to $Re=86.8\pm2.8$, $K=0.71\pm0.04$), is pushed to higher $Re$ as compared to the square cylinder in the classic configuration.
  %, and the incoming upstream flow disymmetrises the K\'arm\'an vortex street to a point that only vortices shed from the high-velocity side are present in the wake
  The advent of three-dimensionality is triggered by a mode-C-type instability at $R\simeq3.1$ ($Re\simeq115$, $K\simeq1.02$) with wavelength $\lambda_z\simeq2.4$, much as reported for open circular rings and square cylinders placed at an incidence. The resulting solution is period-doubled and exhibits a triad of spanwise symmetries: a mirror reflection and two spatiotemporal symmetries involving the evolution by half a period (two vortex-shedding cycles) followed by either specular reflection or a half-wavelength shift. The path towards spatio-temporal chaos is initiated thereafter with a modulational period-doubling tertiary bifurcation at $R\in(3.4,3.8)$ that also doubles the spanwise periodicity. The ensuing nonlinear solution repeats only after four vortex shedding periods and retains only a spatiotemporal invariance consisting in the evolution by half a period (two vortex-shedding cycles) followed by mirror reflection about a streamwise-cross-stream plane. At slighlty higher values of $R\geq4$, the flow has become spatio-temporally chaotic, but the main features of mode C are still clearly distinguishable.
\end{abstract}

\section{Introduction}
	
%%% Opening statements

%% CC as a paradigm of bluff body aerodynamics
The flow past circular cylinders has been extensively studied as a paradigm of bluff body aerodynamics \citep*{WilliamsonARFM1996}, for it involves several interesting phenomena which include, but are not limited to, laminar and turbulent boundary layer separation% \citep*{DovgalPAS1994,SimpsonARFM1989}
, vortex shedding% \citep*{GerrardJFM1966}
, or detached shear layer and wake instabilities% \citep*{Rai2010,WilliamsonJFM1992}
. Flow unsteadiness, spatio-temporal chaos and turbulence are a source of aerodynamic noise % \citep*{HardinAIAAJ1984,InoueJFM2002}%,GerrardPPSLSB1955
 and vortex-induced vibration%s \citep*{WilliamsonARFM2004}
, and crucially affect aerodynamic forces% \citep*{WilliamsonARFM1996}
, to name but a few issues that are relevant from the engineering applications viewpoint. The sole governing parameter, the Reynolds number, is defined as $Re={U D/\nu}$, where $U$ is the free-stream velocity, $D$ the cylinder diameter and $\nu$ the kinematic viscosity of the fluid.
	
%% SC as an archetype of bluff body
Square \citep*{DuraoEF1988,LynJFM1995,LuoPoF2003} and rectangular \citep{OkajimaJFM1982,NorbergJWEIA1993%,SohankarJWEIA1997
} cylinders have also been used, albeit to a lesser extent, as an archetype for bluff body aerodynamics. The same definition as for the circular cylinder is used for the Reynolds number, except that $D$ is taken to be the square cylinder side length (or the cross-stream side length for a rectangle).

%% Cylinder in homogeneous shear
In real-world applications, bluff bodies are often submerged in boundary layers or wakes and the flow past them is decisively modified by the inhomogeneous velocity profiles of the incoming upstream flow. This is the case of a bridge pillar that is close to the river shore, long-span bridges immersed in an atmospheric boundary layer, underwater pipes near the seabed or subject to strong currents, or a building (or motor vehicle) in the wake of an upstream building (vehicle).

%% CC/SC + shear
The simplest model for such a situation is the uniform planar shear flow past a circular \citep*{KiyaJFM1980,TamuraJSME1980,KwonJFE1992} or square \citep*{AyukawaJWEIA1993,HwangIJNMF1997%,AyukawaTJSMEB1985,AyukawaJSME1990
} cylinder, where the homogeneous incoming streamwise velocity is replaced with a linear profile (constant cross-stream gradient of streamwise velocity and, therefore, constant non-null shear). The shear parameter is defined as $K\equiv D G /U_c$, with $D$ the cylinder characteristic length (diameter and side for circular and square cylinders, respectively), and $U_c$ and $G=(\partial u/\partial y)_{y=y_c}$ the upstream streamwise velocity and dimensional cross-stream gradient of streamwise velocity, respectively, at cylinder mid-height.

%\fer{
Ocasionally, the body of interest lies in the way of a thin shear layer rather than a smooth shear profile. This happens, for example, in the near wake of lift-producing devices such as airfoils, stator vanes or rotor blades of compressor and turbines or fans. Struts or rods supporting a structural casing are examples  of objects subject to this type of incoming flows. This is precisely the kind of situations we intend to model here by placing a bluff body, the square cylinder, in the interface of two streams with different velocity. The wake-body interaction problem usually considers a body, streamlined or bluff, placed in the wake of another bluff body. The configuration in which a bluff body is placed in the wake of a streamlined body such as we intend to address here has very seldom been considered in the literature, and then always placing a cylinder or strut in the wake of an airfoil \citep*{Zhang2005,Niu2021}. However distant this problem may seem at first from that of homogeneous upstream shear, the effects associated to the different velocity {\it seen} by the upper and lower sides of the bluff body might be expected -and will indeed be shown- to bear striking resemblance.
%}

%%% Circular cylinder

%% CC primary bifurcation
The symmetric and steady flow past a circular cylinder undergoes a supercritical Hopf bifurcation, the primary instability, at around $Re^H\simeq47$ with frequency $St^H\simeq0.12$ \citep{ProvansalJFM1987,NorbergJFM1994%,RoshkoNACA1954
}, resulting in a time-periodic two-dimensional solution that consists in the alternated shedding of opposite-signed vortices from either side of the cylinder -- a flow configuration commonly referred to as K\'arm\'an vortex street \citep{Karman1911,Karman1912}. Although the spatial Z$_2$ symmetry associated to vertical reflection about a diametral plane aligned with the incoming flow is broken \citep{MarquesPhysD2004}, a spatio-temporal Z$_2$ symmetry persists in the form of flow invariance upon the combined effect of evolution by a half period followed by reflection about the same original reflection-symmetry plane.
	
%% CC secondary bifurcations
The periodic and space-time-symmetric two-dimensional vortex-shedding solution might be observed in experiments all the way up to $Re\lesssim190$, three-dimensionality consistently arising from this point on \citep{WilliamsonPoF1996}. Two distinct three-dimensional vortex-shedding modes have been reported in the so-called wake transition regime, whose inception results in two corresponding discontinuities of the Strouhal number ($St$) dependence on $Re$ \citep{WilliamsonPoF1988a}. The first one, mode A, is characterised by the onset of vortex loops that are stretched by shear into streamwise vortex pairs of spanwise wavelength around $3\sim4D$ and has been shown to persist at flow regimes as low as $Re\gtrsim180$, thus coexisting with the two-dimensional solution over a small range of Reynolds numbers \citep{WilliamsonJFM1996}. Mode A is only regularly patterned at the early stages of inception and then only transiently, but soon after develops intermintent large-scale spot-like wave dislocations that render the spanwise structure rather irregular \citep{WilliamsonJFM1992}. The second, mode B, arises at slightly higher values of the Reynolds number $Re\gtrsim250$ and exhibits a fairly regular spanwise pattern with a shorter characteristic wavelength of about $\sim1D$ \citep{WilliamsonJFM1996}. Mode-B-type vortical structures pervade the near wake even at flow regimes where turbulence has already set in at much higher Reynolds numbers in excess of 1000 \citep{MansyJFM1994}. Floquet stability analysis has shown that mode A emanates from a secondary instability of the two-dimensional periodic vortex-shedding solution at $Re^A=188.5\pm1$ with wavelength $\lambda_z^A=3.96\pm0.02$, and happens to be subcritical \citep{HendersonPoF1996}, hence the hysteretical flow behaviour. Meanwhile, mode B seems to be related to a second instability of the same solution that occurs supecritcally at $Re=259\pm2$ with $\lambda_z^B=0.822\pm0.007$ \citep{BarkleyJFM1996}. The critical values of the parameters at bifurcation have since been further refined to $(Re^A,\lambda_z^A)=(190.2\pm0.02,3.966\pm0.002)$ and $(Re^B,\lambda_z^B)=(261.0\pm0.2,0.825\pm0.002)$ using asymptotically large domains \citep{PosdziechTCFD2001}. A third mode, consisting of a complex-conjugate pair of eigenvalues and dubbed QP on account of its introducing quasi-periodicity into the flow, has been identified as dominant at intermediate wavelengths of about $\sim2D$, in between those characterising modes A and B \citep{BlackburnPoF2003}. Mode QP bifurcates at $Re^{QP}\simeq377$ and generates branches of unstable, and therefore not experimentally realisable, quasi-periodic states \citep{BlackburnJFM2005}.%, which might explain a mode C seen to arise in the wake transition regime instead of modes A and B under appropriate artificially imposed experimental conditions \citep{ZhangPoF1995}.

%% Modes A, B, QP, C
%Circular cylinder experiments: modes A (with dislocations) and B (extending to very high Re) identifiable by discontinuities in the St vs Re, and with distinct wavelengths. [WilliamsonPoF1995, WilliamsonPoF1996, WilliamsonJFM1996, WilliamsonARFM1996]
%Circular cylinder Floquet StAn: modes A and B and QP [BarkleyJFM1996]
%Circular cylinder DNS: mode A is subcritical, B supercritical [HendersonPoF1996, HendersonJFM1997]
%Floq StAn first shows to synchronous modes (A, B) and a subharmonic mode S [RobichauxPoF1999]. The subharmonic mode is in fact quasiperiodic QP, but not previoZusly recognised because of the stability analysis method used [BlackburnPoF2003]
	
%%% Bifurcations under spatio-temporal Z2 and spatial O(2) symmetries
%Spatiotemporal Z2 + spatial O(2): bifurcation types [MarquesPhysD2004]
The spatio-temporal Z$_2$ symmetry of the two-dimensional vortex-shedding regime coexists with the spanwise invariance of the infinite-cylinder flow problem, represented by the orthogonal group O(2)$=$Z$_2\times$SO(2), which includes reflection about any discretionary plane that is orthogonal to the spanwise direction (Z$_2$) and every arbitrary translation along the span (SO(2)). Systems with Z$_2\times$O(2) symmetry, where the Z$_2$ refers to a spatio-temporal symmetry, rather than simply spatial, and O(2) to space invariance, admit two types of synchronous codimension-one bifurcations, one preserving the space-time Z$_2$ symmetry and the other one breaking it \citep{MarquesPhysD2004}. Two-dimensional time-periodic vortex-shedding past a circular cylinder belongs to this symmetry class and, among the secondary instabilities that three-dimensionalise the flow, mode A is triggered by a Z$_2$-preserving bifurcation, while mode B is induced by one that breaks it \citep{BlackburnJFM2005}.
%\fer{Consider removing what follows.}
Whenever the bifurcation in a system with the aforementioned symmetries involves a complex-conjugate pair that is non-resonant with the destabilising two-dimensional time-periodic and space-time symmetric solution, three quasi-periodic solution branches arise, namely a pair of symmetry-conjugate modulated travelling waves and a third of modulated standing waves. Only one of the two distinct types of solution branches -either the pair of travelling or the standing waves- might be stable at a time \citep{MarquesPhysD2004}. The third three-dimensionalising secondary instability of the two-dimensional time-periodic wake past a circular cylinder corresponds precisely to a quasi-resonant quasi-periodic subcritical bifurcation. The two symmetry-conjugate branches of modulated travelling-wave solutions add an unstable eigenmode to the count of the already unstable two-dimensional solution, while the modulated standing wave adds two \citep{BlackburnJFM2005}. The neighbouring 1:4 resonant case, which would correspond to a period-doubling bifurcation, is a codimension-two bifurcation and would therefore require the tuning of a second parameter beside the Reynolds number for a complete unfolding. The symmetry group of the two-dimensional vortex-shedding regime admits also a number of mixed-mode bifurcations and strong 1:1 and 1:2 resonances, all codimension-two, none of which seems to bear any relevance to the cylinder wake problem \citep{MarquesPhysD2004}.
	
%\fer{$C_d$, $C_l$, $St$, $l_r$... trends with $Re$ not explained. Maybe say something, at least of the trends across the wake transition regime. Restrict literature review to $Re\in[0,300]$.}
	
%%% Square cylinder
	
%% SC symmetries and first bifurcation
Square-cylinder wake dynamics bears compelling resemblance to that past a circular cylinder. The symmetries of the problem are the same and the primary instability leads to a two-dimensional time-periodic and space-time symmetric vortex-shedding state at the slighlty lower $Re^H\simeq45$ and $St^H\simeq0.10$ \citep{NorbergUnpublished1996,ParkJFM2016%,SohankarJWEIA1997
}. There are however notable differences that concern the location where the boundary layer separates from the cylinder surface. While separation points can migrate freely on the surface of a circular cylinder, they are bound to coincide with the corners of a square or rectangular cylinder. As it happens, separation occurs from the rear corners only at very low Reynolds number, and then only after reattachment from an initial separation from the front corners \citep{OkajimaJFM1982,RobichauxPoF1999,YoonPoF2010}.
	
%% SC secondary bifurcations
The flow past a square cylinder also exhibits mode A- and B-type structures in the wake transition regime \citep{SohankarPoF1999,SahaIJHFF2003,LuoPoF2003,BaiPoF2018}, but their respective occurrence starts at lower values of the Reynolds number $Re^A\simeq160\pm2$ and $Re^B\simeq204\pm5$ and present somewhat larger wavelengths $\lambda_z^A/D\simeq5.1\pm0.1$ and $\lambda_z^B/D\simeq1.3\pm0.1$ at onset \citep{LuoJFS2007}. While early experiments failed to detect any hysteresis in the inception of mode A and no discontinuity in the Strouhal number dependence on Reynolds number was observed \citep{LuoPoF2003}, later experiments that strived to accurately resolve variations in the driving parameter produced a small hysteretical region \citep{LuoJFS2007,TongJFS2008}.
	
%TongJFS2008 shows the two discontinuities in the St vs Re curve, thus supporting the subcritical A mode hypothesis for the square cylinder.
%Square cylinder: Mode A is supercritical [SheardJFM2009]
	
Linear instabilities akin to modes A and B of the flow past a circular cylinder have been identified through Floquet stability analysis also in the wake transition regime of the flow past a square cylinder \citep{RobichauxPoF1999}, alongside a third, subharmonic, quasi-periodic mode. As for the circular cylinder, modes A and B preserve and break, respectively, the spatio-temporal Z$_2$ symmetry of the two-dimensional vortex-shedding solution, but occur at lower $Re^A\simeq164$ and $Re^B\simeq197$, with slightly longer $\lambda_z^A/D\simeq5.2$ and $\lambda_z^B/D\simeq1.1$ at bifurcation \citep{SheardJFM2009,ChoiPoF2012}. Modulated travelling- and standing-wave solution branches also arise in the wake of a square cylinder following the bifurcation of a complex-conjugate pair \citep{BlackburnPoF2003}, which is the counterpart to quasi-periodic mode QP of the circular cylinder. Mode QP, which bifurcates at $Re^{QP}\simeq215$ with $\lambda_z\simeq2.6$ for the square cylinder \citep{SheardJFM2009}, was originally mistaken for a subharmonic (period-doubling) bifurcation on account of its being of a quasi-resonant quasi-periodic type, and following a shortcoming of the stability analysis method used in finding what they called mode S \citep{RobichauxPoF1999}. Unlike what happens for the flow past a circular cylinder, the symmetry-conjugate branches of modulated travelling-wave solutions issued from the QP bifurcation are supercritical and inherit the stability properties of the two-dimensional solution, while the modulated standing-wave branch, which is also supercritical but with a lesser slope at bifurcation, adds an unstable eigenmode \citep{BlackburnJFM2005}.
	
Nonlinear analysis using the Landau equation as a model for the secondary instability of the flow past a square cyilinder points to a supercritical nature of mode A at bifurcation \citep{SheardJFM2009}, as opposed to what happens for the circular cylinder \citep{HendersonPoF1996}, and in overt contradiction with experimental observation \citep{LuoJFS2007}. This dispute as to the subcritical or supercritical nature of mode A between numerical simulation and experiment has not yet been settled to the authors knowledge.
	
%% CC to SC
The relation between the instabilities (both primary and secondary) in the wake of square and circular cylinders has recently been elucidated by the numerical smooth transformation of the former into the latter by gradual rounding of the corners \citep{ParkJFM2016}. The primary Hopf bifurcation that brings about two-dimensional vortex shedding is initially slightly delayed as the cylinder geometry evolves from circular to square, but the trend is reversed halfway and the critical value brought down to $Re^H\simeq44.7$. The bifurcation remains supercritical all along. Secondary three-dimensionalising instabilities also evolve continuously and uneventfully, gradually advancing the occurrence of the bifurcations of all three-modes, A, B and QP, to lower critical values of $Re$. The order of bifurcation is not altered, but the bifurcation points get closely packed, with mode QP strongly promoted for the square in comparison to the circular shape. Meanwhile, the wavelength at criticality is slightly but steadily increased for all three modes in the morphing from circular to square.

The top-bottom Z$_2$ reflection symmetry of the flow past an infinitely long cylinder might be broken in several possible ways. For a circular cylinder, the symmetry disruption might be achieved by introducing curvature along the spanwise direction turning the cylinder into a ring \citep{MonsonJFE1983,LewekePRL1994,LewekeJFM1995,SheardJFM2003}, by applying rotation about its centerline \citep{KangPoF1999,MittalJFM2003}, by subjecting it to a non-uniform upstream velocity profile (incoming shear flow) \citep{JordanPoF1972,KiyaJFM1980,TamuraJSME1980,ParkPoF2018%,SumnerJFS2003
}, or by combining incoming shear with rotation \citep{YoshinoJSME1984,SungJFE1995}, among other options. % like the presence of a nearby wall .
Upstream shear is also an option for breaking the symmetry of the square cylinder problem \citep{HwangIJNMF1997,SahaJEM1999,SohankarJBSMSE2020}, as also is placing the cylinder at an incidence with respect to the incoming flow \citep{NorbergJWEIA1993,SohankarIJNMF1998,TongJFS2008,YoonPoF2010%,SohankarJWEIA1997
}, or combining both effects together. In most cases, the two-dimensional time-periodic vortex-shedding solution persists upon deliberately breaking the spatial Z$_2$ symmetry of the problem, but the space-time Z$_2$ symmetry is no longer fulfilled \citep{BlackburnPoF2010}.
	
%% rotating circular cylinder?
	
%%% circular ring
%%SheardJFM2003, SheardJFM2005, SheardPoF2005: Modes A, B and C (subharmonic)
%%BlackburnPoF2010: Mode C - Mode QP relationship due to weak symmetry breaking [circular cylinder -> ring]
The primary instability of the flow past an open ring leads to an axisymmetric time-periodic vortex-shedding regime \citep{MonsonJFE1983,LewekePRL1994}, analogous to the two-dimensional vortex-sheding regime past a circular cylinder but obviously lacking its spatio-temporal Z$_2$ symmetry. Numerical studies have shown that, besides the instabilities related to the classic modes A and B, a subharmonic mode C also emerges and is the dominant secondary instability for rings of aspect ratio around $\Gamma\equiv d/D=5$, defined as the quotient between the diameter of the circle described by the cylinder axis ($d$) and the cylinder diameter itself ($D$) \citep{SheardJFM2003}. Both experiments and direct numerical simulation show that the secondary linear instability develops nonlinearly into a period-doubled vortex-shedding solution with a distinct wavelength somewhere in between those predicted by Floquet stability analysis for modes A and B in the ring wake \citep{SheardJFM2005,SheardPoF2005}. Although the solution is period-doubled, aggregate quantities such as aerodynamic forces preserve the original period. Two instants exactly one shedding cycle apart are mutually related by an appropriate symmetry operation (any of either a spanwise/azimuthal rotation by half the angular wavelength or reflection about some collection of appropriately chosen diametral planes) and the period-doubling does not initiate a period-doubling cascade. Instead, the mode-C structures that characterise the wake after the secondary bifurcation are replaced by mode-A-type structures when the Reynolds number is further increased. The scenario that follows seems to be analogous to that for the circular cylinder
%%, where the continuous broadband spectrum of unstable wavenumbers associated to mode A generates a competition that brings about spatio-temporal chaos
	%\citep{LewekePRL1994,LewekeJFM1995,HendersonJFM1997}.
	
%%% square+aoa.
%%BlackburnPoF2010: Mode C - Mode QP relationship due to weak symmetry breaking [square cylinder + inclination]
%%SheardJFM2009, SheardJFS2011: Mode QP becomes C as aoa is increased. Mode A (supercritical) is dominant at aoa<10.5, then Mode C becomes dominant (supercritical), and finally Mode A' (subcritical) for aoa>26.
%%TongJFS2008: Experiments observe the sequence ReH < ReA < ReB at all aoa, mistaking the codim-2 point A-C for a maximum in ReA.
%%SohankarIJNMF1998: 
A subharmonic instability analogous to mode C is also dominant for a square cylinder placed at moderate incidence angles $\alpha$ \citep{SheardJFM2009}. Mode A, which was originally believed to be dominant for all incidences following experiments that unfortunately failed to check the flow topology at intermediate values of $\alpha$ \citep{TongJFS2008}, is in fact overtaken by mode C for $\alpha\gtrsim10.5^\circ$ \citep{YoonPoF2010,SheardJFS2011}, which is in turn outdone by another mode of characteristics similar to those of mode A for $\alpha\gtrsim26^\circ$. This second mode $A^\prime$ is distinct from the one evolving from $\alpha=0^\circ$ in that it evolves from the space-time symmetric mode A corresponding to $\alpha=45^\circ$. Numerical evidence seems to discard any smooth connection between modes A and $A^\prime$ by mere continuous tilting of the cylinder from one incidence to the other, although the physical flow mechanisms at play appear to be the same. Mode-B-type structures have also been detected in experiments above a critical Reynolds number that also evolves smoothly as the incidence angle is changed across the full range \citep{TongJFS2008}. The presence of mode-B structures in the tilted cylinder wake has been confirmed by direct numerical simulation at sufficiently high values of the Reynolds number \citep{SheardJFM2009}, which in this case is usually defined with the cross-stream projected height $D\times(\cos{\alpha}+\sin{\alpha})$ instead of just $D$.

Mode QP of the flow past circular and square cylinders is only the third linear instability, after modes A and B, of the periodic and space-time symmetric two-dimensional vortex-shedding solution, and the associated growth rate is much smaller. It is therefore not to be expected that wake structures related to mode QP might be observed in actual experiments. It does however bear strong resemblance to the mode C observed in the wake behind open rings \citep{SheardJFM2005} of the right moderate aspect ratio and square cylinders at intermediate incidence angles \citep{SheardJFM2009}, regarding both spanwise wavelength and flow topology. As a matter of course, a weak disruption of the spatial Z$_2$ symmetry in problems belonging to the Z$_2\times$O(2) symmetry group will alter the nature and characteristics of the secondary bifurcations. Modes A and B may preserve their respective symmetries only approximately, but the inconmensurate frequencies of mode QP, which did not retain any remnant of the spatio-temporal Z$_2$ symmetry, might experience a locking into some strong resonance. As it happens, mode QP is quasi-resonant for both circular and square cylinder flows and, as the span is curved into a ring or the square tilted into incidence, the associated complex-conjugate pair of Floquet multipliers approaches the negative real axis, collides, and separates into a couple of negative real eigenvalues, \emph{i.e.} two subharmonic/period-doubling modes \citep{BlackburnPoF2010,SheardJFS2011}. It is thus that mode QP evolves into mode C, which eventually overtakes modes A and B in driving the secondary instability once the reflection symmetry across the midplane has been sufficiently broken.

Upstream shear has been shown in experiments to delay the onset of periodic vortex shedding, \emph{i.e.} $Re^H$ increases with $K$, to the point that shedding can be completely suppressed all the way up to $Re<220$ \citep{KiyaJFM1980}. This effect has also been observed in numerical simulation \citep{TamuraJSME1980}. Other computational studies, however, did not detect the phenomenon despite exploring similar values of the parameters \citep{LeiOE2000,CaoJFS2010}. 

Stability analysis seems to favour the notion that $Re^H$ is mostly unaffected by $K$, and that, if anything, it is marginally promoted to slightly lower values, from 46.5 for $K=0$ to 45.5 for $K=0.2$ \citep{ParkPoF2018}. Floquet stability analysis shows that the wake transition regime is instead greatly affected by upstream shear. While modes A, B and QP bifurcate at successively large values $(Re^A\simeq190)<(Re^B\simeq250)<(Re^{QP}\simeq380)$ in the symmetric problem, mode QP locks into subharmonic mode C as $K$ is increased and gradually overtakes mode B and mode A for $K=0.1$ and $0.2$, respectively \citep{ParkPoF2018}. As a matter of fact, modes A and B are pushed to higher $Re^A\simeq240$ and $Re^B\simeq300$, while mode C is advanced to as low as $Re^C\simeq150$ for $K=0.2$. While the critical spanwise wavelength of mode A slightly increases to $\lambda_z^A\simeq4.2$, mode B remains mostly unaltered at $\lambda_z^B\simeq0.8$ and mode C somewhat narrows to $\lambda_z^C\simeq1.6$ for $K=0.2$. Numerical simulation has seen mode-A-type structures supressed, and with them three-dimensionality, at $Re\simeq200$ with $K\gtrsim0.1$, accompanied by the ensuing discontinuity in the $St$ vs $Re$ relationship \citep{CaoJFS2010}. This phenomenon has not been observed in experiments and the emergence of longitudinal vortical structures in the upstream flow may be accounted responsible.

The onset of time-dependence $Re^H$ is slightly advanced and the mean drag coefficient $C_d$ and Strouhal number $St$ reduced for the flow past a square cylinder subjected to increasing $K$ \citep{ChengJFS2007,LankadasuIJHFF2008%,ChengPoF2005
}.
%Blockage acts in the opposite direction by \textcolor{red}{delaying} the first instability and increasing $C_d$, while $St$ is mostly unaffected \citep{LankadasuIJHFF2008}.
At these very low values of $Re$, $C_l$ is negative and its modulus increases with increasing $K$ \citep{ChengJFS2007,LankadasuIJHFF2008%,ChengPoF2005
}, but the trend is reversed at higher $Re$ \citep{LankadasuIJNMF2011}.

The decrease in $Re^H$ has been belied by 2D simulations that explored larger values of $K$ and reported suppression of two-dimensional vortex shedding behind a square cylinder at sufficiently low values of $Re\lesssim200$ whenever $K\gtrsim K_c$ exceeds a certain critical value that increases with $Re$ \citep{RayPoF2017,ChengJFS2007}. As for the circular cylinder, an increasing $K$ renders top vortices stronger and rounder, while bottom vortices become weaker and elongated and dissipate fast in the wake \citep{RayPoF2017}. As for the circular cylinder, the stagnation point systematically rises above the cylinder midplane upon increasing $K$ at any value of $Re$ \citep{CaoCaF2014}.

%While most two-dimensional studies agree that $St$ decreases for increasing $K\leq0.2$ for Reynolds numbers $Re\in[Re^H,250]$ \citep{KumarPE2015,RayPoF2017} and even beyond \citep{SahaJEM1999,HwangIJNMF1997}, experiments and three-dimensional numerical simulation seem to indicate that this trend is overcome for higher values of $Re\in[500-1000]$, for which $St$ is unaffected by $K$ \citep{AyukawaJWEIA1993} or marginaly increases \citep{CaoJEM2012}. Meanwhile, both $\langle C_l\rangle$ and $\langle C_d\rangle$ decrease at low $Re\lesssim100$ (the former pointing towards the low-velocity side) and increases instead at high $Re\gtrsim100$ \citep{SohankarJBSMSE2020}, a behaviour that seems to persist at much higher $Re$ for $C_l$ but not for $C_d$ \citep{AyukawaJWEIA1993}, even in two-dimensional simulations $Re\leq1500$ \citep{HwangIJNMF1997}.
%There exist however notable two- and three-dimensional numerical studies that have reported positive growing $C_l$ for increasing $K$ at moderate $Re$ \citep{LankadasuIJNMF2011,SohankarJBSMSE2020}.
%The rate of increase of $\langle C_l\rangle$ with $K$ sharpens as $Re$ gets larger, while $\langle C_d\rangle$ becomes quite independent of $K$ for $Re>200$ \citep{AyukawaJWEIA1993,CaoJEM2012}. $C_l'$ and $C_d'$ fluctuations increase both with $Re$ and with $K$ at low $Re\leq150$ \citep{SohankarJBSMSE2020} and even beyond \citep{LankadasuIJNMF2011}, but are mostly insensitive to $K$ at higher $Re$ \citep{AyukawaJWEIA1993,HwangIJNMF1997}.

In the presence of upstream shear, the wake behind a square cylinder experiences a unique secondary three-dimensionalising instability characterised by a single mode \citep{LankadasuIJNMF2011}, instead of the two modes A and B that are sequentially observed in the wake transition regime for both circular and square cylinders in the absence of upstream shear. This single mode, which arises at $Re\simeq140\sim150$ when $K=0.2$, was initially mistaken for mode B \citep{LankadasuIJNMF2011} because of its similar wavelength, but the solution seems in fact period-doubled. Its three-dimensional structure appears shifted by half a wavelength after every vortex-shedding cycle, so that the instability would plausibly correspond to mode C, which would have taken precedence over modes A and B following the disruption of the spatial Z$_2$ symmetry. At $Re=200$ and $K=0.2$, three-dimensional simulations have shown mode-A- and mode-B- type structures on the low- and high-velocity halves, respectively, of the wake \citep{LankadasuJFS2009}, although nothing similar has been reported elsewhere.

% Square cylinder in homogeneous shear

%%% Our case
Here we choose to submerge the square cylinder in a thin shear layer by replacing the classic upstream homogeneous shear by a piecewise-constant velocity profile with the discontinuity separating the top and bottom homogeneous streamwise velocities precisely located at cylinder mid-height.
%This arrangement has the virtue of allowing for much larger values of the shear parameter without the need of constraining the domain to high blockage ratios in order to avoid negative velocities at the bottom end of the inlet.
This same setup has been used twice before, but the flow was expressly kept two-dimensional \citep{MushyamMec2017,AnMec2020}. In order to simulate experimentally reproducible conditions, we choose to separate the top and bottom homogeneous velocity streams by a flat plate, such that the shear layer results from the reunion, at its trailing edge, of the top and bottom boundary layers and develops smoothly downstream before reaching the cylinder. This kind of upstream conditions may be obtained experimentally in a wind or water tunnel \citep*{LoucksJFM2012}, and the development of the shear layers thus generated have been thoroughly investigated \citep*{RogersJFM1992,MoserJFM1993%,MoserPoFA1991
}.
The length of the plate, the gap left between its trailing edge and the cylinder and the Reynolds number below the plate are kept constant. The top-to-bottom stream velocity ratio $R\equiv U_T/U_B$ has been varied and the various distinct flow topologies that arise classified. The onset and evolution of three-dimensional structures in the cylinder wake are analysed and discussed in detail.

% Structure of the manuscript
The paper is structured as follows. The mathematical modelling is presented in section \S\ref{sec:math} alongside the numerical approach undertaken. \S\ref{sec:upflow} dissects the flow generated by the interaction of the two different-velocity streams as they flow parallel in the splitter plate wake towards the square cylinder. A temporal characterisation of the resulting flow past the cylinder is then given in section \S\ref{sec:temporal}, followed by an analysis of aerodynamic performances trends in section \S\ref{sec:perfo}.
%Section \S\ref{sec:wake} discusses wake topology and dynamics and section \S\ref{sec:BL} the boundary/shear layers on the top and bottom walls of the cylinder.
Finally, the wake transition regime is dissected in section \S\ref{sec:3D} and the main results summarised and conclusions drawn in section \S\ref{sec:conc}.

\section{Mathematical modelling and numerical approach}\label{sec:math}

\begin{figure}
  \centering
  \includegraphics[width=.8\textwidth]{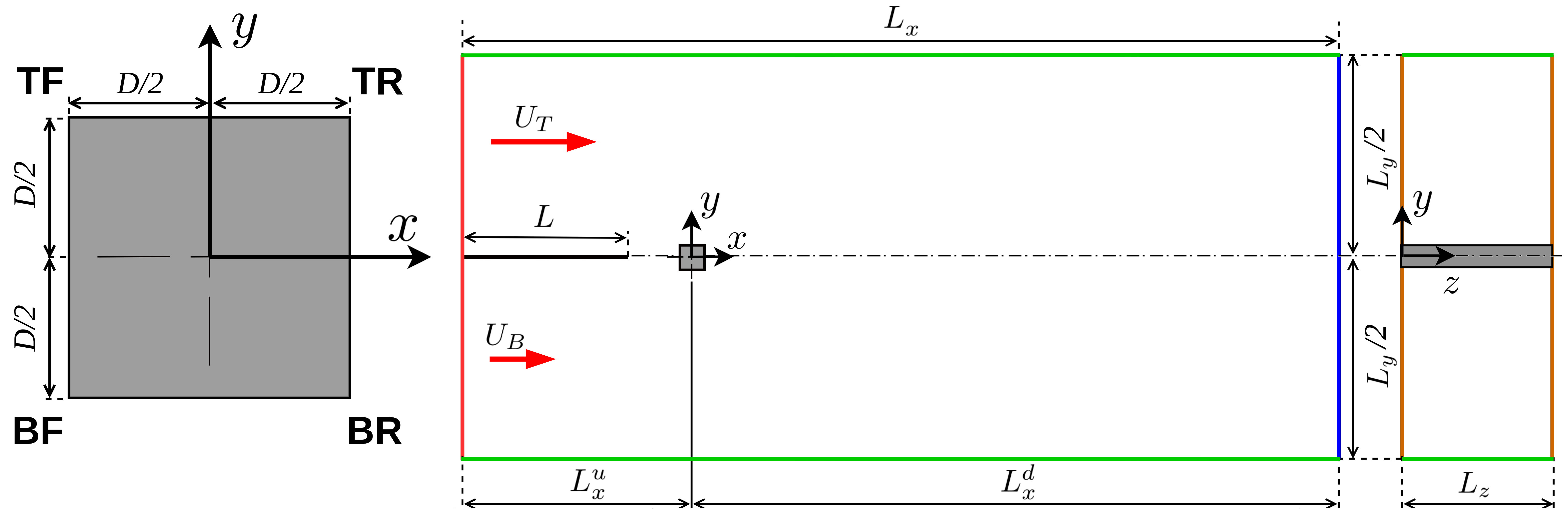}
  \caption{Computational domain. The square cylinder, of side $D$, is located at midheight of the domain, of height $L_y$, at distances $L_x^u$ and $L_x^d$ from the upstream and downstream boundaries, respectively. The cylinder is placed at zero angle of attack and the origin is set at its centre. The flat plate, of negligible thickness and chord $L$, is also at mid-height, horizontal, and starts at the upstream boundary. For three-dimensional simulations the domain is periodic in the spanwise direction with length $L_z$. The colours indicate the boundary condition types: velocity inlet (red, with velocities $U_T$ and $U_B$ for the top and bottom streams, respectively), pressure outlet (blue), slip walls (green), non-slip walls (black) and periodic (orange).
  }\label{fig:domain}
\end{figure}

Figure~\ref{fig:domain} presents the side and front views of the computational domain employed in the simulations.
The square cylinder, of side $D$, is located at the origin of the domain, of size $L_x\times L_y=34.5D\times16D$, at a distance $L_x^u=9D$ downstream from the domain inlet, and at mid-height, such that the top and bottom boundaries are at a distance $L_y/2=8D$ above and below the centreline (blockage ratio $B=0.0625$). All cases were studied with incidence $\alpha=0^\circ$ (all sides are either parallel or orthogonal to the incoming flow direction). The splitter plate, of chord $L=6.5D$ and negligible thickness, extends horizontally from domain inlet at mid-height, thus leaving a $2D$ gap between its trailing edge and the front face of the cylinder.
The small distance left for the development of the shear layer has been checked to be sufficiently short to keep it thin and stable while at the same time long enough for the flow to freely adapt to the presence of the cylinder (see section \ref{sec:upflow}). Meanwhile, the short splitter plate helps smooth the initiation of the shear layer while keeping the boundary layer thin and laminar.
A downstream extent of $L_x^d=L_x-L_x^u=25.5D$ has been allowed to properly capture the wake and avoid interferences of the domain outlet with the flow around the cylinder and in its near-wake. 

The Navier-Stokes equations for incompressible flow, nondimensionalised with the square cylinder side length $D$ and the mean upstream flow velocity $U=(U_T+U_B)/2$ ($U_T$ and $U_B$ are the top and bottom stream constant velocities), read
\begin{equation}\label{eq:NS}
  \begin{array}{rcl}
    \dfrac{\partial \mathbf{u}}{\partial t} + (\mathbf{u} \cdot \nabla) \mathbf{u} & = & -\nabla p + \dfrac{1}{Re} \nabla^{2} \mathbf{u}, \\
    \nabla \cdot {\bf u} & = & 0,\\
  \end{array}
\end{equation}
where $\mathbf{u}(\mathbf{r} ; t)=(u, v, w)$ and $p(\mathbf{r} ; t)$ are the nondimensional velocity and pressure, respectively, at nondimensional location $\mathbf{r}=(x, y, z)$ and advective time $t$ (units of $D/U$). Here, $x$ ($u$), $y$ ($v$), and $z$ ($w$) denote the streamwise, crossflow, and spanwise coordinates (velocity components), respectively.

Neumann boundary conditions for pressure ($\nabla p \cdot \hat{\bf n} = 0$) and Dirichlet for velocity have been prescribed at domain inlet. The velocity profile has been taken as a piecewise-constant function, with ${\bf u} = U_B\, \hat{i} = 2U/(R+1)\, \hat{i}$ below the splitter plate and ${\bf u} = U_T\, \hat{i} = 2U R/(R+1)\, \hat{i}$ above it. No-slip boundary conditions have been enforced on cylinder walls and splitter plate (${\bf u}=0$, $\nabla p \cdot \hat{\bf n} = 0$). The top and bottom domain boundaries have been taken as slip walls with ${\bf u}\cdot \hat{\bf n} = 0$ and $[(\nabla{\bf u})\cdot \hat{\bf n}] \times \hat{\bf n} = {\bf 0}$, and the domain outlet has been set as homogenoeous Dirichlet ($p=0$) for pressure and homogeneous Neumann ($\nabla {\bf u} \cdot \hat{\bf n} = {\bf 0}$) for velocity. Periodic boundary conditions in the spanwise direction have been prescribed for three-dimensional simulations.

The two governing parameters are the Reynolds number and the top-to-bottom stream velocity ratio
\[
Re \equiv \dfrac{U D}{\nu}, \qquad R \equiv \dfrac{U_T}{U_B},
\]
where $\nu$ is the kinematic viscosity of the fluid.

Two separate Reynolds numbers might be defined individually for the top and bottom streams as
\[
Re_T \equiv \dfrac{U_T D}{\nu} = \dfrac{2R}{1+R} Re \qquad \mathrm{and} \qquad Re_B \equiv \dfrac{U_B D}{\nu} = \dfrac{2}{1+R} Re,
\]
and used as an alternative set of governing parameters.

For the somewhat-related problem of a bluff body immersed in upstream shear, the shear parameter is usually defined by nondimensionalising the free-stream velocity gradient as
\[
K \equiv \frac{D}{U}\left(\frac{\partial U}{\partial y}\right)_{\infty}
\]
Here the free stream velocity is a step function and an equivalent shear parameter can be defined with the average free-stream velocity gradient over the characteristic length of the cylinder as $K\equiv2(R-1)/(R+1)$.

%\fer{
As we will see in \S\ref{sec:upflow}, the incoming-flow shear profile is anything but homogeneous, but a parallel can still be drawn on account of the mean shear effects over the cylinder characteristic length, even if shear is concentrated within a thinner layer. This would indicate that part of the phenomenology observed in the presence of homogeneous upstream shear is in fact related to the velocity difference/jump perceived locally by the upper and lower sides of the body under scrutiny, rather than the global effect of the velocity gradient alone.
%}

The bottom stream Reynolds number has been kept fixed to $Re_B=56$ throughout the study, while the top-to-bottom stream velocity ratio has been varied in the range $R\in[1,5.375]$. A $\Delta R=0.2$ step has been used to resolve $R\in[1,3]$ while higher velocity ratios have been tested at a set of discrete values $R\in\{3.1,3.2,3.3,3.4,3.8,4,5.375\}$. The equivalent Reynolds number and shear parameter, which are linked to $R$, have been simultaneously varied in the ranges $Re\in[56,178]$ and $K\in[0,1.37]$.

% Definition of force coefficients

Wall shear ($\tau_w$) and pressure ($p$) have been non-dimensionalised into the friction and pressure coefficients as
\begin{equation}\label{eq:taupcoeff}
  C_f \equiv \dfrac{\tau_w}{\frac{1}{2}\rho U^2} \qquad \mathrm{and} \qquad C_p \equiv \dfrac{p-p_{\infty}} {\frac{1}{2}\rho U^2},
\end{equation}
with $p_{\infty}$ the far field pressure, while their added integral over the cylinder surface, the aerodynamic force, has been projected into the classical lift ($F_l$) and drag ($F_d$) components per unit length and non-dimensionalised into the lift and drag force coefficients following
\begin{equation}\label{eq:forcecoeff}
  C_l \equiv \dfrac{F_l}{\frac{1}{2}\rho D U^2} \qquad \mathrm{and} \qquad C_d \equiv \dfrac{F_d}{\frac{1}{2}\rho D U^2}.
\end{equation}

% Solver

The incompressible Navier-Stokes equations have been evolved in time using Nektar++, an opensource code based on the spectral/hp element method framework \citep{CantwellCPC2015}. This method combines the geometric flexibility of the finite element method with the high-order accuracy of spectral methods. We have employed the velocity correction scheme, which is made to be consistent with its overall temporal accuracy by an appropriate high-order discretisation of the pressure on all boundaries \citep{CantwellCPC2015,MoxeyCPC2020}.

% Mesh

\begin{figure}
  \begin{center}	
    \begin{tabular}{cc}
      (a) & (b)\\
      \includegraphics[height=0.28\columnwidth]{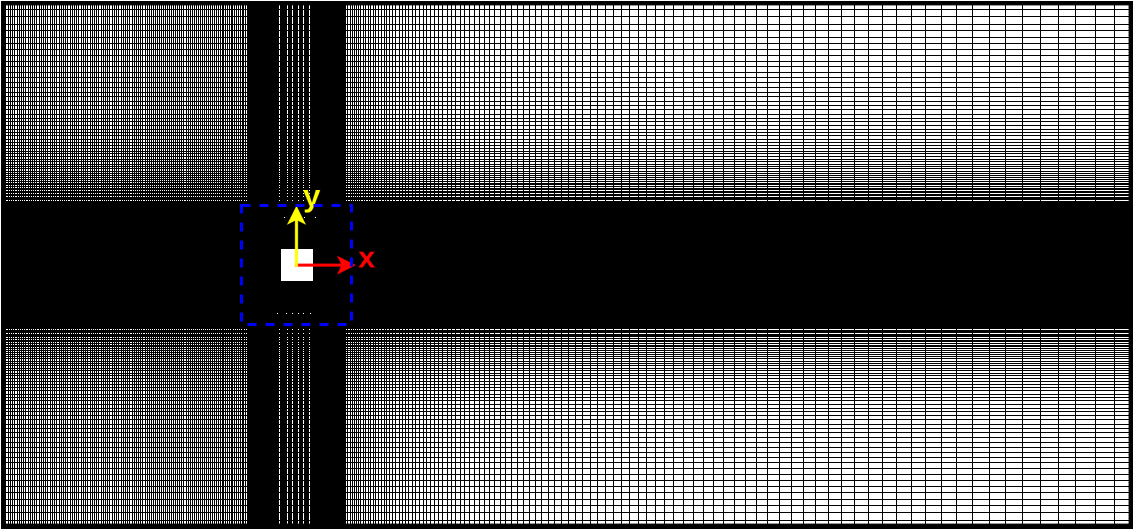} &
      \includegraphics[height=0.28\columnwidth]{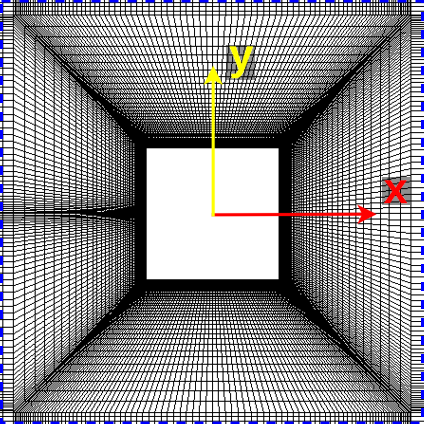}
    \end{tabular}
  \end{center}	
  \caption{Computational mesh. (a) Full domain and (b) detail around the square cylinder.
  }\label{fig:mesh}
\end{figure}

The in-plane two-dimensional structured mesh, consisting exclusively of quad elements, is shown in figure \ref{fig:mesh}.
A particularly high resolution has been employed on all no-slip wall surfaces to properly resolve boundary layers, as well as along the splitter plate and cylinder wakes. Away from these regions, the mesh density has been allowed to relax. Tensor-products of Lagrange polynomial bases of order 6 have been deployed within each quadrilateral element and spanwise Fourier expansions have been used in the homogenous spanwise direction for three-dimensional computations. Polynomial expansions of orders 4 to 8 were initially tested for the highest velocity ratio, but 6 was chosen for the exploration as it provided a fair compromise between accuracy and computational burden.

An in-plane mesh consisting of $N_{xy}=95294$ 6$^{\mathrm{th}}$-order quadrilateral elements has been used for all simulations with velocity ratios in the range $R\in[1,3.4]$, with the time step gradually reducing from $\Delta t=0.0028$ to 0.0012. A second finer mesh with $N_{xy}=125204$ 6$^{\mathrm{th}}$-order quad elements has been employed for $R>3.4$ and the time step reduced from $\Delta t=0.0007$ at $R=3.8$ to $\Delta t=0.0004$ at the highest $R=5.357$. A thorough code validation alongside a resolution study is presented in \ref{app:validation}.

% choice of spanwise domain extent.

The requirements for the spanwise size of the computational domain at any given value of the Reynolds number and velocity ratio can be assessed by determining the range of wavenumbers that are unstable. This can be done with a full-fledged Floquet analysis, but given that only the dominant eignmode for each spanwise wavenumber, rather than the full spectrum (or even a subset of it), is actually required, the stability analysis has been carried out by simply time-stepping on a quasi 3D domain, adding to the two-dimensional field a unique spanwise Fourier mode of the chosen wavelength, randomly initialised to a very low energy level (See appendix \ref{app:StAn}).

Figure~\ref{fig:floquet}a illustrates the time-stepping-based stability analysis for $(R,\lambda_z)=(3.4,2.5)$.
\begin{figure*}
  \centering
  \begin{tabular}{c@{\hskip 0em}cc@{\hskip 0em}c}
    \raisebox{0.31\linewidth}{(a)} &
    \includegraphics[height=0.3\linewidth]{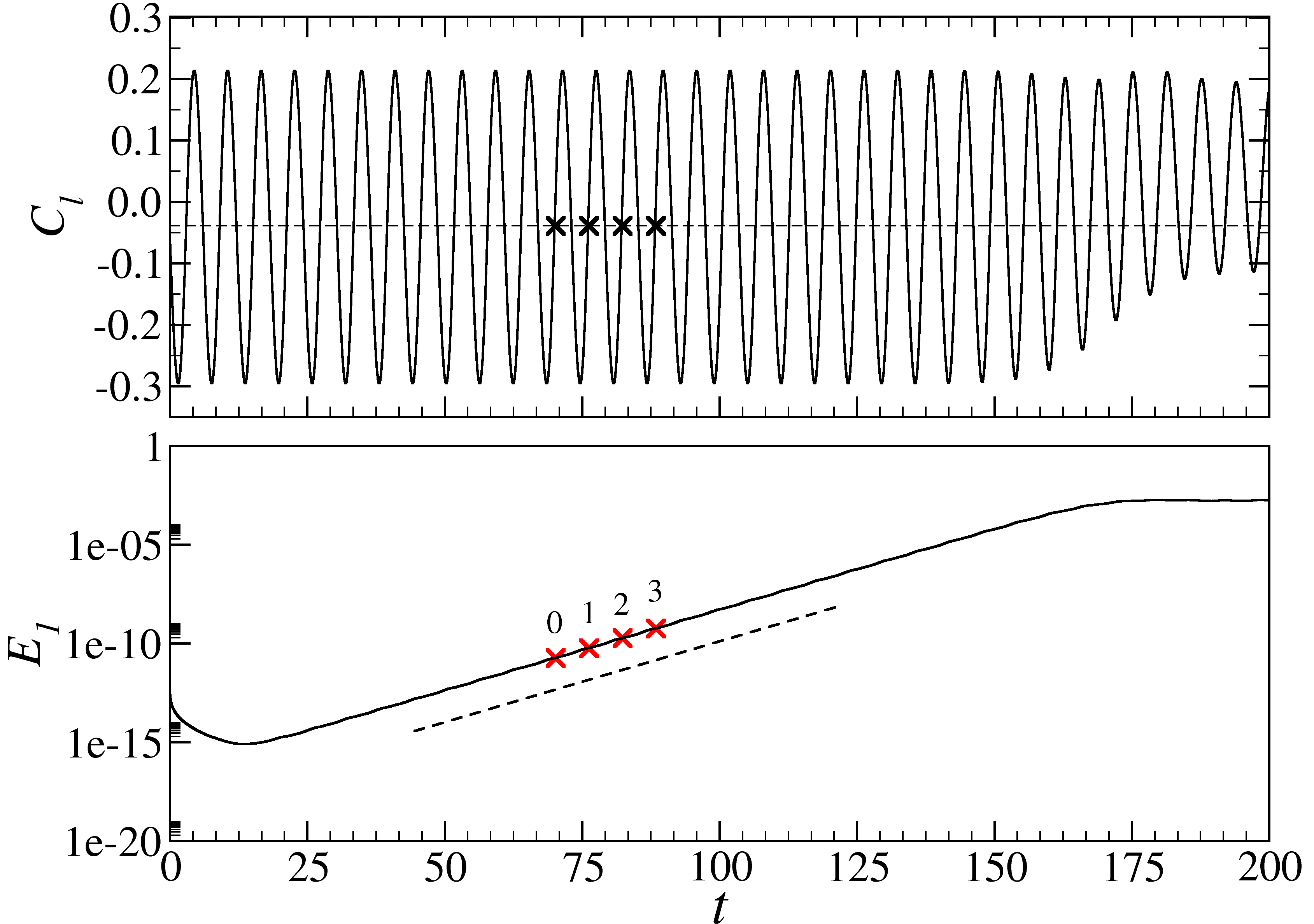} & 
    \raisebox{0.31\linewidth}{(b)} &
    \includegraphics[height=0.3\linewidth]{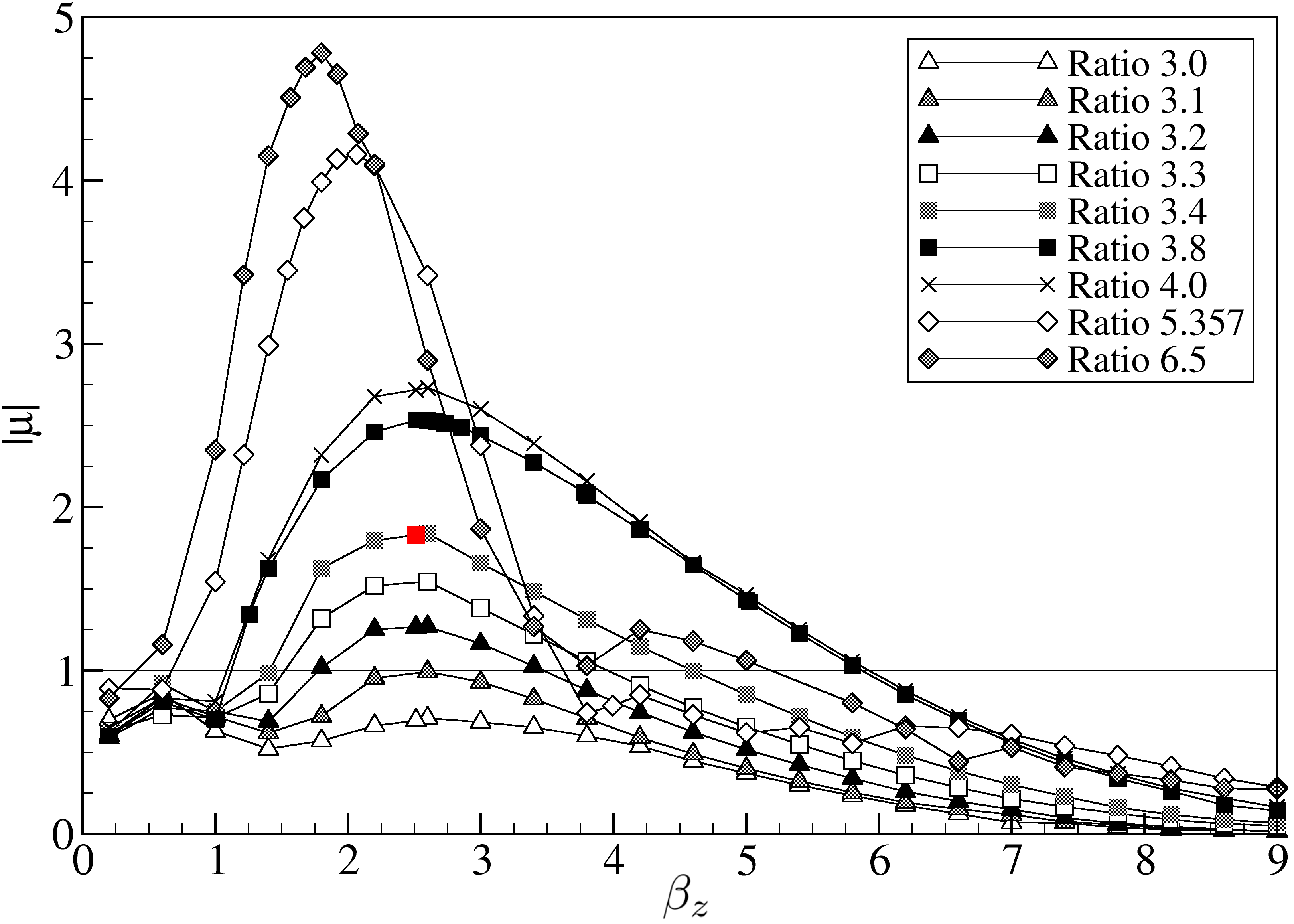}\\
  \end{tabular}
  \caption{Stability analysis of the flow past a square cylinder in the interface of two different-velocity streams with $Re_B=56$. (a) Time-evolution of the lift coefficient $C_l$ (top) and of the modal energy associated to the only spanwise-dependent Fourier mode $E_1$ considered (bottom) for $(R,\lambda_z)=(3.4,2.5)$. The red crosses indicate $\{C_l=\langle C_l\rangle_t; \dot{C}_l>0\}$ Poincar\'e crossings in the linear regime used for the exponential fit (red dashed line). (b) Least stable Floquet multipliers for velocity ratios in the range $R\in[3,6.5]$. The red square corresponds to the case analysed in panel (a).
  }
  \label{fig:floquet}
\end{figure*}
After some initial transients, the energy associated to the only spanwise mode considered  ($E_1(t)$, bottom) starts growing exponentially, with a barely perceptible superimposed oscillation that is inherited from the underlying two-dimensional vortex-shedding periodicity ($C_l$, top). An exponential fit (dashed line) to the Poincar\'e crossings in the linear growth regime (red crosses) provides an estimation of the modulus of the unstable multiplier $|\mu|$.

The results of the stability analysis for varying $R$ and $\lambda_z$ are presented in figure~\ref{fig:floquet}b. The actual onset of flow three-dimensionality occurs for $R\simeq3.1$. Below this value, all three-dimensional perturbations decay (the Floquet multiplier has $|\mu|<1$) and the flow remains two-dimensional. The largest modulus Floquet multiplier corresponds to a fairly constant spanwise wavenumber of about $\lambda_z =2\pi/\beta_z\simeq2.4$ all the way up to $R\leq4$ and then steadily increases to $2.8$ for $R=6.5$. Consequently, a spanwise domain size $L_z=2.5$ (and integer multiples of it) constitutes a fair choice if the fastest-growing three-dimensional mode is to be represented in the simulation. The smooth evolution of the spectrum indicates that, despite its evolution in wavelength, the same mode is responsible for the instability over the whole range of $R$ explored. At $R=6.5$, however, a second unstable mode, of much shorter associated wavelength $\lambda_z\simeq1.5$ has destabilised. The first mode is clearly dominant, but the presence of this second mode may play some role in shaping the small-scale vortical sturcutres that are present in the wake once fully developed turbulence has kicked in.

In accordance with the Floquet stability analysis, spanwise lengths $L_z=5$ have been employed in most three-dimensional computations, deploying resolutions in the range $N_z\in[28,40]$ to guarantee six orders of magnitude energy decay in the spanwise Fourier spectrum. A few cases with $L_z=10$ and $N_z=80$ have been run at the highest explored values of $R$ to confirm that the observed solution features are not an artifact of overly limited domain size.

Table~\ref{tab:CompDetails} summarises the space and time discretisations used for the several computations run in the present study.
\begin{table}
  \centering
  \begin{tabular}{r|rrrrrrr}
    \multicolumn{1}{c|}{$R$} & \multicolumn{1}{c}{Mesh} & \multicolumn{1}{c}{$L_z$} & \multicolumn{1}{c}{$N_{xy}$} & \multicolumn{1}{c}{$N_z$} & \multicolumn{1}{c}{$N=N_{xy}\times N_z$} & \multicolumn{1}{c}{$L_z/N_z$} & \multicolumn{1}{c}{$\Delta t$}   \\\hline
    %         1-3.1 & Mesh B &  \multicolumn{1}{c}{-} &  95294 &  \multicolumn{1}{c}{-} &                     \multicolumn{1}{c}{-} &     \multicolumn{1}{c}{-} & 0.0028-0.0014 \\
    1     & Mesh B &  \multicolumn{1}{c}{-} &  95294 &  \multicolumn{1}{c}{-} &                     \multicolumn{1}{c}{-} &     \multicolumn{1}{c}{-} & 0.0028 \\[-0.5em]
    $\vdots$   &  &  &  &  &  &  & $\vdots$ \\
    3.1   & Mesh B &  \multicolumn{1}{c}{-} &  95294 &  \multicolumn{1}{c}{-} &                     \multicolumn{1}{c}{-} &     \multicolumn{1}{c}{-} & 0.0014 \\
    3.4   & Mesh B &  5 &  95294 & 28 &  $2.668 \times10^{6}$ & 0.179 & 0.0012        \\
    3.8   & Mesh C &  5 & 125204 & 34 &  $4.256 \times10^{6}$ & 0.147 & 0.0007        \\
    4.0   & Mesh C &  5 & 125204 & 36 &  $4.507 \times10^{6}$ & 0.138 & 0.0006        \\
    5.357 & Mesh C & 10 & 125204 & 80 & $10.016 \times10^{6}$ & 0.125 & 0.0004        \\
    %%7.143 &Mesh C&4&125204&64&8.013&0.063\\
  \end{tabular}
  \caption{Summary of the space and time discretisations used in the present computations.}
  \label{tab:CompDetails}
\end{table}

For completeness, the code and mesh validation are presented in full in appendix~\ref{app:validation}.

\section{Incoming upstream flow}\label{sec:upflow}

The initially flat interface of two parallel viscous streams flowing at different speeds is not stable. The step profile of purely streamwise velocity at best diffuses into an increasingly wide shear layer or, more generally, destabilises following a Kelvin-Helmholtz instability. The aim here is to subject the square cylinder to as close a step profile as experimentally feasible by keeping the distance over which the two streams interact to a minimum while still allowing the upstream flow to adapt to the incoming obstacle. In order to smooth the development of the shear layer, the two streams have been allowed to flow over a flat plate for some distance before meeting, sufficiently long for a Blasius boundary layer to develop but at the same time short enough for it to remain thin and laminar.

\begin{figure}
  \centering
  \includegraphics[width=0.9\linewidth]{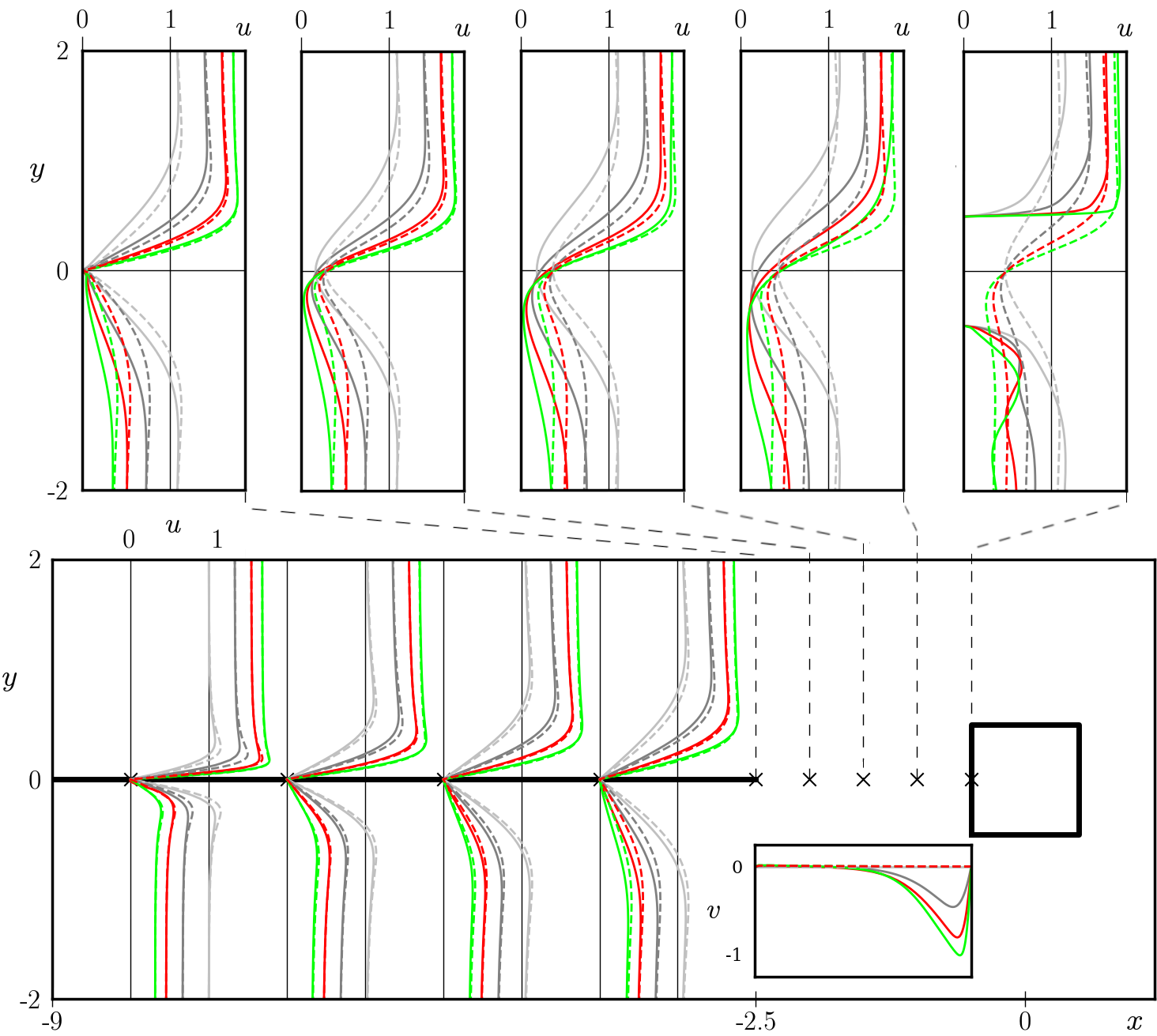}\\
%  \begin{tabular}{cc}
%    \raisebox{0.25\linewidth}{(a)} &
%    \includegraphics[width=0.8\linewidth]{plate_all.png}\\
%    \raisebox{0.22\linewidth}{(b)} &
%    \includegraphics[width=0.8\linewidth]{U_profile.png}\\
%    \raisebox{0.25\linewidth}{(c)} &
%    \includegraphics[width=0.8\linewidth]{V_profile.png}\\
%  \end{tabular}
  \caption{Span- and time-averaged characterisation of the incoming flow as it develops on the splitter plate and downstream along the near wake, covering the gap with the square cylinder. Shown are Cross-stream profiles of streamwise velocity over the flat plate (main panel) the plate-cylinder gap (top panels), along with centre-line distribution of cross-stream velocity in the gap (inset). Velocity ratios $R=1$ (light gray), 2 (dark gray), 3.4 (red), and 5.357 (green) have been represented both in the presence (solid) and absence (dashed) of the cylinder.
  }
  \label{fig:velocity_profiles}
\end{figure}

Figure~\ref{fig:velocity_profiles} depicts cross-stream profiles of streamwise velocity as the flow develops over the splitter plate and downstream along its wake, which constitutes a shear layer for $R>1$.
The flow is essentially two-dimensional and steady in this region for all cases considered, but spanwise and time averaging has been applied nonetheless.

In the absence of the square cylinder (dashed lines), laminar boundary layers of the Blasius type naturally develop on both surfaces of the splitter plate. For $R=1$ (light gray) top and bottom boundary layers are symmetric and generate a top-down symmetric velocity profile in the wake that diffuses gradually as the centre-line velocity, which coincides with the streamline issued from the trailing edge of the flat plate, progressively recovers. The asymmetry is already evident for $R=2$ (dark gray) and becomes increasingly marked for $R=3.4$ (red) and 5.357 (green), but the centreline velocity defect is recovered all the same as the velocity profile adopts a smoothed step-like shape that connects the top- and bottom-stream velocities fairly linearly over about half the square cylinder height by the time the flow reaches the (absent) cylinder location. The higher the value of $R$, the faster and greater the velocity defect recovery with respect to the slow stream, such that the negative shear associated to the lower side of the shear layer becomes barely detectable. The resulting region of roughly homogeneous shear is noticeably shifted towards the high velocity side. Meanwhile, the trailing-edge streamline remains horizontal independently of $R$, as ascertained by the zero cross-stream velocity distribution along the wake centre line in the inset of figure~\ref{fig:velocity_profiles}. None of the cases run for the splitter plate alone resulted in any instability of its wake or trailing shear layer, such that the flow fields remained two-dimensional and steady.

We will be placing the square cylinder at a downstream distance from the splitter plate trailing edge where the unperturbed flow is characterised by a parallel profile of purely streamiwse velocity that is neither linear nor a step function, but can be assimilated to a step function smoothed over about half the cylinder height. Anyhow, whatever the profile looks like across the cylinder height, it can be considered as fairly flat above and below the cross-stream locations where the top and bottom surface of the cylinder will be, with respective velocities those of the fast and slow streams. The streamwise velocity can therefore be taken as approximating to some extent the intended step profile.

The presence of the cylinder (solid lines), introduces a blockage that obviously affects the incoming flow. The boundary layers on the top and bottom surfaces of the flat plate are somewhat thickened, particularly on the low velocity side and for low values of $R$, albeit only slightly. The massflow blockage becomes all the more prominent as we dive into the plate wake. The shear layer remains fairly thin while the presence of the cylinder is still not decisively felt, but the streamwise velocity recovery along the wake is hindered and the lowest velocity is shifted towards the slow stream as $R$ is increased. The thickening of the shear layer as the cylinder is approached and its downwards bias results in something like a (not-quite) homogeneous shear profile that spreads over the cylinder height, but still flattens to about constant velocity above and below the cylinder. In addition, a net downwards cross-stream velocity builds up along the wake that transfers massflow from the high velocity stream to the low velocity stream. An increasingly larger portion of the fast stream flies below the cylinder as the velocity ratio is increased.

\section{Temporal characterisation of the flow}\label{sec:temporal}

\begin{figure}
  \centering
  \includegraphics[width=0.875\textwidth]{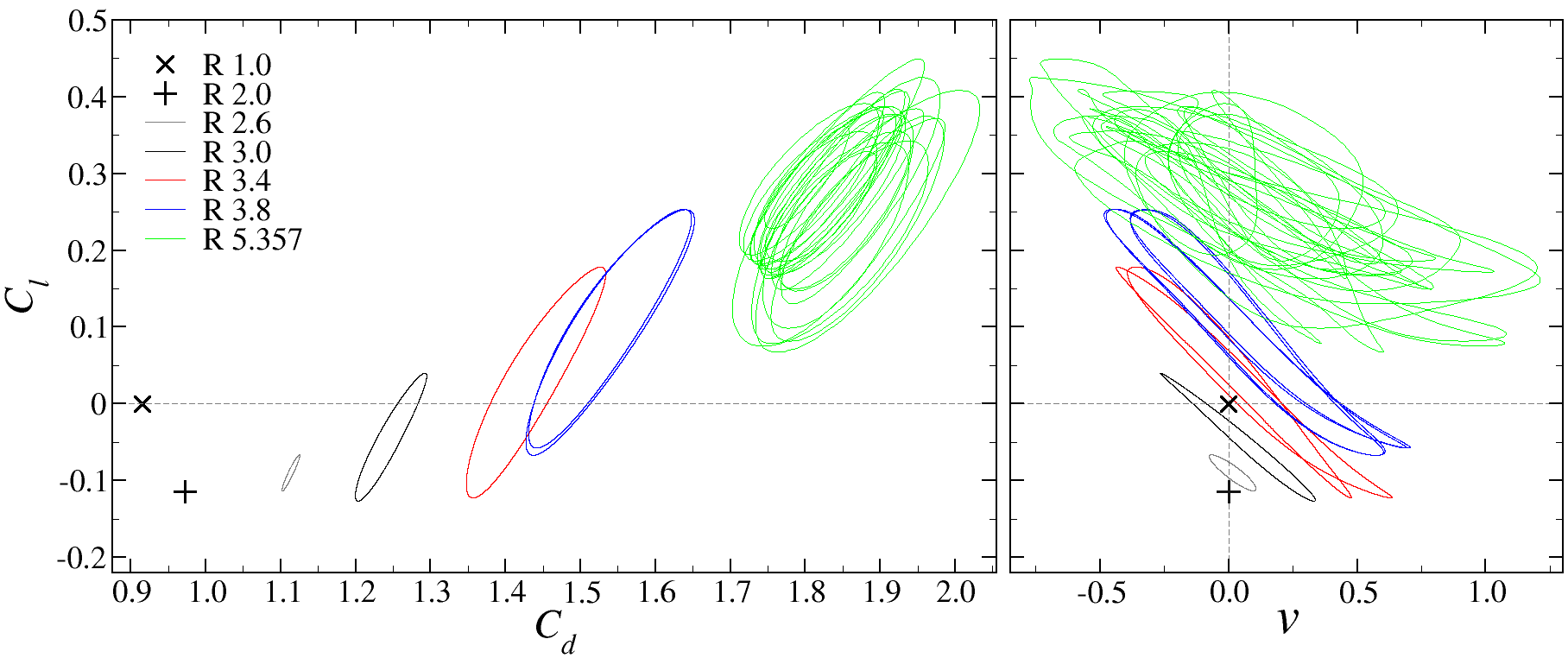}
  \caption{Phase maps projections on the $(C_l,C_d)$ and $(C_l,v)$, with $v$ the cross-stream velocity at $(x,y,z)=(2,0,z_m)$, where $z_m$ has been chosen in each case to pick the largest amplitude in the evolution of $v$ for consistency.
  }
  \label{fig:PhaseMaps}
\end{figure}

\begin{figure}
  \setlength\tabcolsep{1pt}% see you need to put a definition at the beginning of the document
  \begin{tabularx}{\textwidth}{lc}
    \raisebox{2.7cm}{(a)} & \includegraphics[ width=0.9\linewidth, height=\linewidth, keepaspectratio]{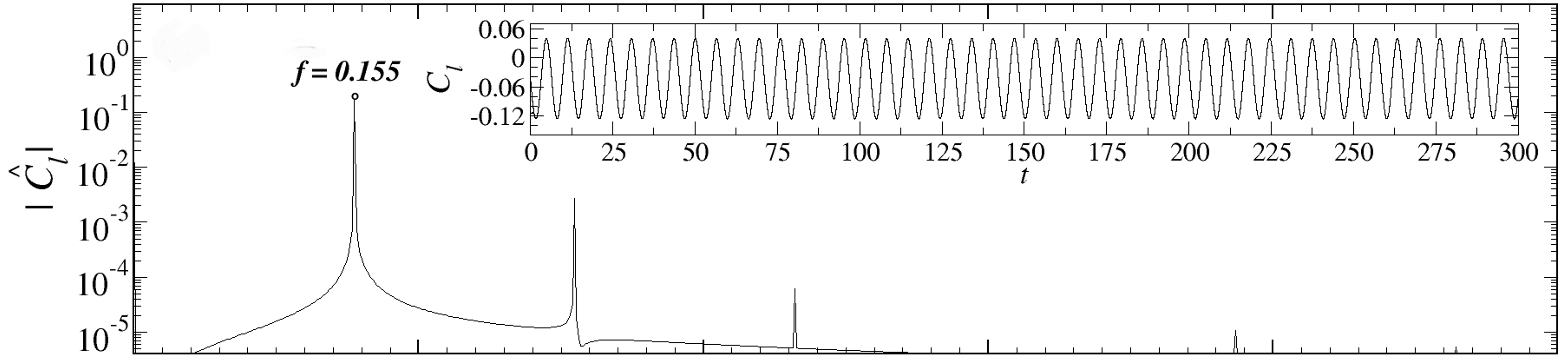}\\
    \raisebox{2.7cm}{(b)} & \includegraphics[ width=0.9\linewidth, height=\linewidth, keepaspectratio]{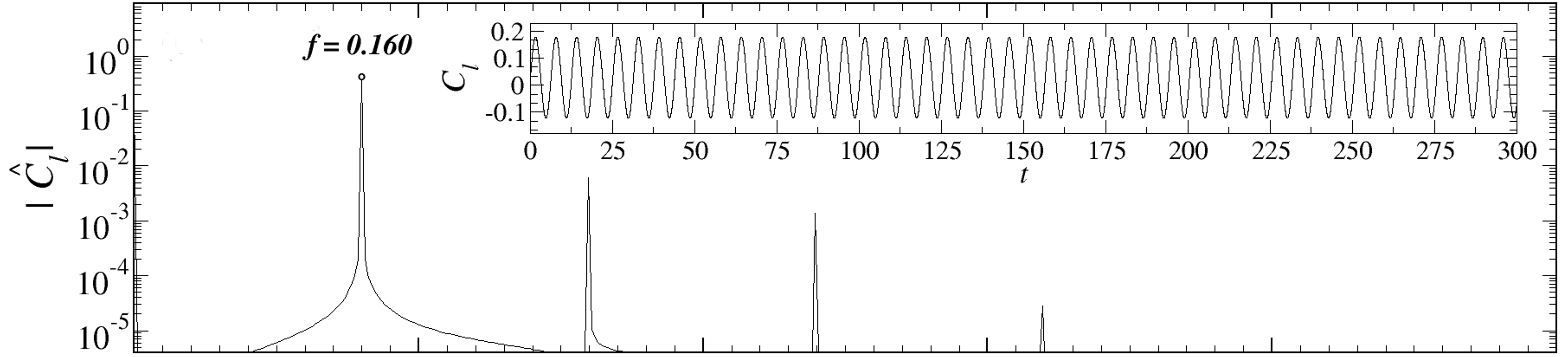}\\
    \raisebox{2.7cm}{(c)} & \includegraphics[ width=0.9\linewidth, height=\linewidth, keepaspectratio]{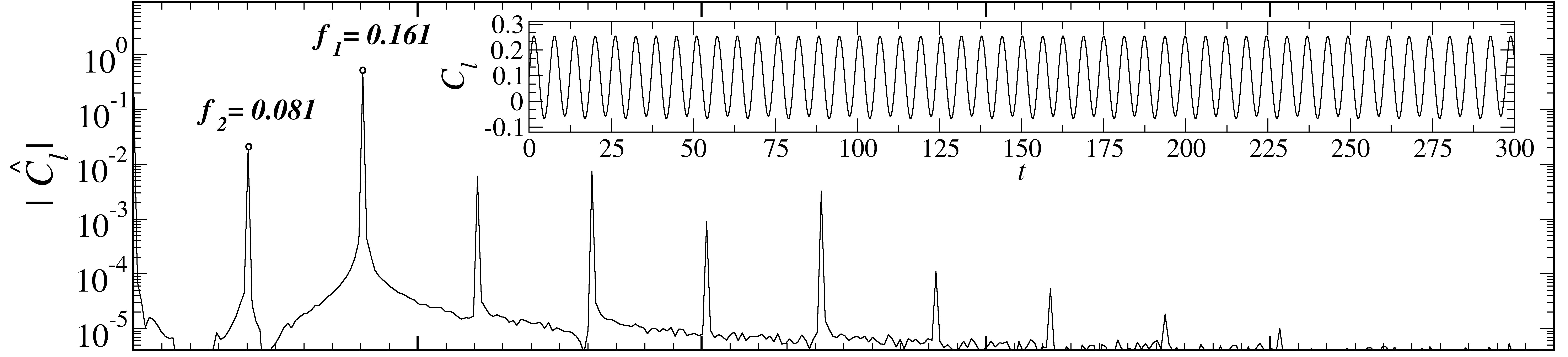}\\
    \raisebox{3.3cm}{(d)} & \includegraphics[ width=0.9\linewidth, height=\linewidth, keepaspectratio]{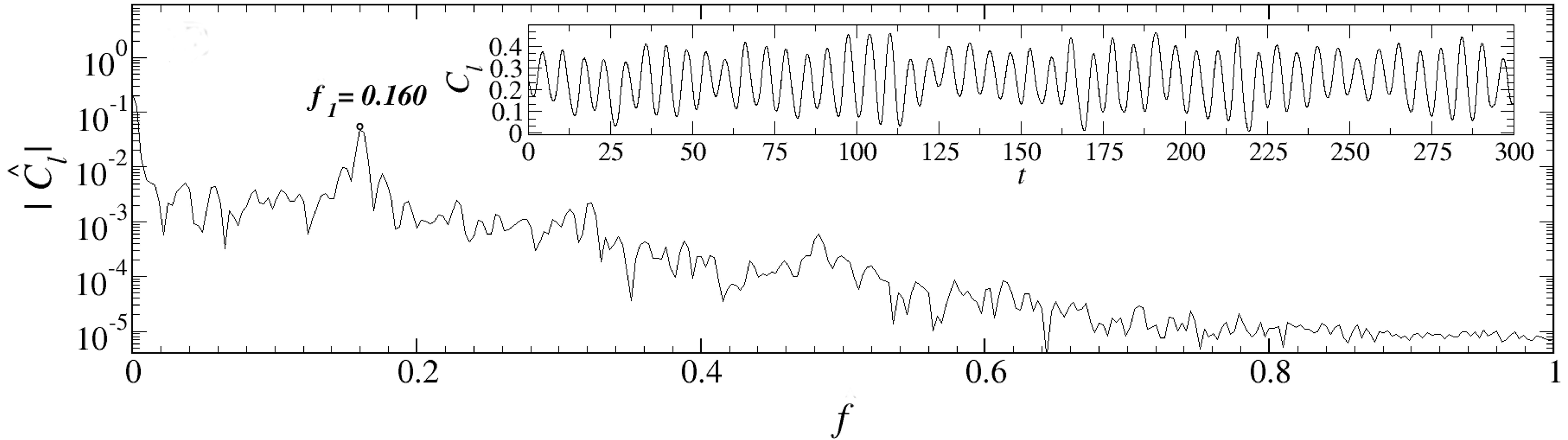}\\
  \end{tabularx}
  \caption{$|\hat{C}_l|$ Power spectra of the lift coefficient for different velocity ratios. The insets shows the time series of drag coefficient versus advective time and the phase diagram for drag and lift coefficients. (a) $R=3$, (b) $R=3.4$, (c) $R=3.8$, (d) $R=5.357$.}
  \label{fig:15_power_spectrum_2D_3D}
\end{figure}

The starting case $(Re_B,R)=(56,1)$, which roughly corresponds to the homogeneous flow past a square cylinder at $Re=56$, is steady and reflection symmetric with respect to the horizontal mid-plane ($y=0$). The presence of the splitter plate reduces the effective incoming velocity through viscous effects and, consequently, has a stabilising effect on the flow past the square cylinder, which is otherwise expected to display periodic vortex-shedding already at this $Re$ \citep{NorbergUnpublished1996}. The flow remains steady but the symmetry is broken as the velocity ratio is increased at constant $Re_B=56$ all the way up to $R\leq2$. This is illustrated by the point phase-map projections for $R=1$ (cross) and $R=2$ (plus sign) in figure~\ref{fig:PhaseMaps}.
The symmetric case $R=1$ is characterised by $C_l=v=0$, while $R=2$ has evidently lost the symmetry (also $v\neq0$, but small enough to be imperceptible to the naked eye). Soon after, a Hopf bifurcation somewhere in the range $R^{\mathrm H}\in(2,2.2)$ introduces time-dependence, such that solutions are characterised by asymmetric time-periodic vortex shedding. The periodicity shows up in the phase-map projections of figure~\ref{fig:PhaseMaps} as closed loops, and the oscillation amplitude grows fast as $R$ is increased. The onset of time dependence occurs for a bulk Reynolds number and shear parameter $(Re^H,K^H)\simeq(87\pm3,0.71\pm0.04)$, way larger than the critical threshold for the square cylinder in homogeneous flow, which undergoes a Hopf bifurcation at $Re^H\simeq45\sim53$ \citep{KelkarIJNMF1992,SohankarIJNMF1998,LankadasuIJHFF2008%,SohankarJWEIA1997
}.
%The critical Reynolds number for the onset of time dependence is reported as $Re^H=47\pm2$, $46\pm1$, $50$ or $53$, according to the several sources cited.
The critical threshold found here corresponds to fairly high $K$ and moderately low $Re$, which is consistent with previous observation that vortex-shedding might be suppressed/delayed by upstream shear \citep{ChengJFS2007,RayPoF2017}. Upstream shear has therefore a stabilising effect, as previously reported also for the circular cylinder \citep{KiyaJFM1980,TamuraJSME1980}, although some studies attest to the contrary both for circular \citep{ParkPoF2018} and square cylinders \citep{LankadasuIJHFF2008}. Figure~\ref{fig:15_power_spectrum_2D_3D}a shows the spectrum ($|\hat{C}_l|$) of the lift coefficient signal ($C_l(t)$, portrayed in the inset) for $R=3$ (black line in figure~\ref{fig:PhaseMaps}).
A clear peak is clearly identifiable at $f_1=0.155$, along with a few harmonics of decaying amplitude at integer multiples of the main frequency. The solution remains periodic and the fundamental peak and harmonics move to higher $f_1$ as the velocitiy ratio is increased to $R=3.4$ (figure~\ref{fig:15_power_spectrum_2D_3D}b). Besides the slight drift of the Strouhal frequency, an essential alteration has occurred that is not pereceptible in the $C_l$ spectrum, nor in the $(C_l,C_d)$ phase-map projection for that matter, but becomes obvious instead from the $(C_l,v)$ phase-map projection (red line in figure~\ref{fig:PhaseMaps}). The actual period of the solution is not that of the $C_l$ or $C_d$ signals but double, and only shows in time-series of local quantities such as point velocity probe readings. The period doubling is related to the three-dimensionalisation of the flow, which has earlier been shown to occur at $R\simeq3.1$, correponding to $(Re,K)\simeq(115,1.02)$. The spanwise invariance of the two-dimensional vortex-shedding flow is disrupted in a way that preserves a triad of remaining spanwise symmetries (spanwise reflection, half period evolution followed by spanwise reflection and half period evolution followed by half spanwise wavelength shift) that will be discussed later. All three symmetries retain the original period for all variables that aggregate/average over the spanwise direction, as is the case of force coefficents, but the actual invariance after a full vortex-shedding cycle requires the further composition with a space operation and the original solution is only fully recovered after a second vortex-shedding cycle, which makes the solution qualify as period-doubled.

At $R=3.8$, a second period doubling of an altogehter different nature has occurred. Now the bifurcation affects not only local variables but also aggregates, as clearly shown by the double loop in the $(C_l,C_d)$ and the four-fold loop in the $(C_l,v)$ phase-map projections (blue line in figure~\ref{fig:PhaseMaps}). As a matter of fact, local variables repeat only after four vortex-shedding cycles. The fundamental peak $f_1$ in the $C_l$ spectrum is still dominant in figure~\ref{fig:15_power_spectrum_2D_3D}c, but a subharmonic peak at the exact half frequency $f_{1/2}=f_1/2$ has arisen along with its harmonics. The flow topology only repeats every four vortex-shedding cycles, but a space-time symmetry operation still exists that renders the flow invariant after every two vortex-shedding cycles (half a period) provided an appropriate spanwise reflection is also applied.

At $R=5.357$ the flow has become temporally chaotic (green line in figure~\ref{fig:PhaseMaps}) and though the main fundamental peak and its first few harmonics are still discernible in the spectrum of figure~\ref{fig:15_power_spectrum_2D_3D}d, they are accompanied by high-energy broadband noise. The second period-doubling bifurcation identified here is suggestive of a period-doubling cascade as the origin of chaotic dynamics, although this cannot be concluded from the rather coarse discretisation of parameter space investigated in the present study.

\section{Aerodynamic performances}\label{sec:perfo}

\begin{figure}
  \setlength\tabcolsep{0.01pt}% see you need to put a definition at the beginning of the document
  \begin{tabularx}{\textwidth}{lc}
    \raisebox{2.7cm}{(a)} & \includegraphics[width=\linewidth,height=0.9\linewidth, keepaspectratio]{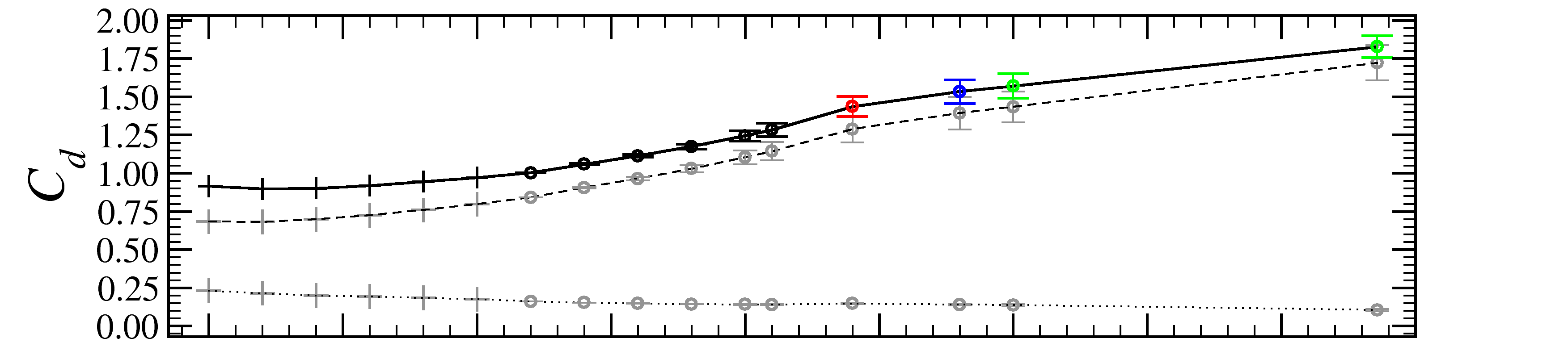}\\
    \raisebox{2.7cm}{(b)} & \includegraphics[width=\linewidth,height=0.9\linewidth, keepaspectratio]{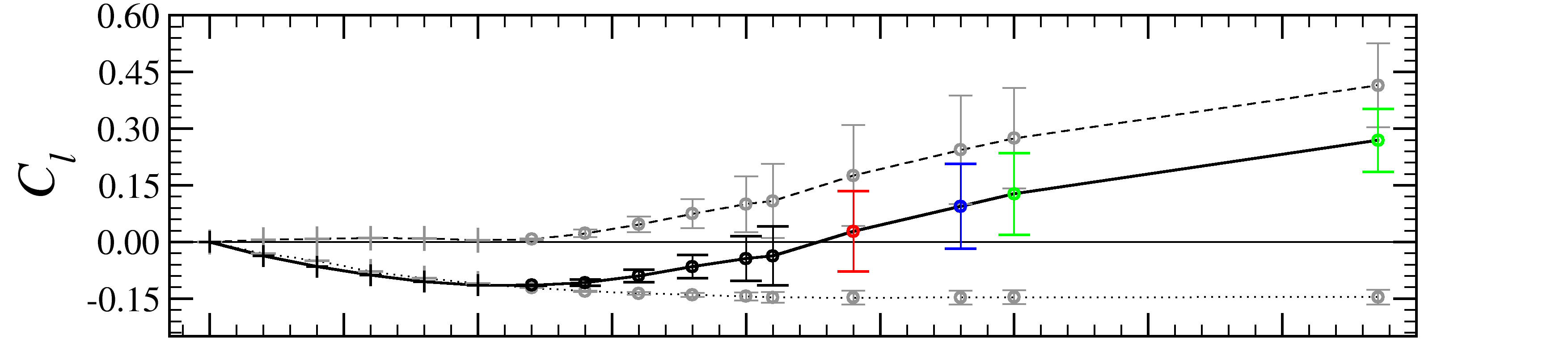}\\
    \raisebox{2.7cm}{(c)} & \includegraphics[width=\linewidth,height=0.9\linewidth, keepaspectratio]{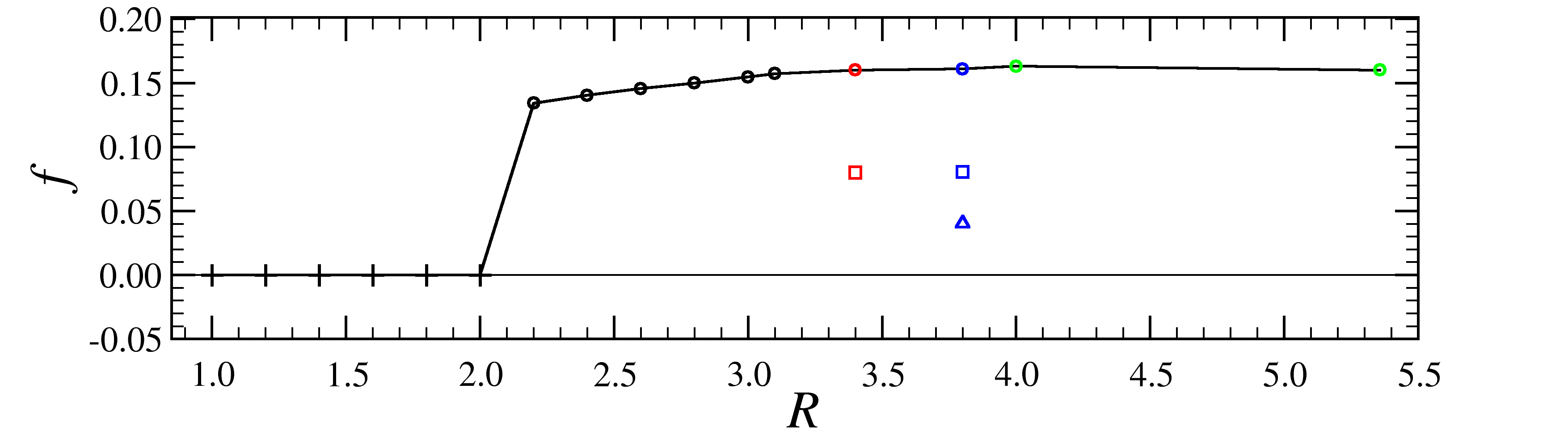}
  \end{tabularx}
  \caption{Performance parameters as a function of the velocity ratio $R$ at constant bottom Reynolds number $Re_B=56$. (a) Drag coefficient $C_d$. (b) Lift coefficient $C_l$. (c) Strouhal number $f$ and, for period-doubled solutions, subharmonic frequencies (squares for the half and triangle for the quarter frequency). Different symbols denote steady (plus signs) and unsteady (circles) solutions. Colour coding separates 2D (black), 3D periodic (red), 3D period-doubled (blue) and chaotic (green) solutions. Total force coefficient trends (solid lines) are split into their pressure (dashed) and friction (dotted) components. Error bars denote oscillation amplitude.)
  }
  \label{fig:ClCdf}
\end{figure}

The evolution of the drag ($C_d$) and lift ($C_l$) coefficients as the velocity ratio $R$ is increased at a constant bottom Reynolds number $Re_B=56$ is presented in figures~\ref{fig:ClCdf}a and b, respectively.
The oscillation amplitude is indicated by error bars for unsteady regimes, and different symbols and colours denote different types of solutions (plus sign: steady; circle: unsteady; black: 2D periodic; red: 3D periodic; blue: 3D period-doubled; green: chaotic). After a barely perceptible decrease, $C_d$ starts growing steadily as $R$ is increased through the various flow regimes. This dependence is consistent with the combined effect of increasing $K$, which tends to reduce $C_d$ \citep{SahaJEM1999,LeiOE2000,ChengJFS2007%,ChengPoF2005
}, and also increasing $Re$, which makes it grow fast \citep{DavisJFM1982,FrankeJWEIA1990,SahaIJHFF2003,MahirIJTS2017}. The combined effect of both parameters shows that the dependence of $C_d$ on $K$ changes from a declining trend at low $Re\lesssim100$ to increasing beyond this value \citep{SohankarJBSMSE2020}. Some simulations, both two- and three-dimensional, at very low $Re<200$ and $K<0.5$ predict a decreasing trend of $C_d$ \citep{ChengJFS2007%,CaoJEM2012
} also with $Re$. The additive decomposition of $C_d=C_d^p+C_d^f$ into its pressure ($C_d^p$, dashed) and friction ($C_d^f$, dotted) components, reveals how pressure forces dominate over viscous forces across the full regime, as is typical of bluff body aerodynamics. As a matter of fact, $C_d^f$ declines with $R$, so that $C_d^p$ takes a larger and larger fraction of the total $C_d$. For circular cylinders, $C_d^p$ has been shown to take a constant proportion of total $C_d$ independently of $K$, but this is only valid for constant $Re$ and the trend expected for rising $Re$ is actually increasing \citep{TamuraJSME1980,LeiOE2000}. The initial decline of $C_l$ is much more pronounced and results in a clear downforce as $R$ is increased from symmetric. This follows from the fast drop of $C_l^f$ for low $R$, with friction downforce taking the lead over pressure lift, which remains negligible while the symmetry disruption is moderately low. The decreasing trend is reversed as the Hopf bifurcation renders the flow time-periodic at $R\in(2,2.2)$ and $C_l^p$ starts growing fast, but actual postitive average lift is not achieved until the flow has become three-dimensional for $R\in(3.1,3.4)$. Prior to that, positive lift is recovered over some parts of the vortex-shedding cycle, even if on average it remains negative. Once the initial reduction is overcome, the increasing trend is sustained over the full range of $R$ explored because while friction downforce $C_l^f$ saturates, pressure lift $C_l^p$ keeps growing. Net downforce has been reported for the square cylinder in homogenous shear at $Re\leq100$ and moderate $K\leq0.5$ \citep{ChengJFS2007,LankadasuIJHFF2008%,ChengPoF2005
}, while positive lift is recovered for $Re\geq150$ \citep{LankadasuIJNMF2011,SohankarJBSMSE2020%,CaoJEM2012
}, in agreement with present results. As for $C_d$, the $C_l$ trend with $K$ is a decreasing one for $Re\lesssim120$ and a growing one above this threshold \citep{SohankarJBSMSE2020}. The oscillation amplitude increases fast after the initial Hopf bifurcation for both $C_d$ and $C_l$, as expected from 2D numerical literature results \citep{SohankarJBSMSE2020,SahaJEM1999}, but it almost saturates by the time three-dimensionality has kicked in. The escalating fluctuations that follow from increasing $Re$ might be partially or totally compensated by a decline associated to increasing $K$ \citep{LankadasuIJNMF2011}. Most of the force oscillation can be ascribed to presure fluctuation, as friction behaves rather stably.

The evolution of the Strouhal number (vortex-shedding frequency $f$) across the full exploration, measured at several locations in the near wake at $(x,y,z)=(2.5,0,0), (2.5,0,2.5)$ and $(3.5,0,0)$, is shown in figure~\ref{fig:ClCdf}c. After the debut of vortex-shedding at the Hopf bifurcation, the frequency drifts to slightly higher values but remains mostly unaltered thereafter. The strouhal number $f$ is expected to grow fast with $Re$ while vortex-shedding remains two-dimensional and to stagnate for a while thereafter across the wake transition regime before starting a slow decline \citep{OkajimaJFM1982,DavisJFM1982,KelkarIJNMF1992,NorbergJWEIA1993,SohankarPoF1999,SahaIJHFF2003}. The trend with $K$ is a slightly decreasing one, at least for sufficiently low $Re<200$ \citep{AyukawaJWEIA1993,SahaJEM1999,ChengJFS2007,KumarPE2015%,CaoJEM2012
}, so that it cannot be discarded that the combined effect produces the slowly increasing and then saturating behaviour observed here.
%A similar trend has been observed experimentally for the planar shear flow past a circular cylinder \citep{kwon1992experimental}.

The onset of three-dimensionality, introduces a new frequency (see figure~\ref{fig:ClCdf}c) that is exactly half the Strouhal number (squares), and the ensuing period-doubling adds yet a third quarter-frequency to the solution (triangle).

\section{Three-dimensionalisation of the flow: Wake transition regime}
\label{sec:3D}

%%% Symmetries

The reflection symmetry about the horizontal midplane is broken by setting $R\neq1$. In these conditions, the only remaining symmetries of the problem concern the spanwise direction. The problem is thus {\it only} invariant under the orthogonal group O(2)$=$SO(2)$\times$Z$_2$ symmetry, involving both arbitrary spanwise shifts (SO(2)) and mirror reflections about all planes orthogonal to the spanwise direction (Z$_2$). As long as the solutions remain two-dimensional, spanwise invarariance is preserved, and the onset of time dependence is a matter of no moment.

The bifurcation that triggers three-dimensionality, though, necessarily breaks the translational invariance SO(2) and restricts the mirror symmetry Z$_2$ to a collection of $z$-planes equispaced at intervals $\lambda_z/2$, where $\lambda_z$ is the fundamental spanwise wavelength of the arising three-dimensional solutions. Two-dimensional periodic solutions may still destabilise following both synchronous or quasiperiodic bifurcations, as was the case in the presence of the space-time Z$_2$ symmetry \citep{MarquesPhysD2004,BlackburnJFM2005}, but none of the two possible types of synchronous bifurcations can preserve a symmetry that was not there in the first place. If the quasiperiodic bifurcation becomes resonant and turns into a period-doubling bifurcation, the original time-periodicity is broken but still retained in the form of a space-time symmetry that recovers the solution after evolution over the original period (half the new period) followed by reflection about any $z$-plane located halfway between consecutive mirror-symmetry planes. In this guise, the period of the three-dimensional solution is double that of the destabilising two-dimensional solution, but shall appear to be the same whenever aggregate quantities are monitored. The phase maps in figure~\ref{fig:PhaseMaps} suggest that the advent of three-dimensionality follows precisely this path, as has been shown to occur for other systems breaking the space-time symmetry of the two-dimensional solution \citep{BlackburnPoF2010} such as circular rings \citep{SheardPoF2005,SheardJFM2005} or inclined cylinders \citep{SheardJFM2009,SheardJFS2011}.

%\fer{We will use spanwise velocity $w$ instead of displacement thickness $\delta_1$.}

%In order to put this hypothesis to test, the instantaneous displacement thickness %of (\ref{eq:delta1})
%is computed at the TR and BR corners and additively decomposed into their spanwise average $\langle\delta_1\rangle_z$ and spanwise modulation $\hat{\delta}_1$ components as
%\begin{equation}
%\delta_{1}^{TR,BR}(z;t) = \delta_{1}^{T,B}(0.5,z;t) = \langle \delta_{1}^{TR,BR} \rangle_{z}(t) + \hat{\delta}_{1}^{TR,BR}(z;t),
%\end{equation}\label{eq:delta1R}
%with
%\begin{equation}
%\langle \delta_{1}^{TR,BR} \rangle_{z}(t) = \frac{1}{L_z}\int_0^{L_z}{\delta_{1}^{TR,BR}(z;t)\;dz}.
%\end{equation}\label{eq:delta2R}
%\fer{and the displacement thickness, which quantifies viscous blockage in terms of massflow reduction, given by}
%\begin{equation}
%  \delta_1^{T,B}(x,z;t) = \int_{y_{T,B}=\pm0.5}^{\pm0.5(1+2\delta^{T,B})}{1-\frac{u(x,y,z;t)}{u_e^{T,B}(x,z;t)}\;dy}.
%\end{equation}\label{eq:delta1}
%The maximum streamwise velocity $u_e^{T,B}(x,z;t)=\max_y u(x,y^{\pm},z;t)$ is used as the inviscid outer flow velocity at location $(x,z)$, and the boundary layer thickness $\delta^{T,B}$ interpreted as the cross-stream distance from the wall to the point where this maximum velocity is reached.

\begin{figure*}
  \setlength\tabcolsep{2pt}% see you need to put a definition at the beginning of the document
  \begin{tabularx}{\textwidth}{@{}c*{1}{C}@{}}
    \raisebox{0.09\linewidth}{(a)} &
    \includegraphics[width=1\linewidth]{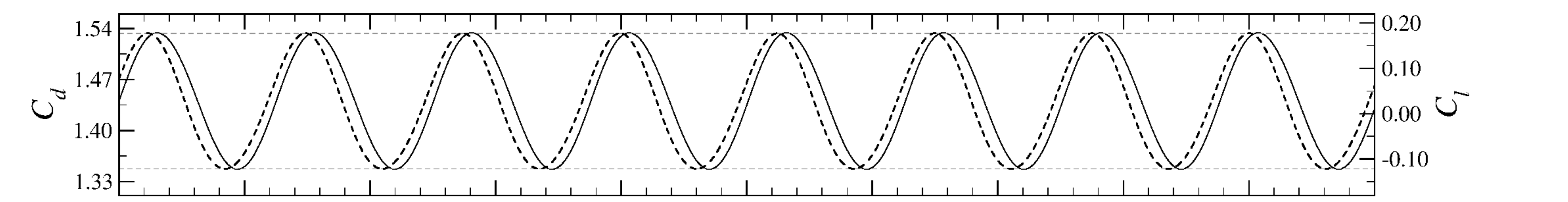}\\
    \raisebox{0.16\linewidth}{(b)} &
    \includegraphics[width=1\linewidth]{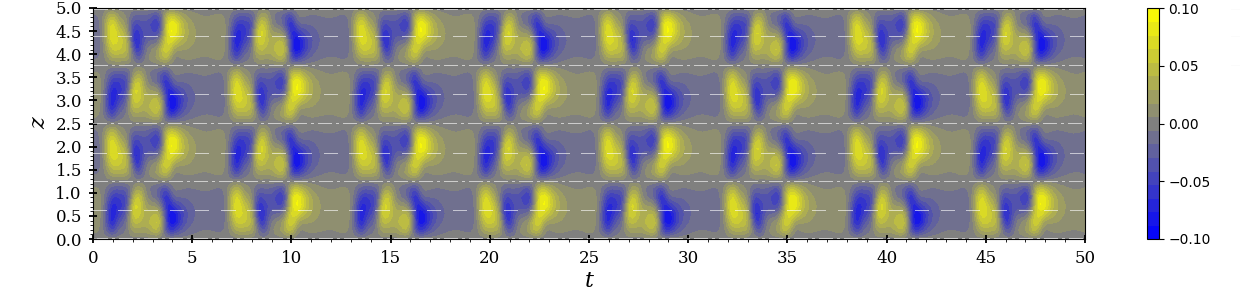}
  \end{tabularx}
  \caption{Space-time properties of the $R=3.4$ solution.  (a) Dynamic evolution of the lift $C_l$ (solid) and drag $C_d$ (dashed) coefficients. (b) Space-time diagrams of spanwise velocity $w(x,y,z;t)$ at probe location $(x,y)=(6.0,0.5)$. Reflection (dash-dotted white lines) and time-shift plus reflection (dashed white) symmetry planes are indicated.}
  \label{fig:delta1ztR34}
\end{figure*}

Figure~\ref{fig:delta1ztR34}b contains space-time diagrams of spanwise velocity $w(x,y,z;t)$ along a probe array located at $(x,y)=(6.0,0.5)$ alongside corresponding time series of $C_l$ and $C_d$ (panel a) for the three-dimensional periodic solution at $R=3.4$.
The space-time diagrams correspond to 8 vortex-shedding cycles, as clear from the $C_l$ and $C_d$ time series, and their periodicity corresponds exactly with that of vortex shedding. The space-time diagrams show instead that the actual periodicity of the solution is twice that of vortex shedding, as previously unraveled by the corresponding phase maps in figure~\ref{fig:PhaseMaps}. The nature of the period doubling, which does not affect aggregate quantities, is now clearly explained as a result of the way in which the spanwise invariance has been disrupted upon three-dimensionalisation of the flow. The reflection Z$_2$ symmetry is only preserved at all times about planes located at $z=z_0+(2j)\lambda_z/4$ (white dash-dotted lines), where the origin for the spanwise coordinate has been chosen to enforce $z_0=0$, $\lambda_z=2.5$ here and $j\in\mathbb{Z}$. Additionally, a space-time symmetry operation consisting in the evolution by half a period $T/2$, where $T$ is the actual period of the solution corresponding to two vortex-shedding cycles, followed by reflection about any plane located at $z=z_0+(2j+1)\lambda_z/4$ (white dashed lines) also leaves the solution invariant. The appropriate composition of the two symmetries shows that the solution is also invariant to evolutions by half a period $T/2$, followed by a spanwise shift by a half wavelength $\lambda_z/2$.

\begin{figure*}
  \setlength\tabcolsep{2pt}% see you need to put a definition at the beginning of the document
  \begin{tabularx}{\textwidth}{@{}c*{1}{C}@{}}
    \raisebox{0.09\linewidth}{(a)} &
    \includegraphics[width=\linewidth]{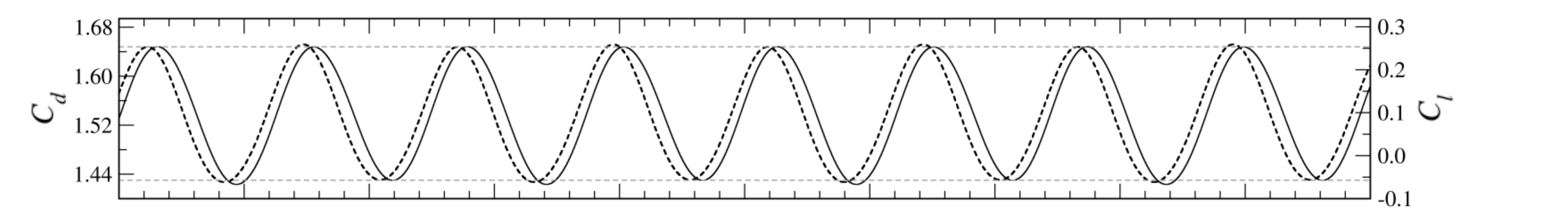}\\
    \raisebox{0.16\linewidth}{(b)} &
    \includegraphics[width=\linewidth]{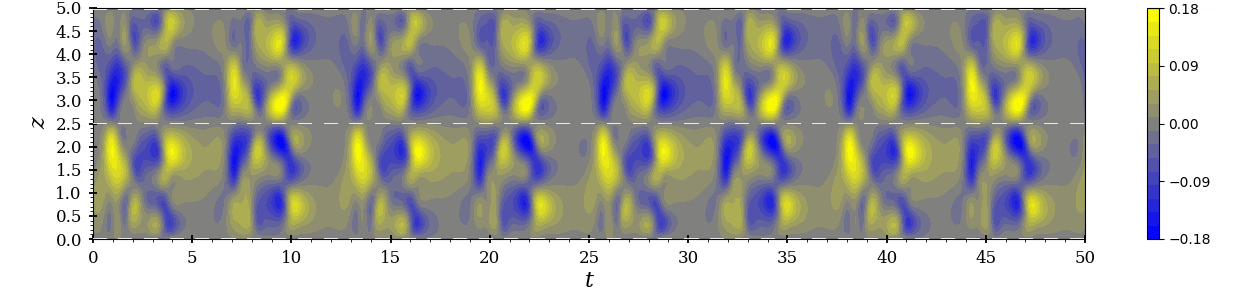}
  \end{tabularx}
  \caption{Space-time properties of the $R=3.8$ solution.  (a) Dynamic evolution of the lift $C_l$ (solid) and drag $C_d$ (dashed) coefficients. (b) Space-time diagrams of spanwise velocity $w(x,y,z;t)$ at probe location $(x,y)=(6.0,0.5)$. Time-shift plus reflection symmetry planes (dashed white) are indicated.}
  \label{fig:delta1ztR38}
\end{figure*}

The ensuing period-doubling bifurcation is of an altogehter different nature. The solution at $R=3.8$ has an actual period of four vortex-shedding cycles, yet aggregate quantities repeat every two vortex-shedding cycles. Close inspection of figure~\ref{fig:delta1ztR38}a shows how the $C_l$ and $C_d$ time series have a period that is about twice that for $R=3.4$.
The space-time diagram for $w$ provide the full picture. The bifurcation that produces the period-doubled solution is spatially subharmonic in the sense that the spanwise wavelength is double that at $R=3.4$. It is a modulational instability to perturbations of a wavelength twice that of the bifurcating solution. This breaks the mirror symmetry and the only remaining symmetry leaves the solution invariant to evolution for half a period (two vortex-shedding cycles) followed by reflection about planes located at $z=z_0+j\lambda_z/2$ (white dashed line), where $\lambda_z=5$ now.

\begin{figure}
  \centering
  \begin{tabular}{ccc}
    k & (a) & (b) \\
    \raisebox{0.25\linewidth}{0} &
    \includegraphics[width=0.45\textwidth]{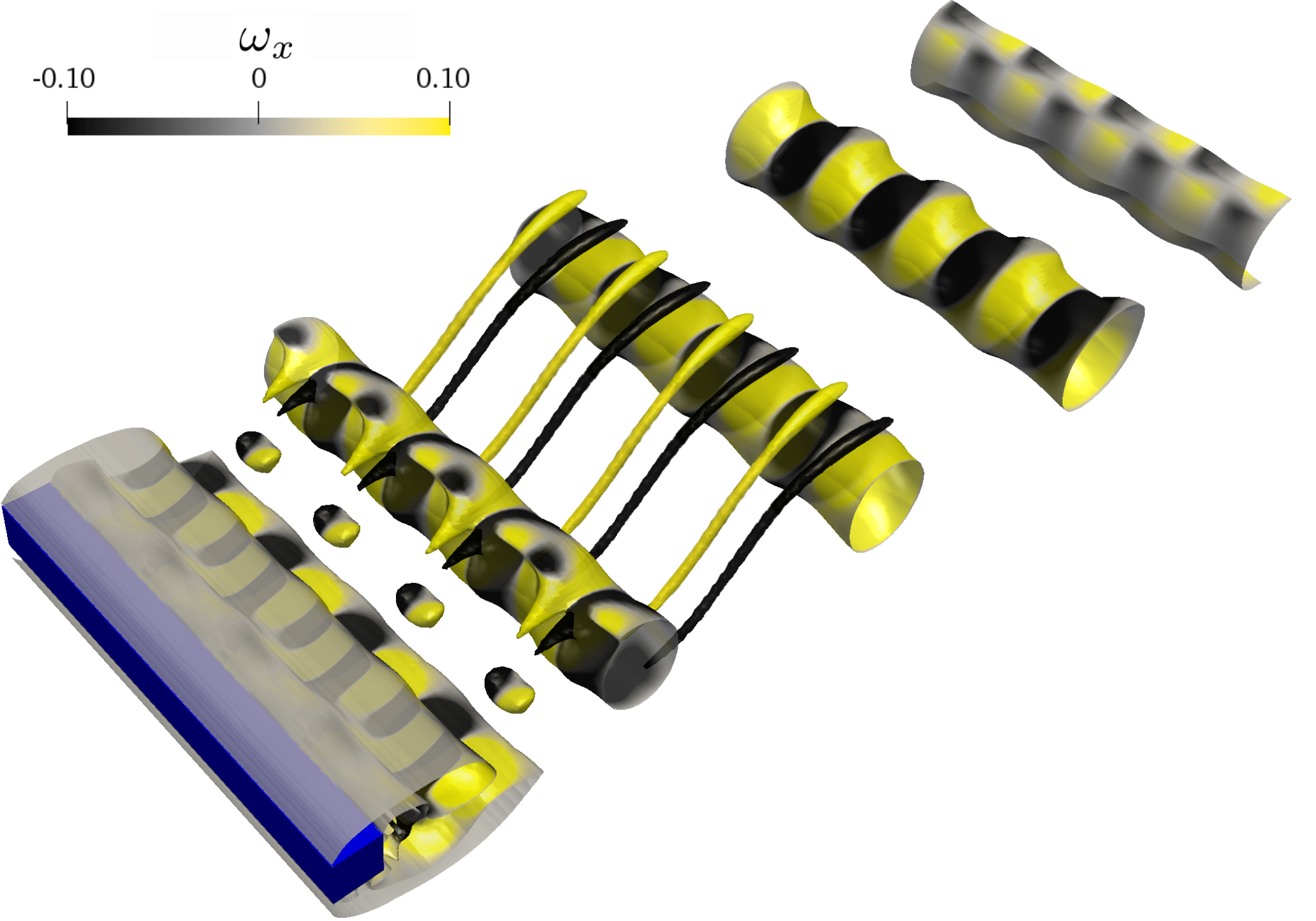} &
    \includegraphics[width=0.45\textwidth]{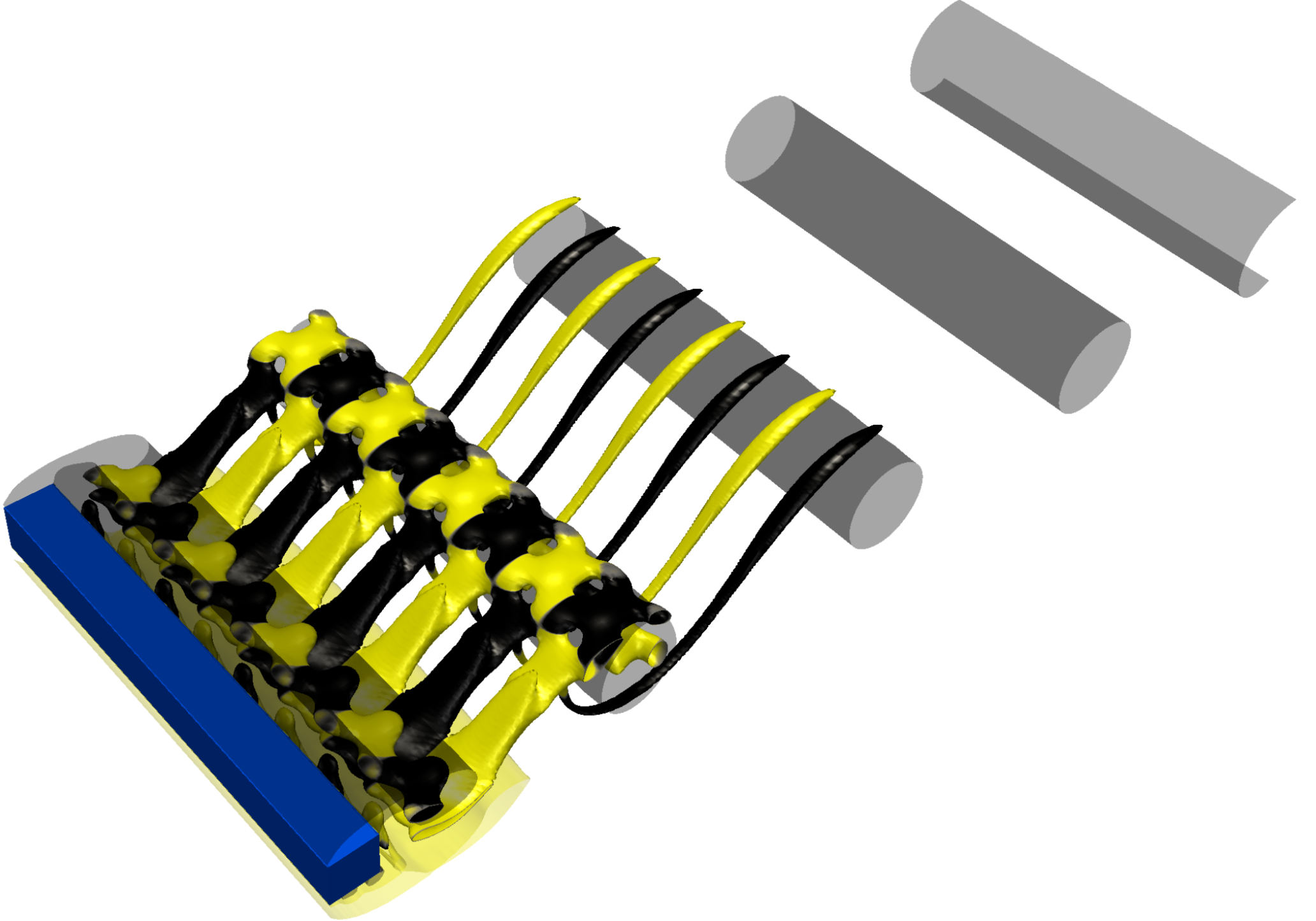} \\
    \raisebox{0.25\linewidth}{1} &
    \includegraphics[width=0.45\textwidth]{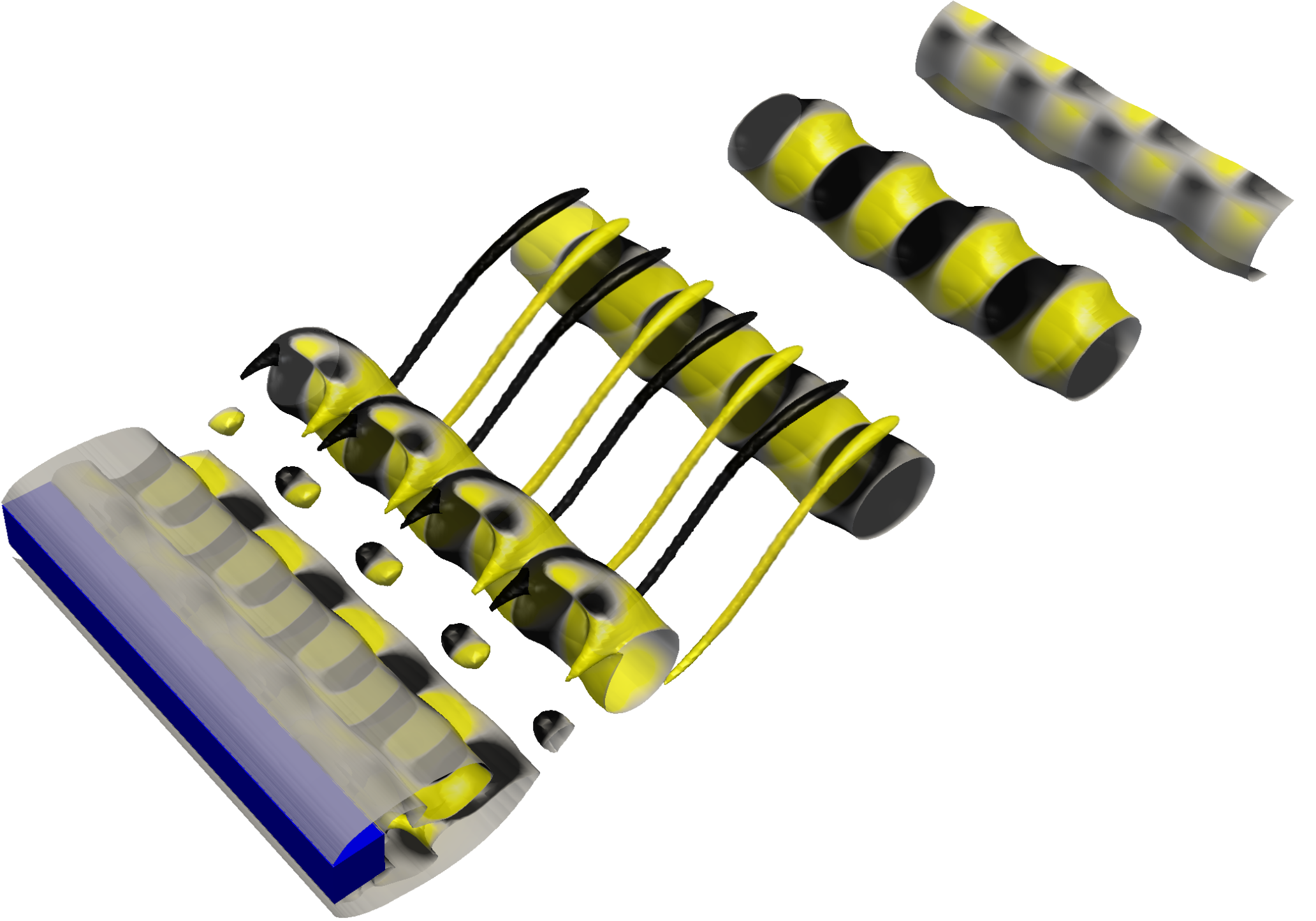} &
    \includegraphics[width=0.45\textwidth]{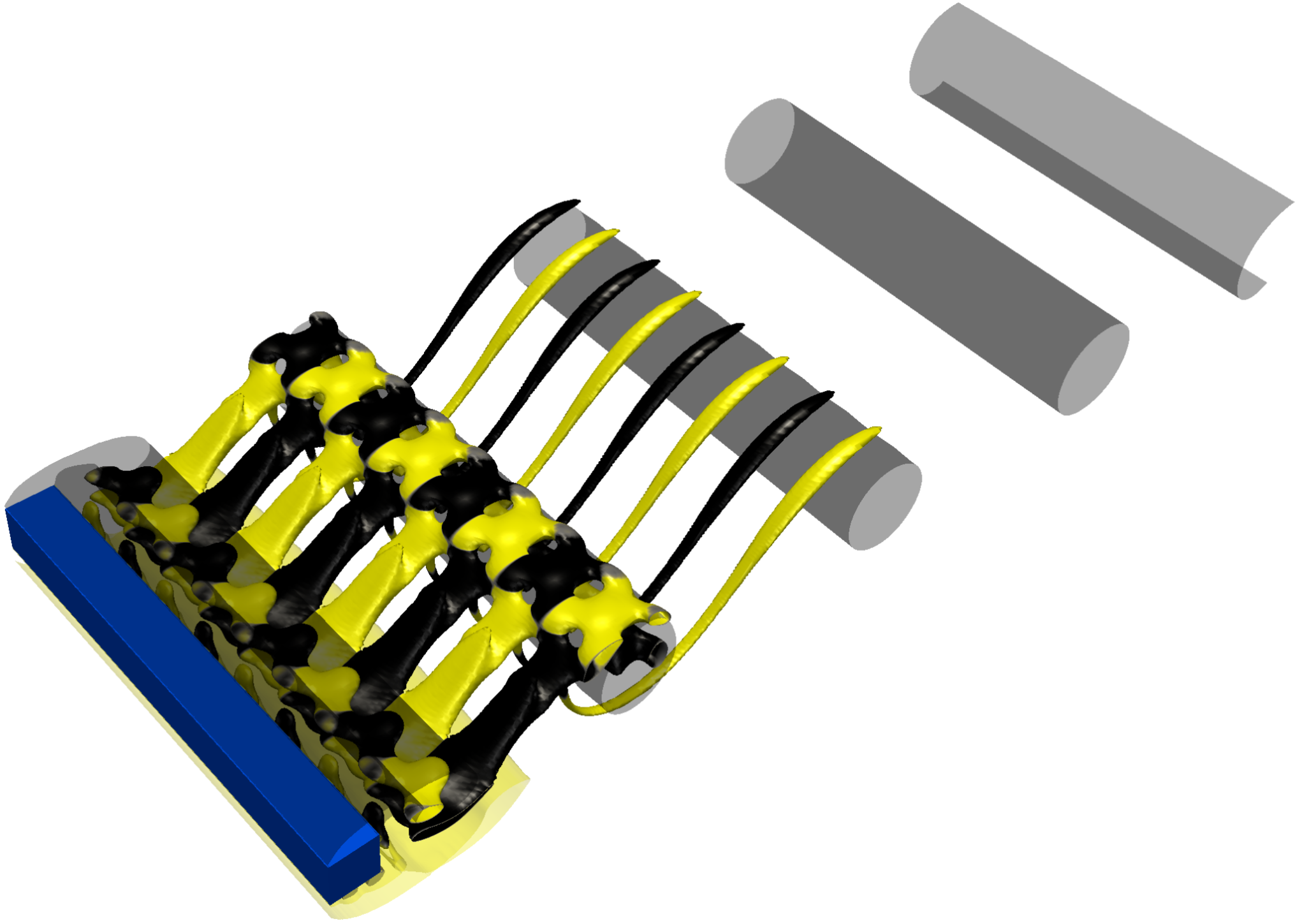} \\
  \end{tabular}
  \caption{Instantaneous snapshots of (a) the three-dimensional solution at $R=3.4$ (Movie 1) and (b) corresponding dominant eigenmode of the underlying two-dimensional solution (Movie 2), at two consecutive crossings of the Poincar\'e section. Shown are iso-surfaces of the Q-criterion ($Q=0.0002$ for three-dimensional solution, arbitrarily chosen for the dominant eigenmode to ease comparison), coloured by streamwise vorticity (symmetric arbitrary scale for the eigenmode). Spanwise vortices are shown with transparent iso-surface for $\omega_z=\pm0.1$. The solution is replicated twice in the spanwise direction to $L_z=10$ only for visualisation purposes.
  }
  \label{fig:3DQR34}
\end{figure}

A three-dimensional representation of the solution at $R=3.4$ is shown in figure~\ref{fig:3DQR34}a at two consecutive crossings of the Poincar\'e section defined by $C_l=\langle C_l\rangle_t$ and $\dot{C}_l>0$.
Spanwise mode zero, shown with a transparent iso-surface for $\omega_z=\pm0.1$, is taken to represent the K\'arm\'an vortices. The three-dimensional vortical structures present in the flow are exposed by means of the Q-criterion, which identifies vortical structures as regions where the second invariant of the velocity gradient tensor becomes positive ($Q>0$), \emph{i.e.} the rotation rate overcomes the shear rate \citep{HuntCTR1988}. Streamwise-cross-stream vortical structures are made visible through $Q=0.0002$ isocontours coloured by streamwise vorticity $\omega_x$. A first visible effect of three-dimensionalisation is that of introducing spanwise modulation to K\'arm\'an vortices, whose vorticity is no longer merely spanwise but has become slightly tilted and alternates positive and negative values of the streamwise component. Elongated streamwise-cross-stream vortices extend along the braids connecting consecutive K\'arm\'an vortices. The three-dimensional flow topology is very different from mode-A- and mode-B-type structures observed for circular \citep{WilliamsonARFM1996} and square \citep{BaiPoF2018} cylinders in the wake transition regime. Instead, the streamwise-cross-stream vortices, in particular their wavelength and time periodicity, is highly reminiscent of mode QP in the wake of circular \citep{BlackburnJFM2005}, square \citep{RobichauxPoF1999,BlackburnJFM2005} and rounded-square \citep{ParkJFM2016} cylinders, provided that the positive K\'arm\'an vortices emanating from the bottom shear layer are overlooked and the elongated vortices in the braids connecting them to the preceding and following negative K\'arm\'an vortices duly spliced. As a matter of fact, the parallel is all the more convincing when a comparison is drawn between the present three-dimensional wake vortices and mode C structures in the wake of open rings \citep{SheardJFM2005}, squares at incidence \citep{SheardJFM2009} or cylinders submerged in upstream shear \citep{ParkPoF2018}.
%These elongated vortical structure pairs take a maximum slope bordering $45^\circ$ and seem to connect as horse-shoe or hairpin vortices on top of the most recently formed K\'arm\'an vortex, as previously observed in the instability of planar sher layers \citep{jimenez1983spanwise}. Linear stability theory predicts the instability to grow in the direction of maximum strain rate, which is found precisely in the braids. \fer{(Not sure we want to leave this last sentence...)}

Comparing two snapshots exactly one Poincar\'e section crossing apart, the aforementioned space-time symmetries become apparent. The solution looks exactly the same but shifted by exactly half a wavelength in the spanwise direction or reflected about appropriate symmetry planes as previously observed. The fully developed three-dimensional solution bears strong resemblance and shares the same symmetries with the dominant eigenmode of the underlying two-dimensional periodic solution at the same value $R=3.4$, as clear from figure~\ref{fig:3DQR34}. Here the most unstable mode of figure~\ref{fig:floquet}b for $\lambda_z=2.5$ has been represented using the Q-criterion and coloured by $\omega_x$ (symmetric arbitrary range). The unstable two-dimensional solution is indicated by transparent $\omega_z=\pm0.1$ iso-surfaces that clearly show the K\'arm\'an vortices. The topologies of the dominant eigenmode and of the three-dimensional solution are evidently related and so is the space-time symmetry dynamics as illustrated by snapshots taken one Poincar\'e crossing apart.

The symmetries and wavelength of the dominant eigenmode (and, with it, the nonlinear solution) make it analogous to the mode C that destabilises the axisymmetric/two-dimensional periodic solutions characteristic of the flow past a circular ring \citep{SheardPoF2005,SheardJFM2005} or an inclined square cylinder \citep{SheardJFM2009,SheardJFS2011}, which evolve from the respective modes QP of the top-bottom symmetric cases and take precedence over modes A and B when the symmetry has been sufficiently disrupted. Mode C, its evolution from mode QP in a codimension-2 bifurcation, and its overtaking modes A and B, has been recently exposed in the somewhat related case of a circular cylinder in homogeneous upstream shear \citep{ParkPoF2018}. Here we are only varying one parameter $R$, so that modes A and B are never encountered along the particular $Re-K$ path followed. The rapid increase of the upstream velocity ratio and its associated mean shear is probably suppressing them before any trace can be even found in the Floquet spectrum.

The dominant eigenmode of the underlying two-dimensional solution is the same for all values of $R$ investigated here. With minor variations, it remains topologicaly identical from as low as $R=3$, which corresponds to a stable case, to as high as $R=6.5$, where the three-dimensional solution has evolved into chaotic dynamics. The most unstable (or least stable) wavelength remains around $\lambda_z\simeq2.5$ for low values of $R$ and then gradually grows as $R$ is increased. In particular, the most unstable mode remains the same for $R=3.8$, for which it has been seen that a second bifurcation has taken place and the space-time symmetry of $R=3.4$ is no longer present. It is to be surmised that the period doubling bifurcation, that appears to arise from a spanwise subharmonic/modulational instability of the $R=3.4$ solution to perturbations of double its spanwise wavelength, may also affect the underlying two-dimensional flow, but since this is already unstable to the eigenmode of figure~\ref{fig:3DQR34}b, computing the second unstable mode is not within the reach of mere time-stepping. Besides, it may also be the case that this second instability is related to the interactions of unstable wavelengths across the continuous eigenspectrum. For $L_z=5$ the domain admits instability to perturbations of wavelength $\lambda_z=L_z/l$, with $l\in\mathbb{Z}$, or, equivalently, wavenumber $\beta_z=2\pi l/L_z$. Now, for $R=3.4$, the unstable dominant mode can only fit twice or thrice within the domain considered and the system seems to choose $\lambda_z=2.5$ as the prevailing instability. At $R=3.8$, instead, the unstable band has grown large enough to allow for the dominant mode to fit either twice, thrice, four times or even just once. The three-dimensional solution has $\lambda_z=L_z=5$ but it retains a $\lambda_z\simeq2.5$-dependence to a large extent. This suggest that the mode with $\lambda_z=2.5$ and $\lambda_z=5$, which in fact are one and the same mode except for their different wavelengths, might be competing and that none completely prevails over the other.

\begin{figure}
  \centering
  \begin{tabular}{cccc}
    k & (a) & (b) & (c)\\
    &&&\\
    \raisebox{0.2\linewidth}{0} &
    \includegraphics[width=0.3\linewidth]{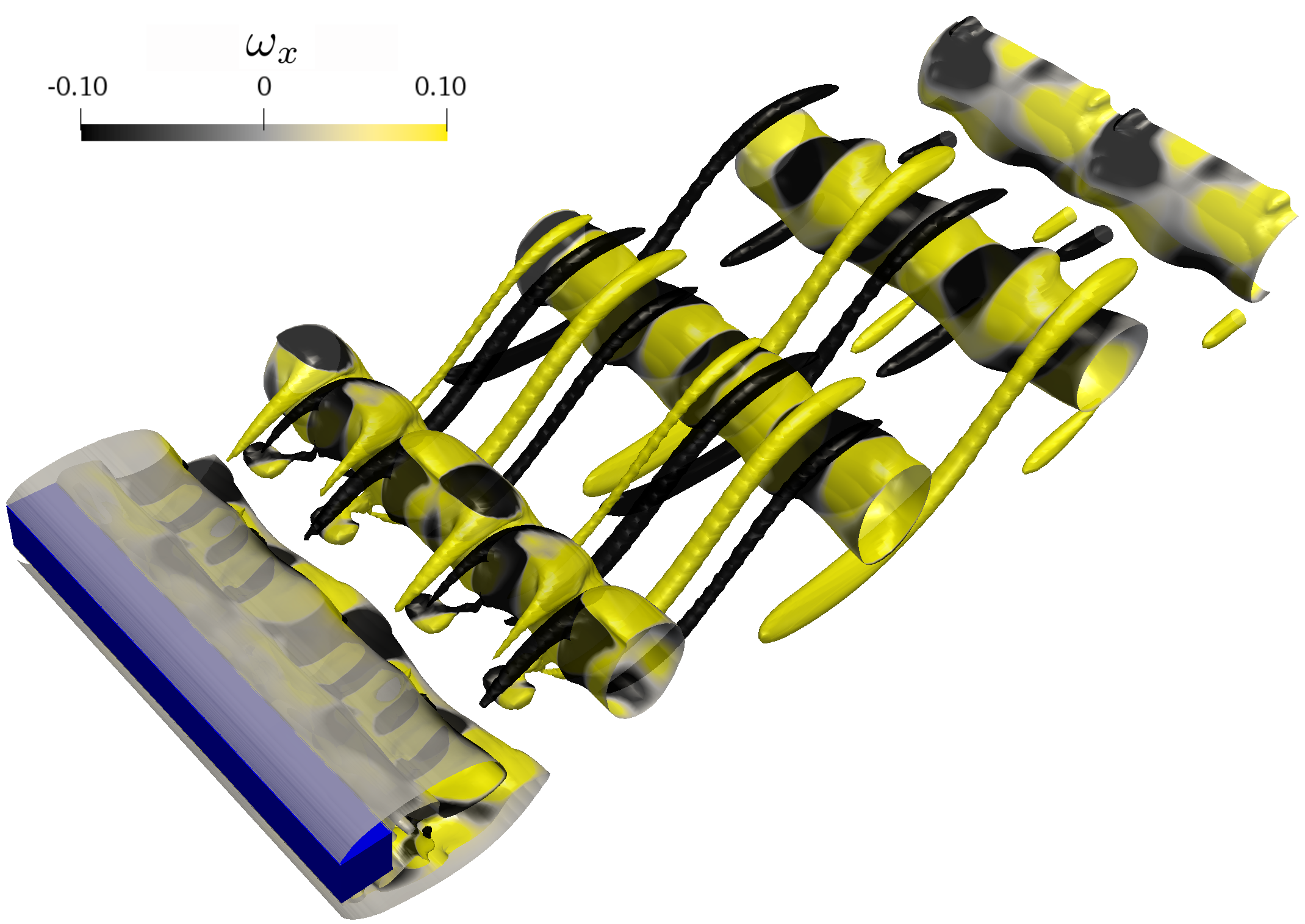} &
    \includegraphics[width=0.3\linewidth]{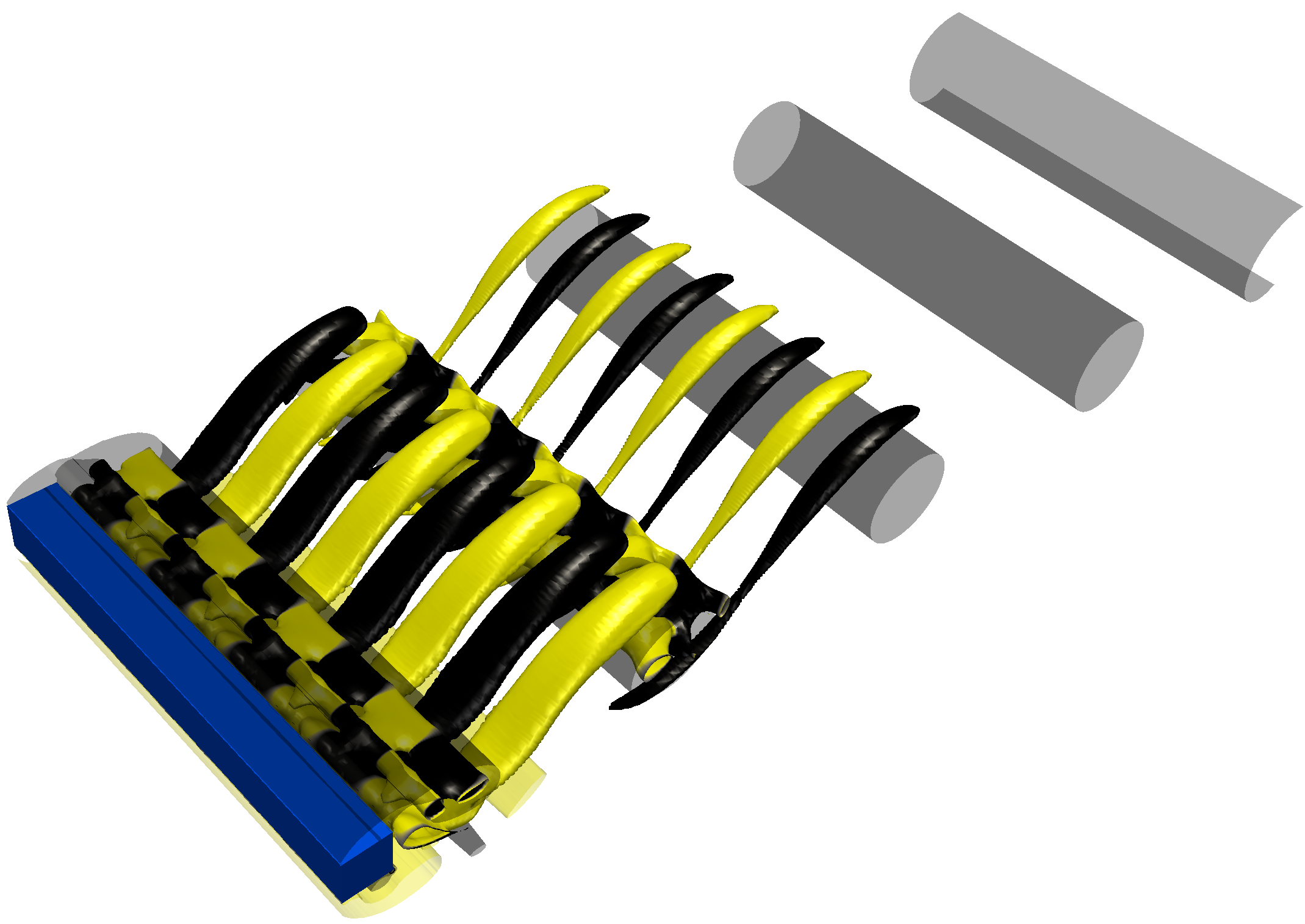} &
    \includegraphics[width=0.3\linewidth]{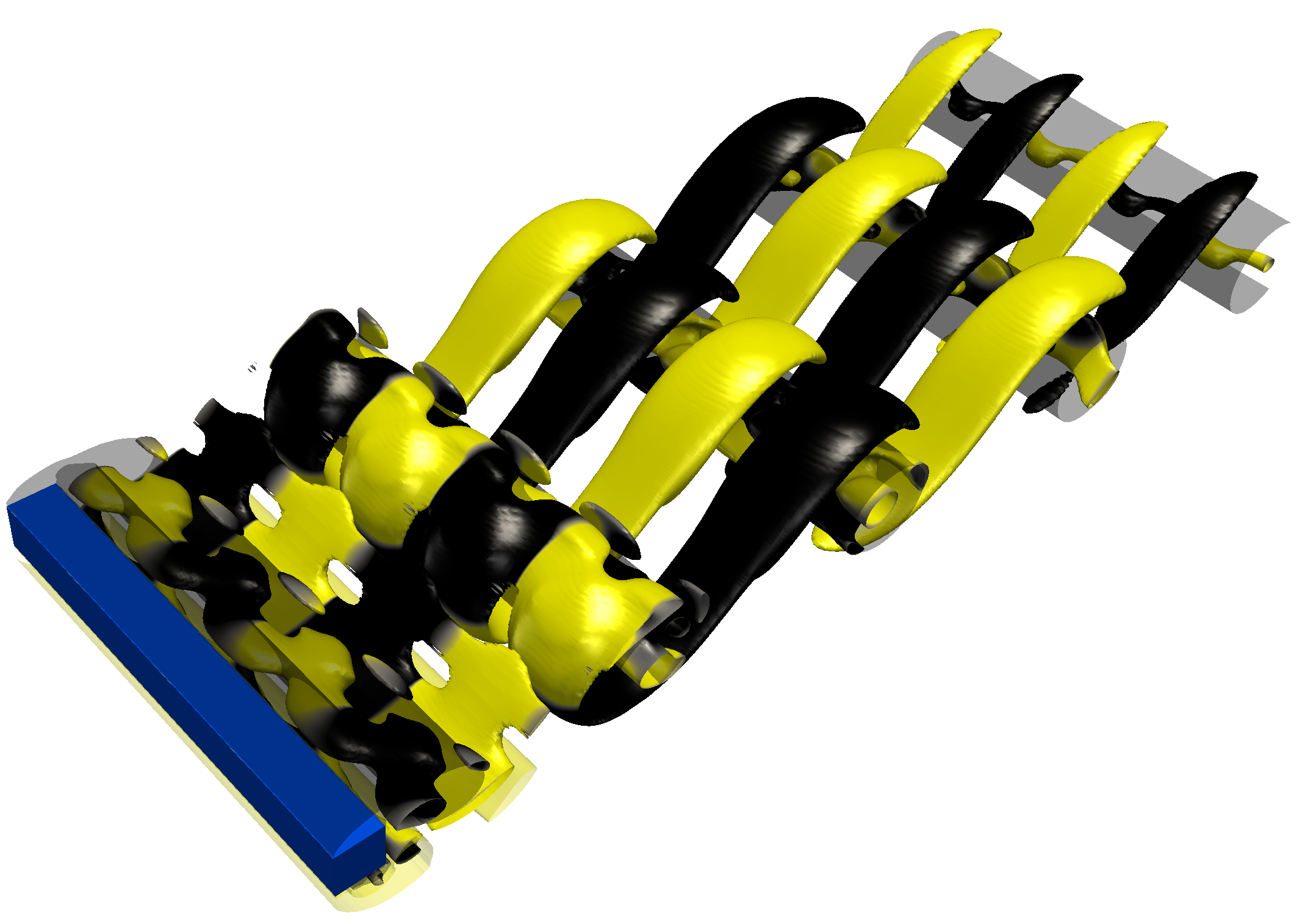} \\
    \raisebox{0.2\linewidth}{1} &
    \includegraphics[width=0.3\linewidth]{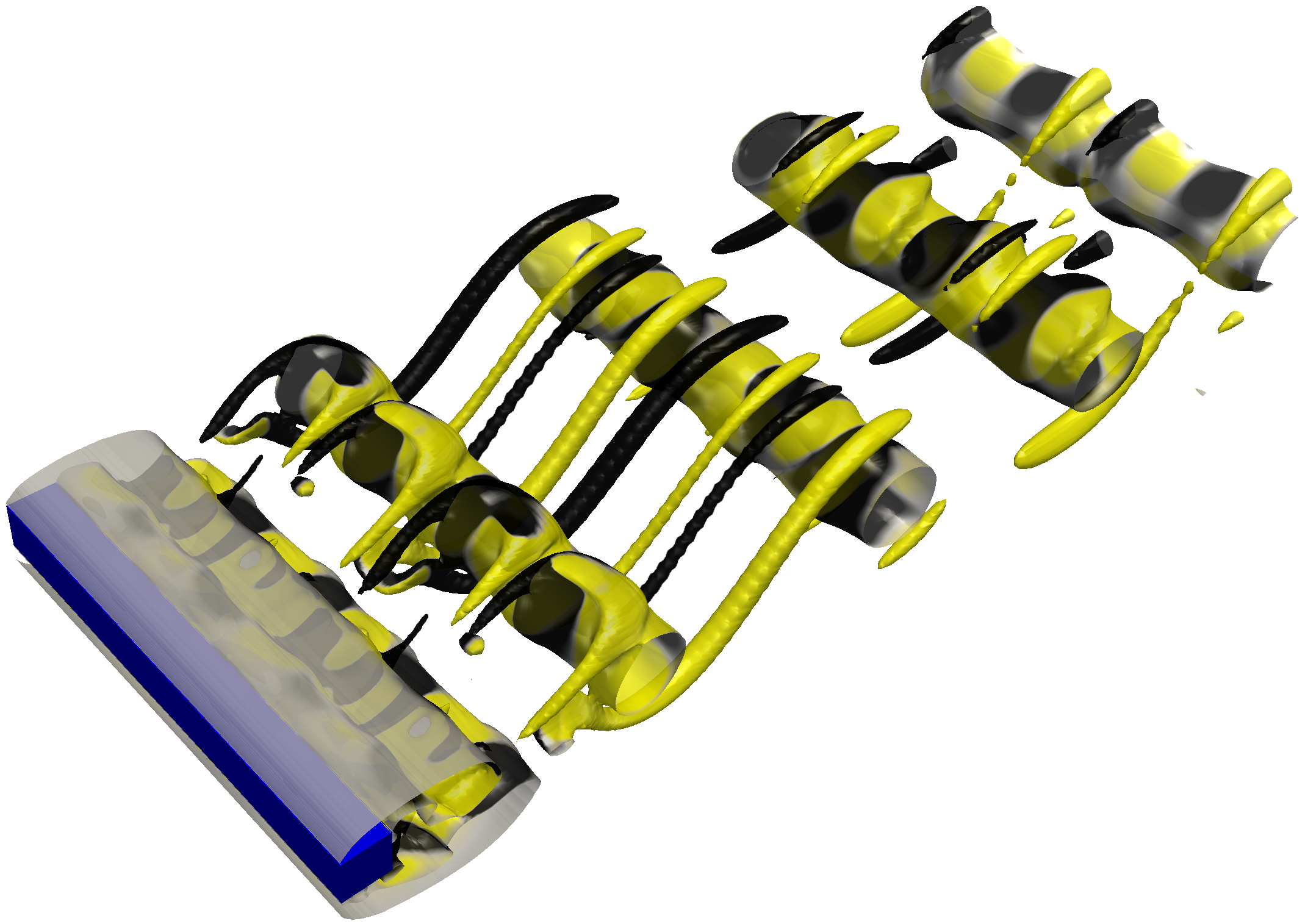} &
    \includegraphics[width=0.3\linewidth]{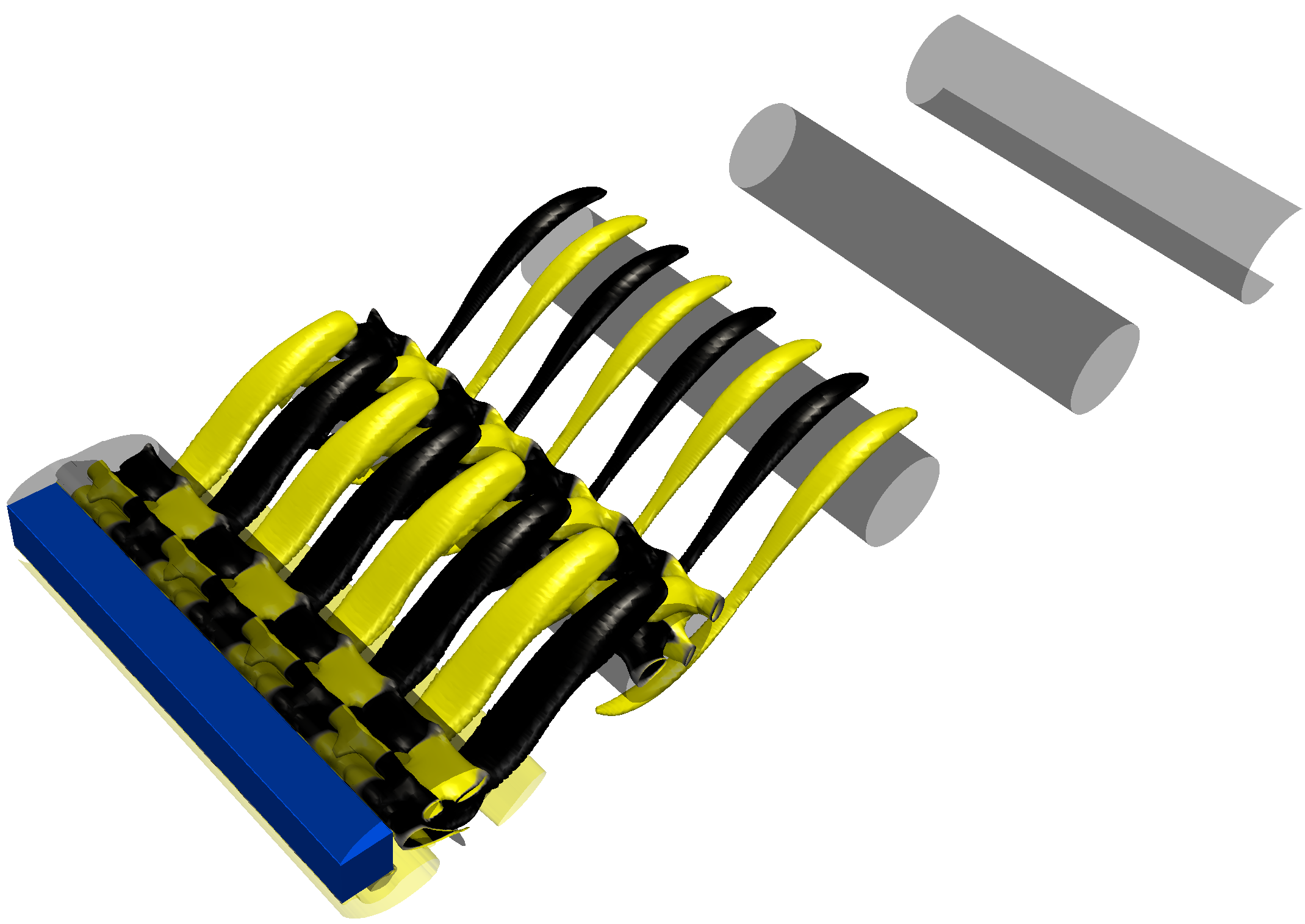} &
    \includegraphics[width=0.3\linewidth]{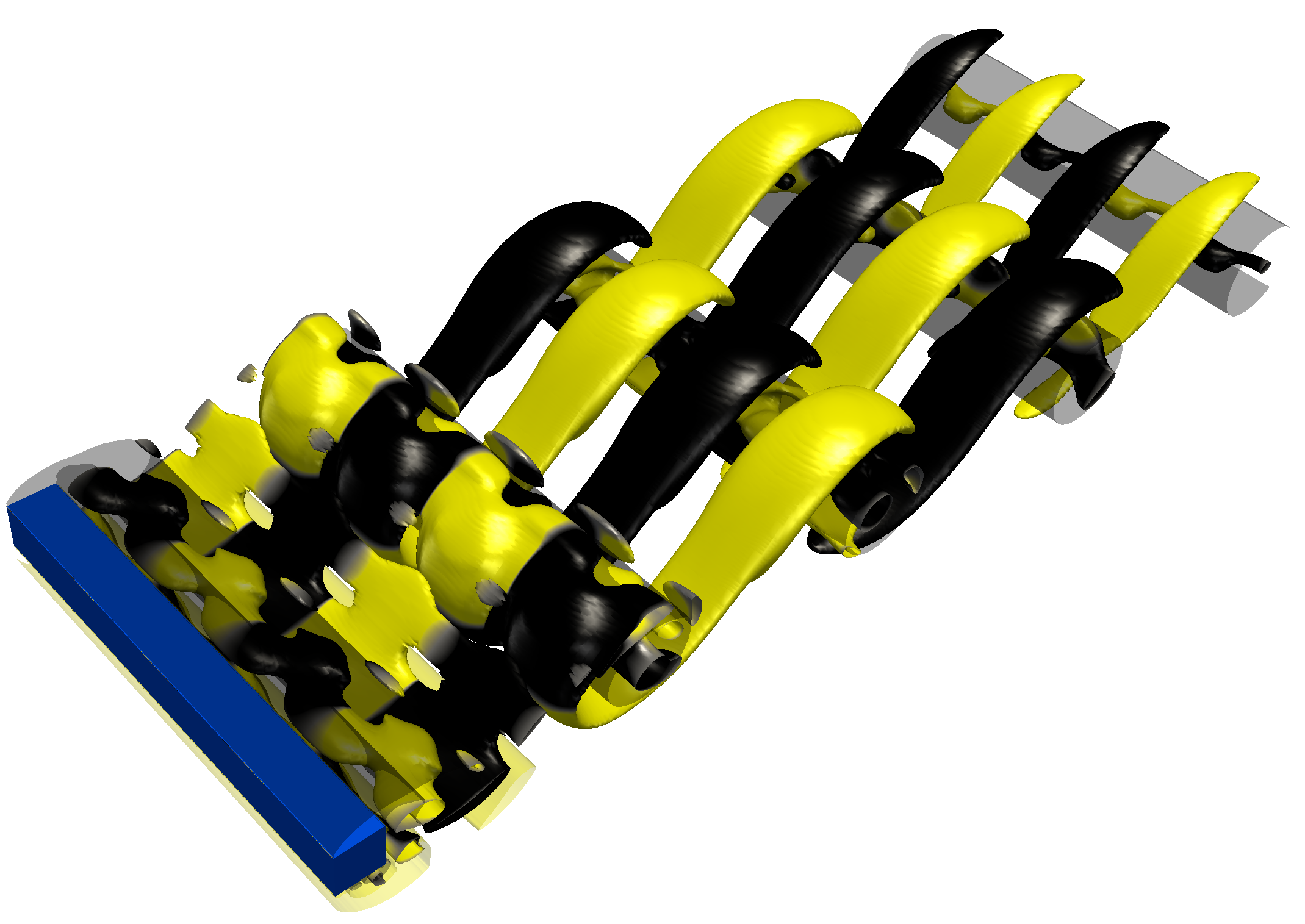} \\
    \raisebox{0.2\linewidth}{2} &
    \includegraphics[width=0.3\linewidth]{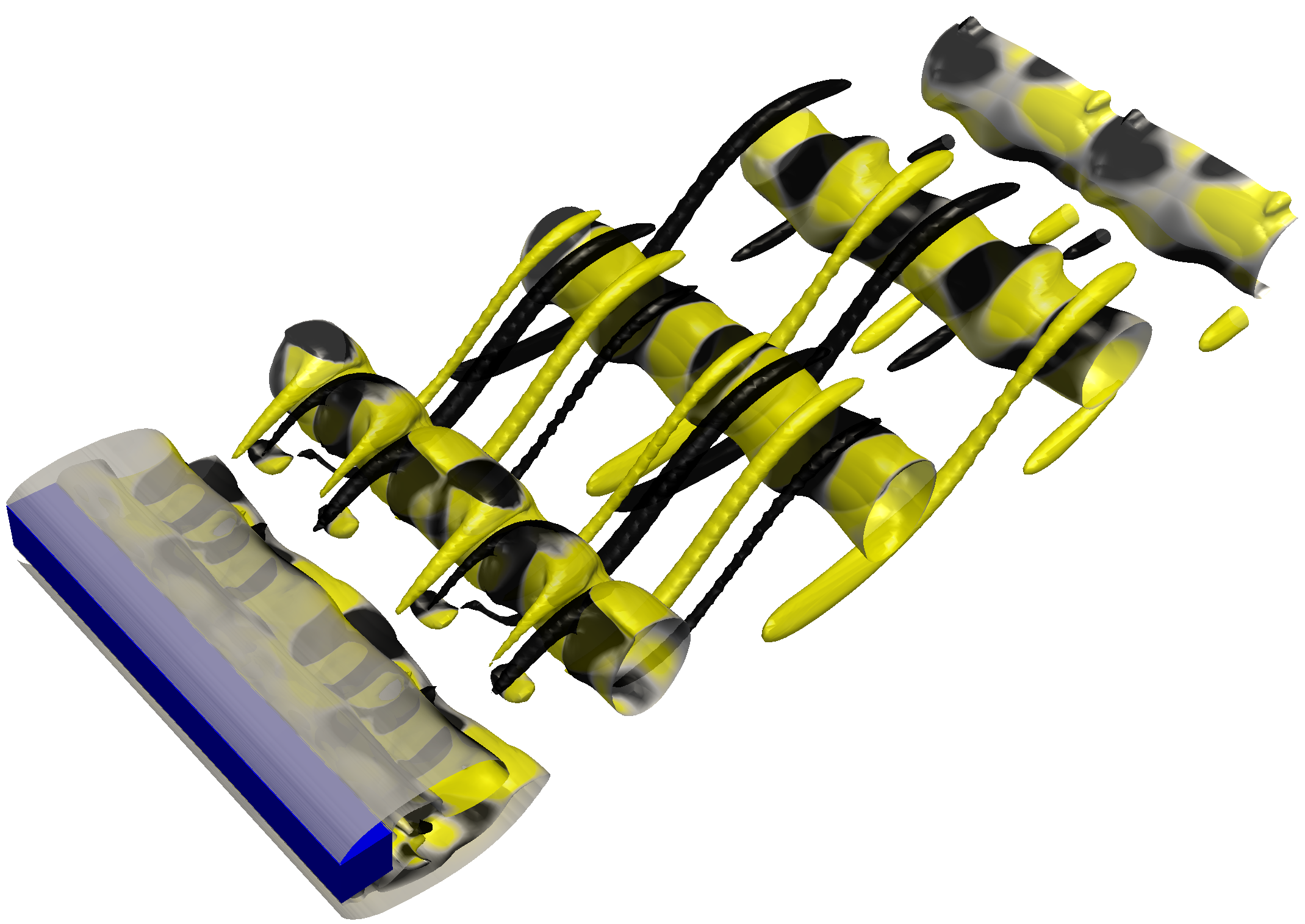} &
    \includegraphics[width=0.3\linewidth]{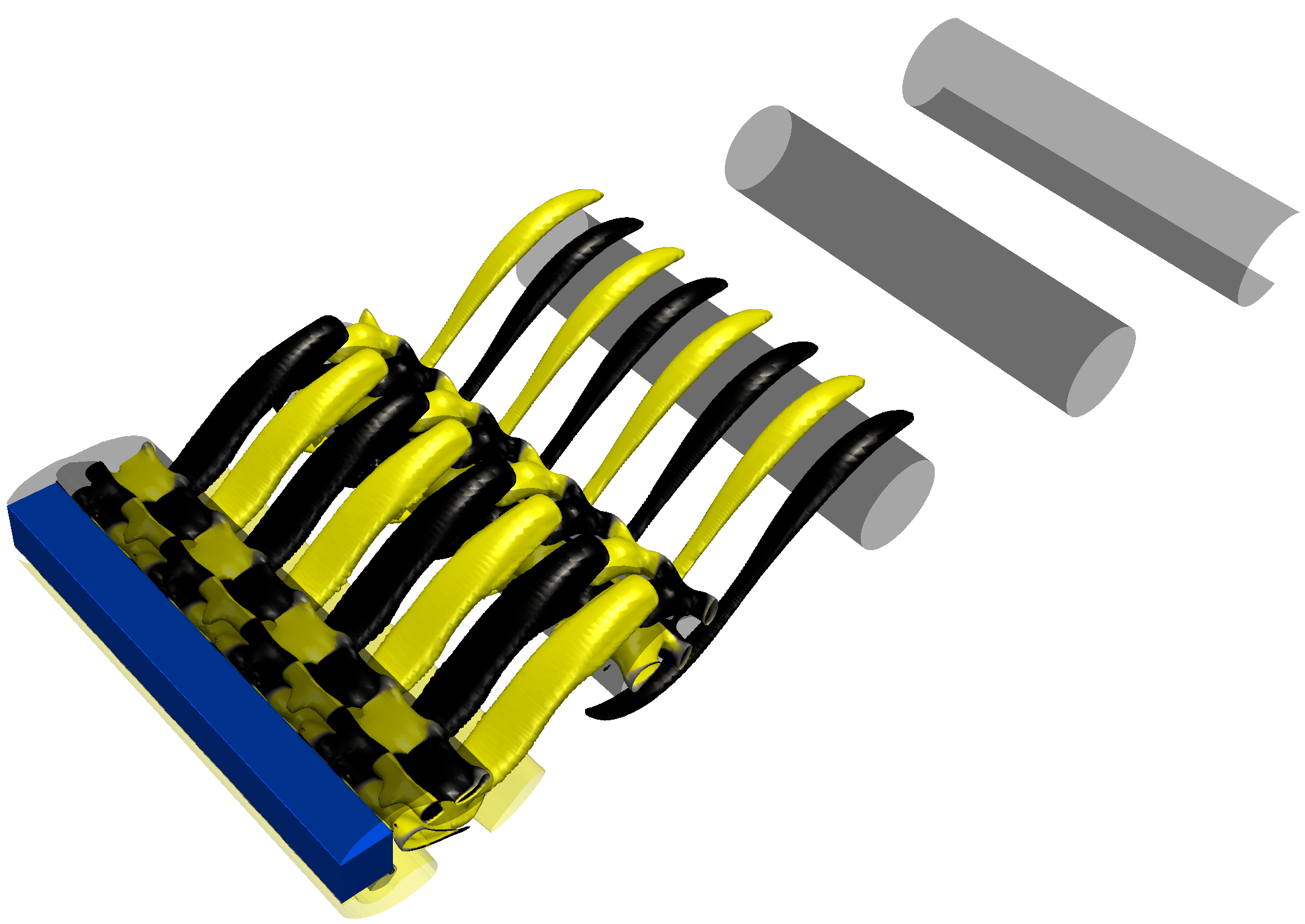} &
    \includegraphics[width=0.3\linewidth]{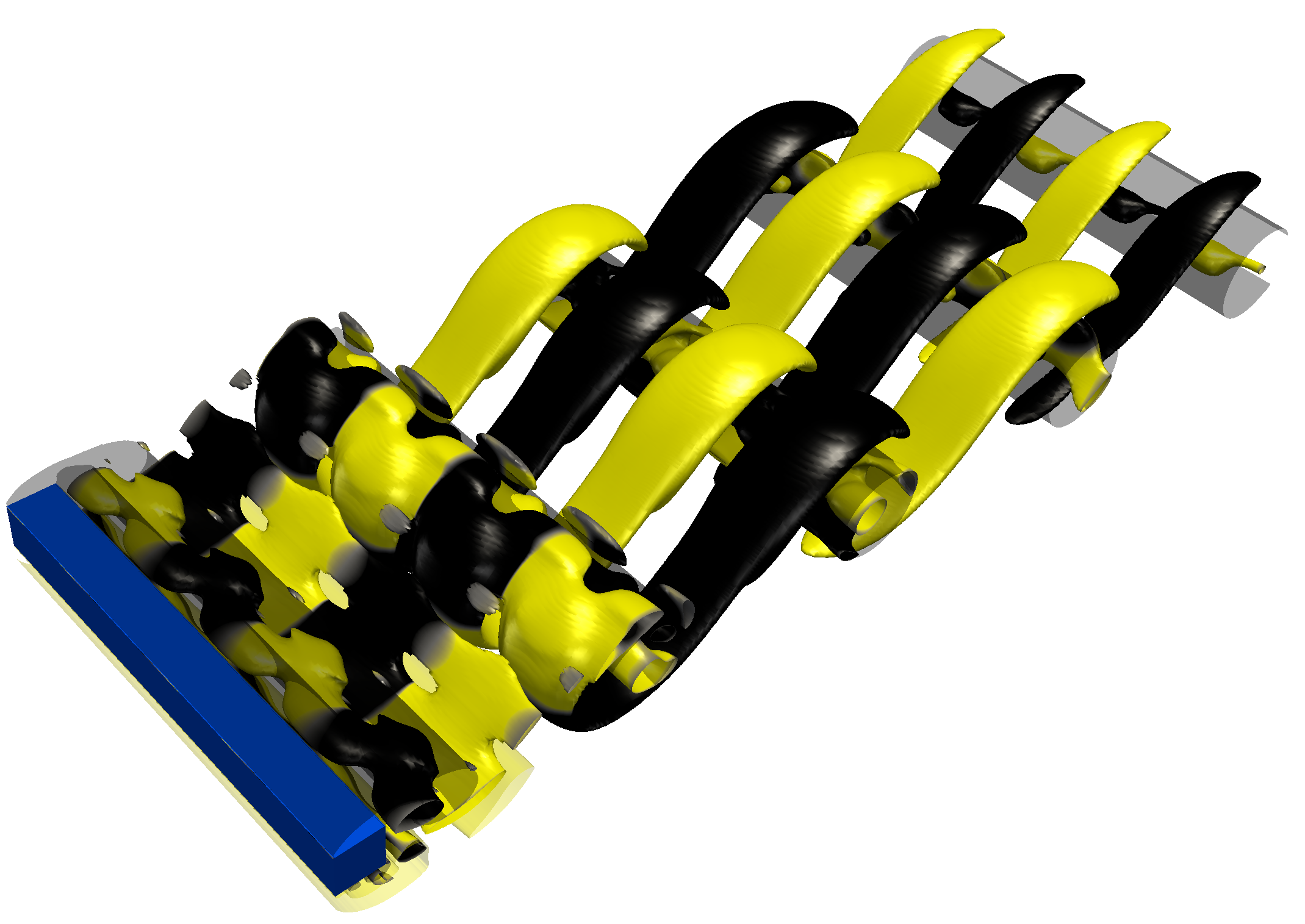} \\
    \raisebox{0.2\linewidth}{3} &
    \includegraphics[width=0.3\linewidth]{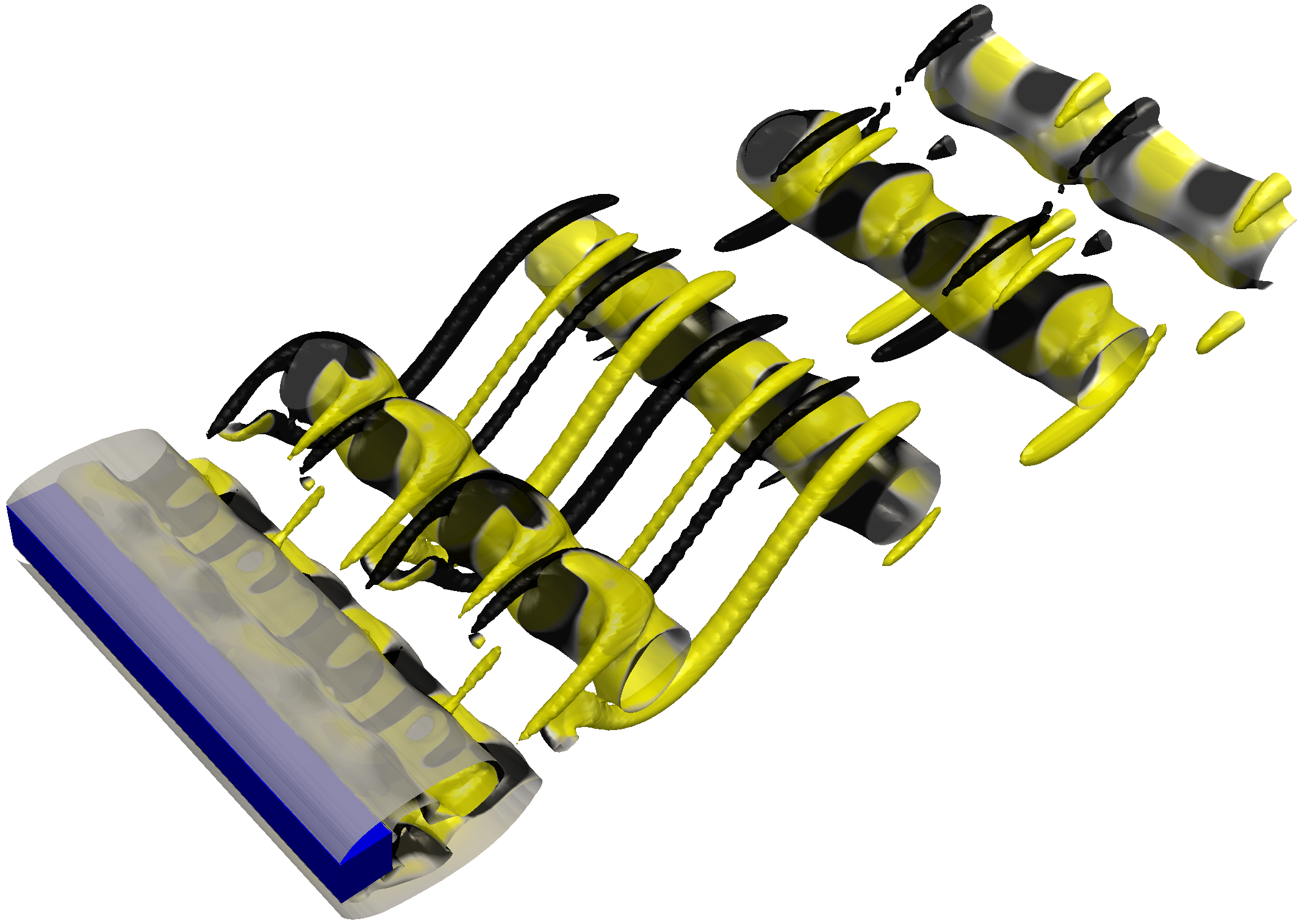} &
    \includegraphics[width=0.3\linewidth]{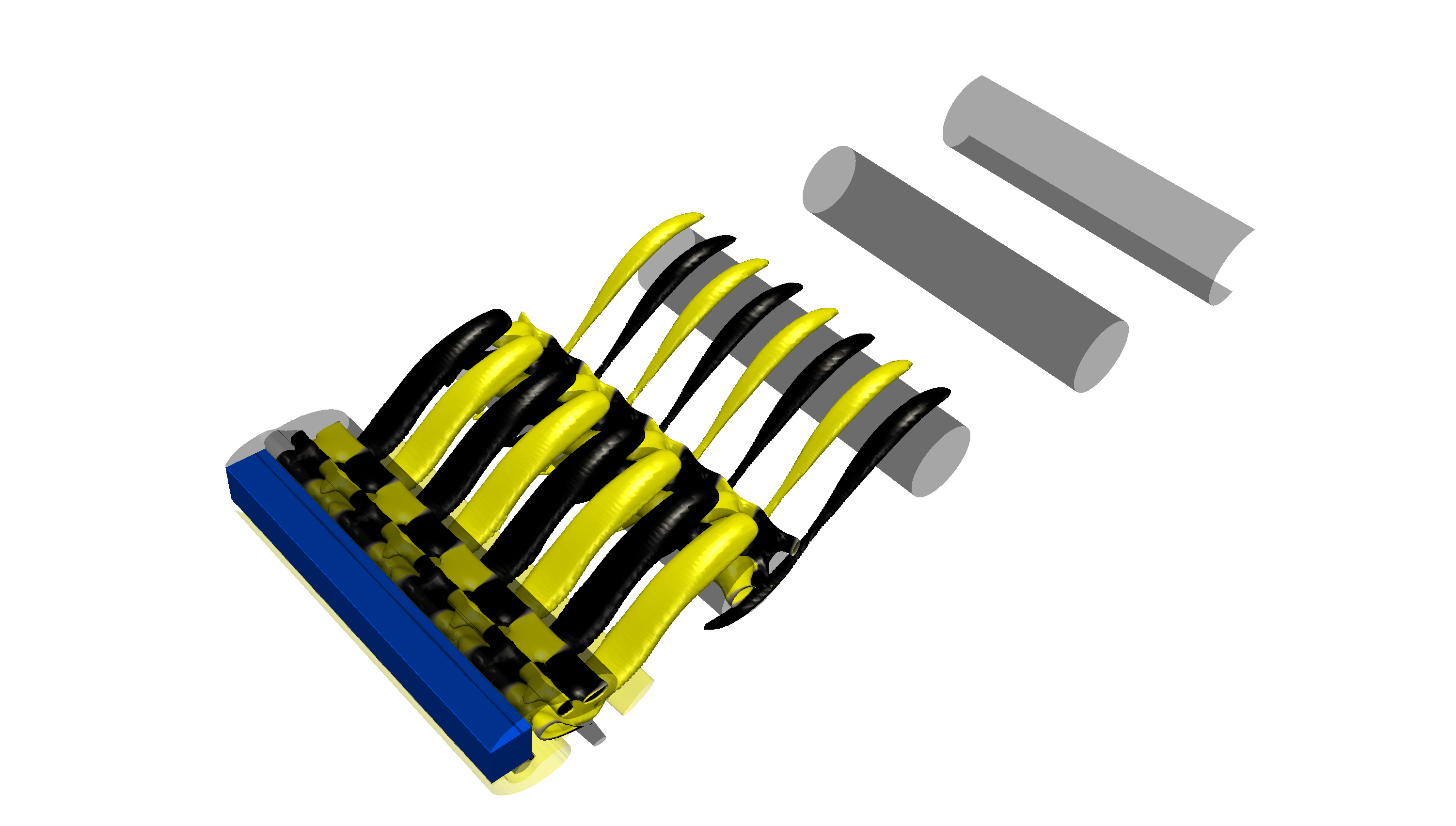} &
    \includegraphics[width=0.3\linewidth]{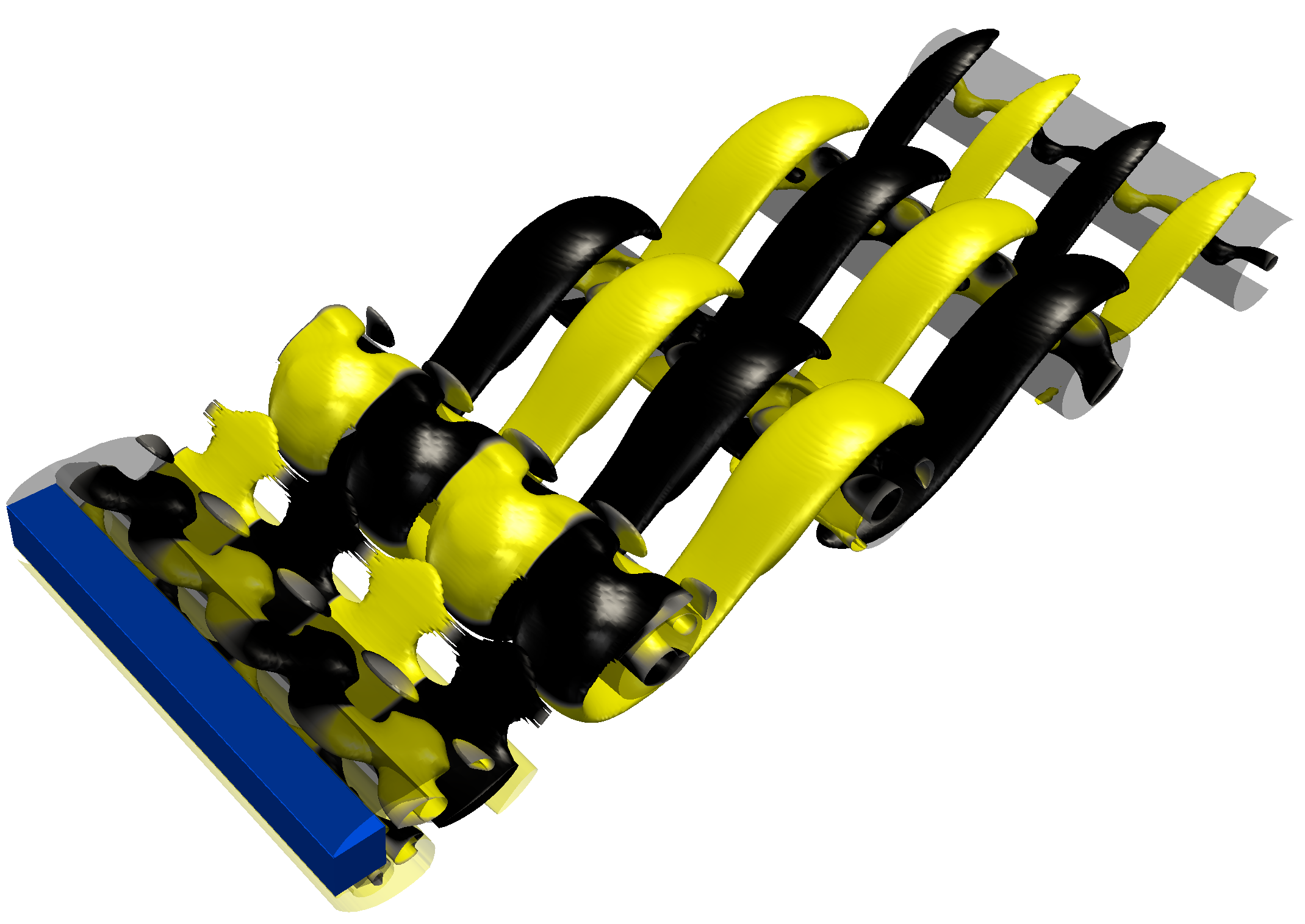} \\
  \end{tabular}
  \caption{Instantaneous snapshots of (a) the three-dimensional solution at $R=3.8$ (Movie 3), (b) dominant eigenmode for $(R,\lambda_z)=(3.8,2.5)$ (Movie 4), and (c) dominant eigenmode for $(R,\lambda_z)=(3.8,5)$ (Movie 5), at four consecutive crossings of the Poincar\'e section. Iso-surfaces and colours as for figure~\ref{fig:3DQR34}, except that $Q=0.0001$ is used to display nonlinear three-dimensional vortical structures.
  }
  \label{fig:3DQR38}
\end{figure}

Figure~\ref{fig:3DQR38}a shows snapshots of the three-dimensional solution for $R=3.8$ at four consecutive crossings of the Poincar\'e section.
The original spanwise wavelength $\lambda_z=2.5$ is preserved to a large extent, but the actual periodicity is now the full $\lambda_z=L_z=5$. The similarity with $R=3.4$ is apparent, but the elongated vortices along the braids reach further downstream into the wake. Although crossings 0 and 2 (also 1 and 3) look very much alike, careful inspection reveals that very slight differences exist that are related with the period-doubled nature of the solution. Also hard to notice, but nevertheless present, is the symmetry that leaves the solution invariant after evolution over two consecutive crossings of the Poincar\'e section followed by reflection about appropriately chosen streamwise-cross-stream planes.

The spanwise periodicity $\lambda_z=5$ is doubtless unrelated to a mode A instability despite the compatible wavelength. The typical features of mode A are absent from the wake, while mode C vortical structures clearly prevail, albeit with a two-fold subharmonic spanwise modulation. In order to discard any involvement of mode A in the solution observed at $R=3.8$, 4 consecutive normalised snapshots of the dominant mode for wavelengths $\lambda_z=2.5$ and $5$, both unstable according to figure~\ref{fig:floquet}, are shown in figures~\ref{fig:3DQR38}b and c, respectively. The spatial structure of the dominant eigenmode is clearly the same at both wavelengths, with minor differences that are perfectly imputable to a continuous transformation from one to the other. Moreover, the two modes are definitely of the C type, that is subharmonic, as normalised snaphots taken exactly two vortex-shedding cylces apart are identical in all respects, and those taken only one vortex-shedding cycle apart are related by the space-time symmetries described earlier. Therefore, the actual frequency and symmetry-breaking of the nonlinear solution at $R=3.8$ cannot be explained by a period-doubling bifurcation affecting also the already unstable two-dimensional vortex-shedding solution. The origin of the period-doubling must probably be sought in some modulational instability of the nonlinear solution at $R=3.4$.

\begin{figure*}
  \centering
  \begin{tabular}{ccc}
    $k$ & (a) $R=3.4$ & (b) $R=3.8$\\
    \raisebox{0.05\linewidth}{0} &
    \includegraphics[width=0.45\linewidth]{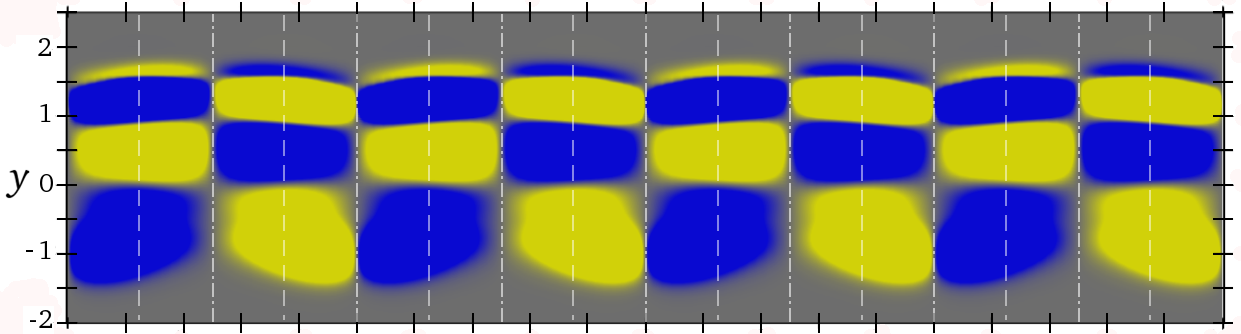} &
    \includegraphics[width=0.45\linewidth]{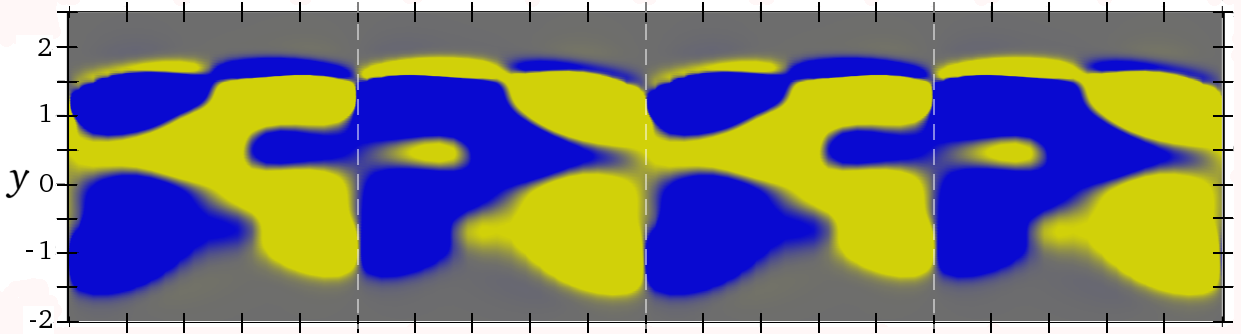}\\
    \raisebox{0.05\linewidth}{1} &
    \includegraphics[width=0.45\linewidth]{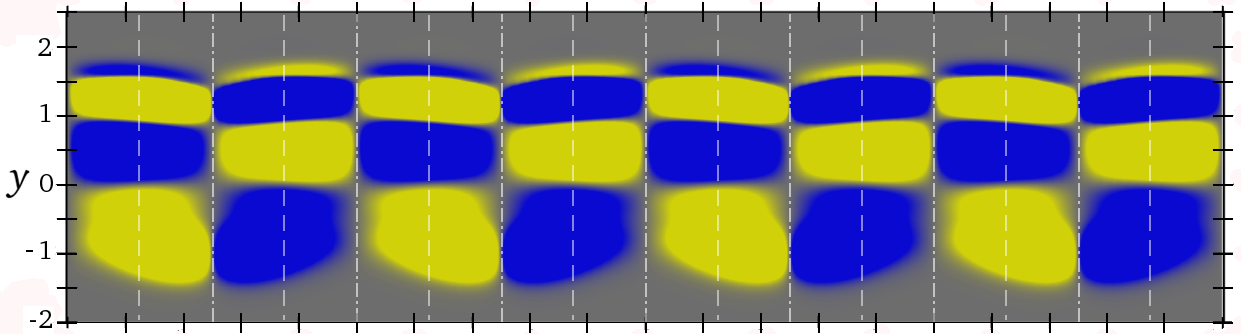} &
    \includegraphics[width=0.45\linewidth]{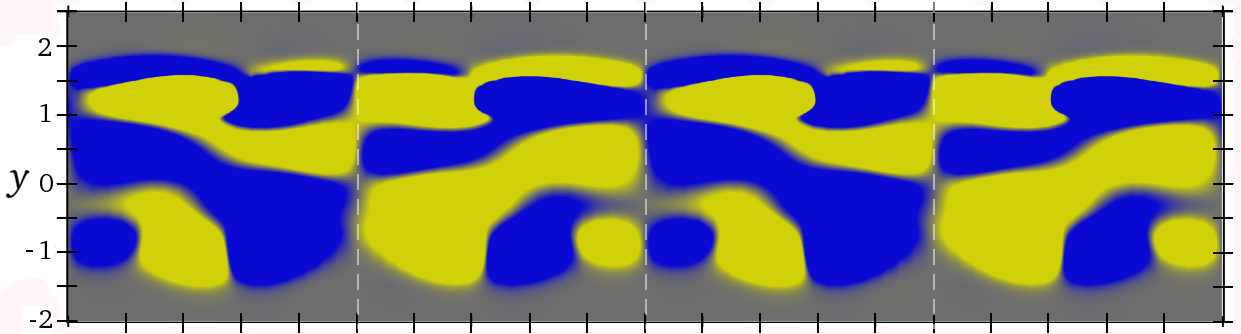}\\
    \raisebox{0.05\linewidth}{2} &
    \includegraphics[width=0.45\linewidth]{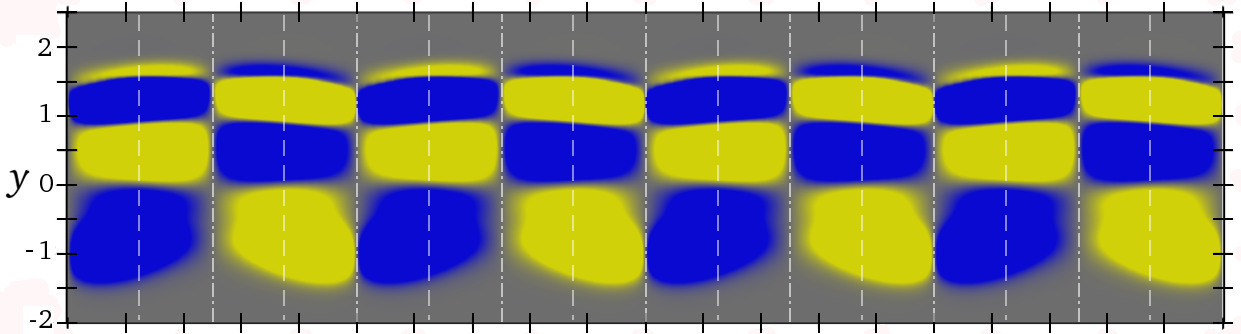} &
    \includegraphics[width=0.45\linewidth]{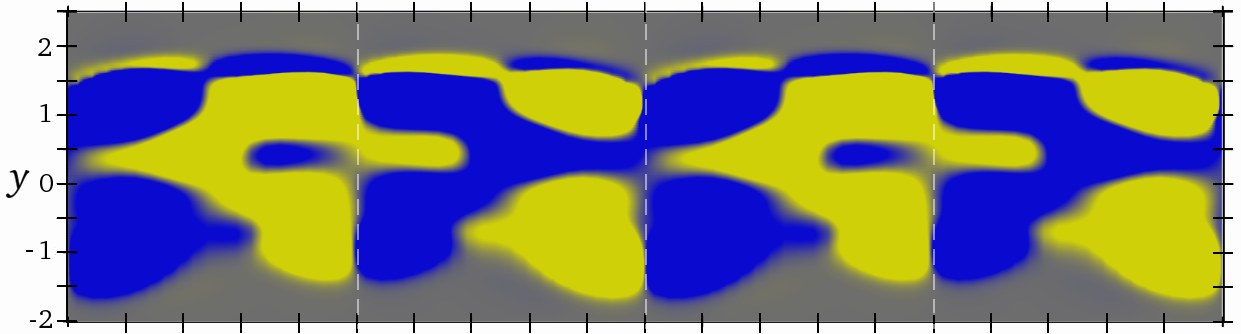}\\
    \raisebox{0.05\linewidth}{3} &
    \includegraphics[width=0.45\linewidth]{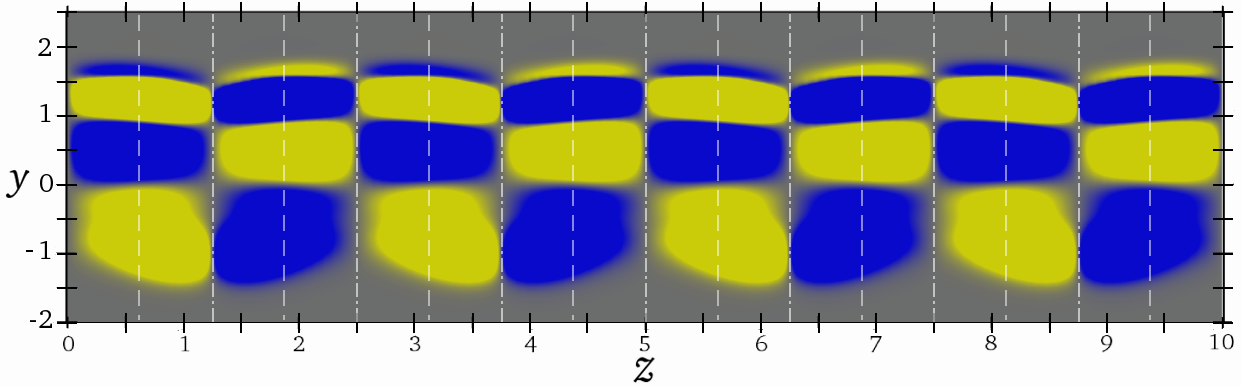} &
    \includegraphics[width=0.45\linewidth]{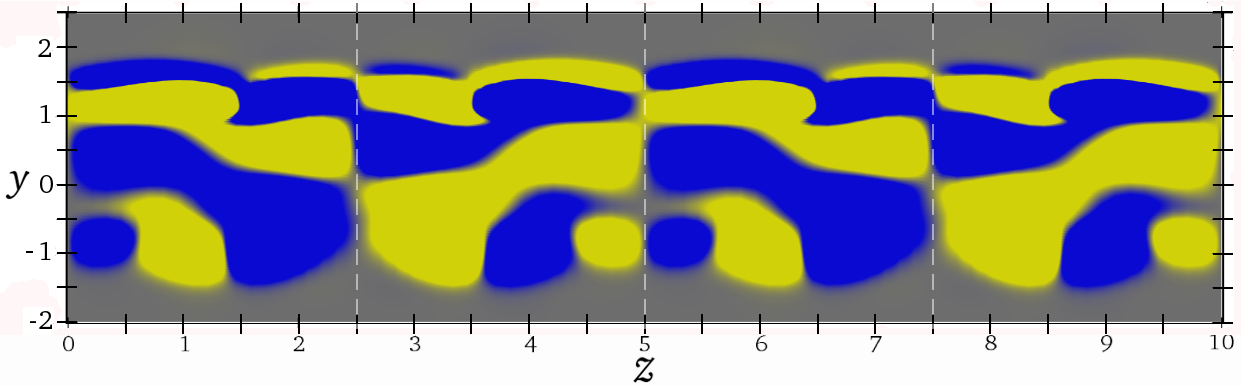}\\
  \end{tabular}
  \caption{Streamwise vorticity colourmaps $\omega_x\in[-0.1,0.1]$ (yellow for positive, black for negative) at $x=6$ for velocity ratios (a) $R=3.4$ and (b) $R=3.8$. Four consecutive crossings of the Poincar\'e section have been represented, respectively.}
  \label{fig:Omxx6R34R38}
\end{figure*}

Colourmaps of streamwise vorticity $\omega_x$ on a spanwise-cross-stream plane located at $x=6$ in the cylinder wake -the precise location of the K\'arm\'an vortex centre at the instants chosen for the snapshots- help pinpoint the actual spanwise wavelength and space-time symmetry of the three-dimensional solutions for $R=3.4$ and $R=3.8$ in figures~\ref{fig:Omxx6R34R38}a and b, respectively.
The exact repetition of the solution every two crossings of the Poincar\'e section is now evinced by the exact matching of crossings 0-2 and 1-3 for $R=3.4$. The symmetry planes at $z=z_0+(2j)\lambda_z/4=2.5\,j$ are perfectly identifiable, as is also clear that even/odd Poincar\'e crossings are mutually related by a reflection about planes at $z=z_0+(2*j+1)\lambda_z/4=1.25+2.5\,j$. Note that reflection symmetries, as is the case of the two symmetries discussed here, imply a change of sign of $\omega_x$, hence the change of colour. Both symmetries combined result in a third symmetry that is detectable as a shift by $\lambda_z/2=1.25$ from one Poincar\'e crossing to the next.

For $R=3.8$, the planes $z=z_0+(2j)\lambda_z/4=2.5\,j$ are no longer symmetry planes, although the solution keeps the shadow of a reflectional symmetry to a certain extent. Remnants of the original spanwise periodicity $\lambda_z=2.5$ are also still perceptible, but the actual periodicity has indeed doubled to $\lambda_z=5$. The only remaining symmetry, which is of a spatio-temporal nature, relates even/odd crossings by a reflection about planes at $z=z_0+(2j)\lambda_z/4=2.5\,j$.

\begin{figure}
  \centering
  \begin{tabular}{cc}
    (a) & (b) \\
    \includegraphics[width=0.45\linewidth]{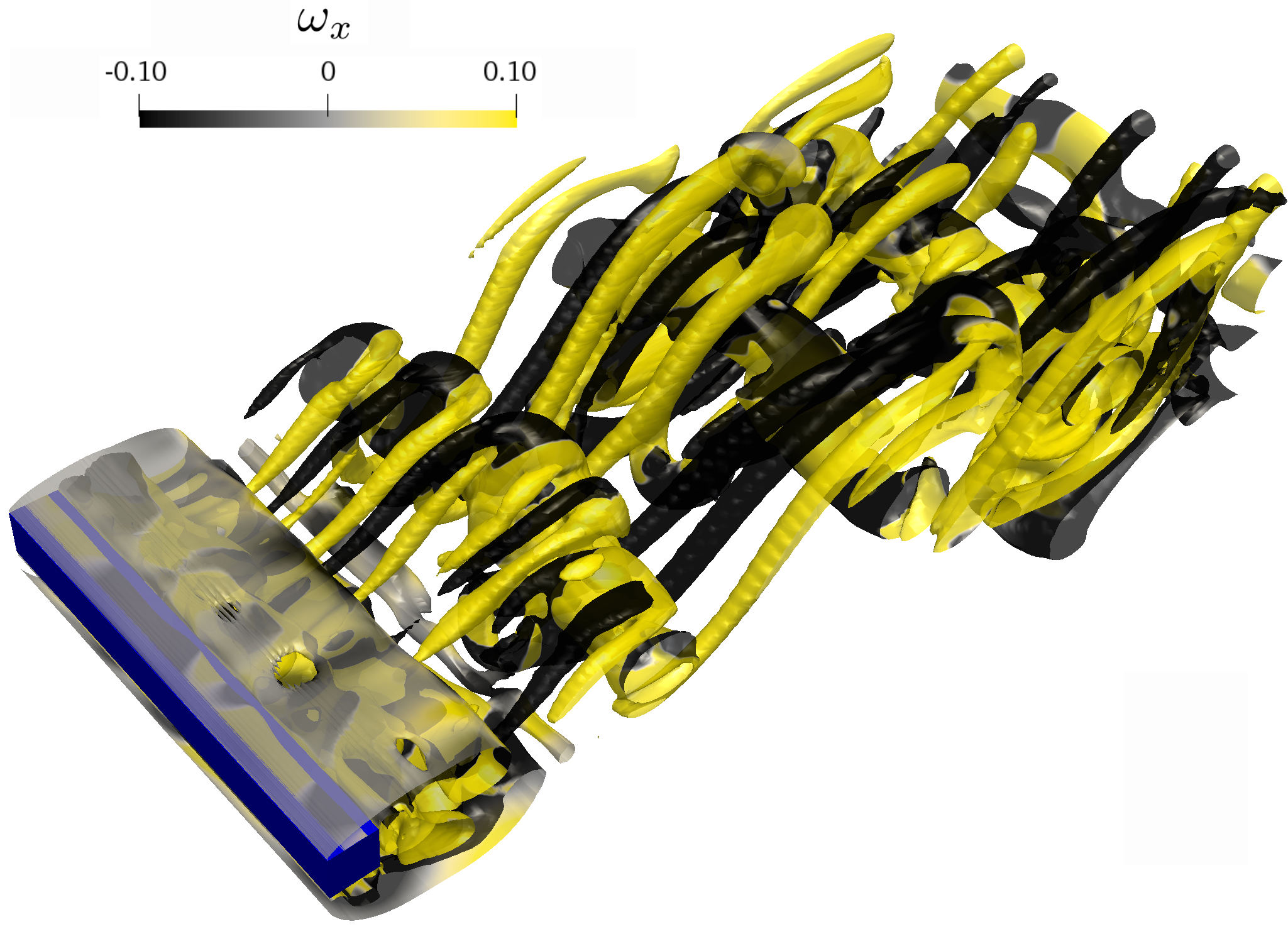} &
    \includegraphics[width=0.5\textwidth]{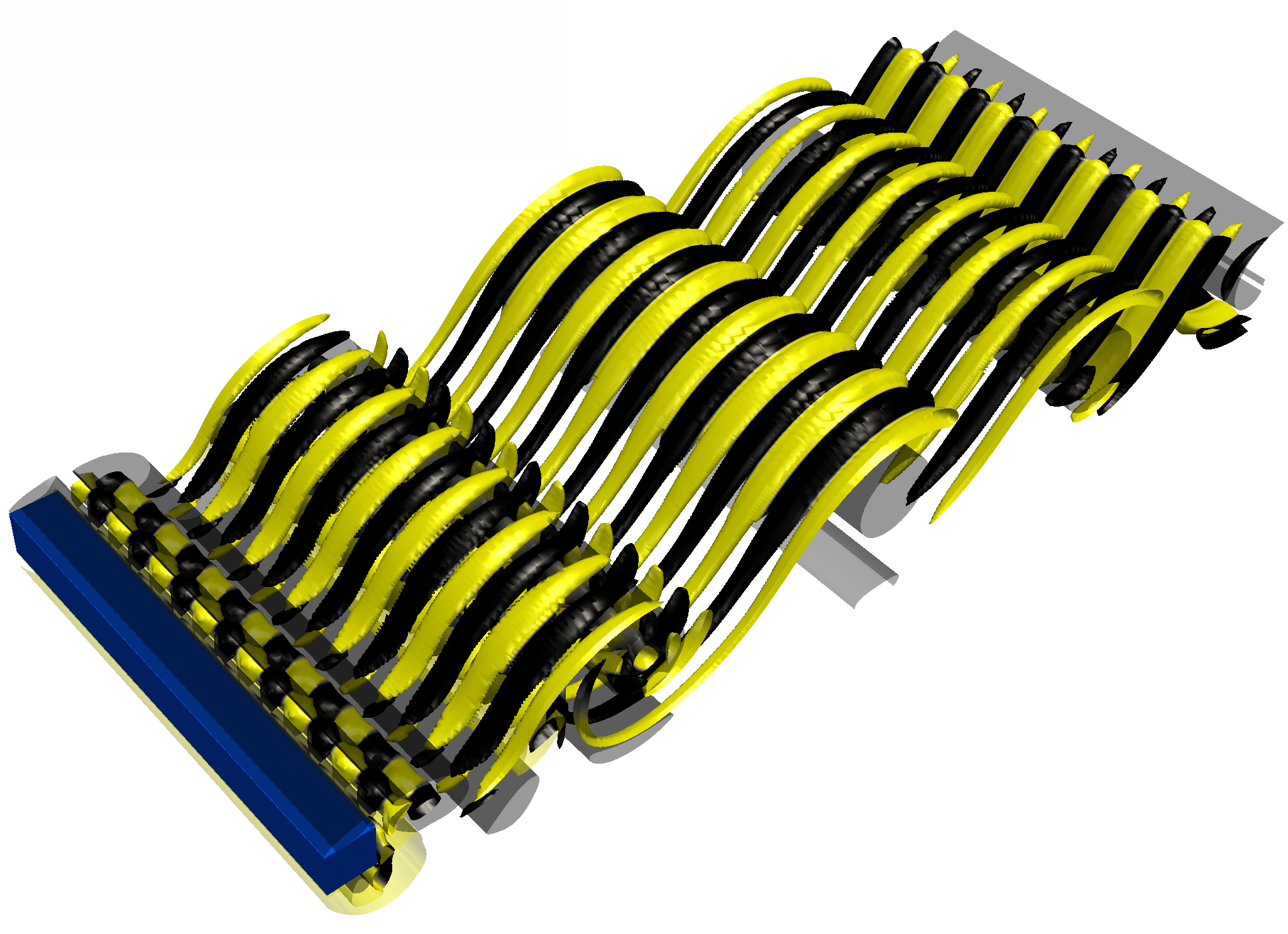}
  \end{tabular}
  \caption{Instantaneous snapshot of (a) the chaotic three-dimensional solution at $R=5.357$ at a random crossing of the Poincar\'e section, and (b) short wavelength dominant eigenmode at $(R,\lambda_z)=(5.357,1.429)$. Iso-surfaces and colours as for figure~\ref{fig:3DQR38}.}
  \label{fig:3DQR54}
\end{figure}

A single snapshot of a random crossing of the Poincar\'e section for the chaotic solution at $R=5.357$ is shown in figure~\ref{fig:3DQR54}a.
The K\'arm\'an vortices are still clearly identifiable as are also the elongated vortical structures along the braids, which are clearly arranged in pairs forming horse-shoe vortices on top of the primary spanwise vortices. The flow remains farily well organised in the near wake, but clearly breaks into turbulent dynamics a little further downstream. The typical wavelength of the vortical structures remains of about $\lambda_z\sim2.5$, but there is no spanwise periodicity other than that artificially imposed by the fundamental Fourier mode employed in the simulation, here $L_z=10$. A second emerging sub-dominant mode of a much shorter spanwise wavelength, already visible at the far end of the Floquet spectrum of figure~\ref{fig:floquet}, is shown in figure~\ref{fig:3DQR54}b. The mode is still stable at $R=5.357$, but bifurcates with critical $\lambda_z\simeq10/7$ for $R<6.5$. Its characteristically short wavelength and synchronous nature identify it as mode-B type. It is however not detectable in the turbulent snapshot and its growth rate once unstable remains well below that of the dominant eigenmode for $\lambda_z\simeq2.5$, which in fact has shifted to even larger values for these high values of $R$. It is therefore improbable that this other mode plays any important role in the transition to turbulence of the wake past the square cylinder in upstream shear.

A quasi-static variation of $R$, which is out of the scope of this study, would be required to cast light on the actual path leading to turbulence, but the presence of a period-doubling bifurcation suggests that a period-doubling cascade might play a role. There is also evidence suggesting that the complexification of the dynamics might arise from the competition of an increasing number of unstable wavelengths of the same instability type (actually a continuum if an infinite-span cylinder was to be considered), so that an instability of the Eckhaus or Benjamin-Feir type might in fact be held responsible, both for the period-doubling bifurcation reported here and the route to chaotic dynamics \citep{LewekePRL1993,LewekePRL1994,LewekeJFM1995}.

%%% Further discussion of 3D modes   
%%% Linear

\begin{figure*}
  \centering
  \begin{tabularx}{\textwidth}{cccc}
    $k$ & (a) $(R,\lambda_z)=(3.4,2.5)$ & (b) $(R,\lambda_z)=(3.8,2.5)$ & (c) $(R,\lambda_z)=(3.8,5)$\\
    &&&\\
    \raisebox{0.06\linewidth}{0} &
    \includegraphics[width=0.31\linewidth]{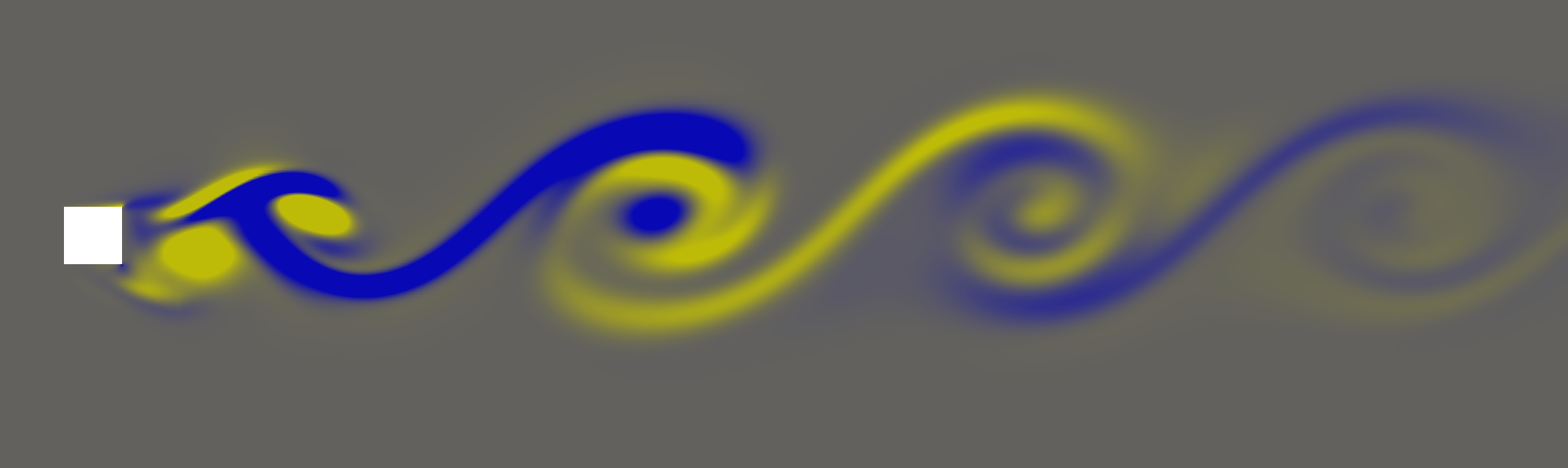} &
    \includegraphics[width=0.31\linewidth]{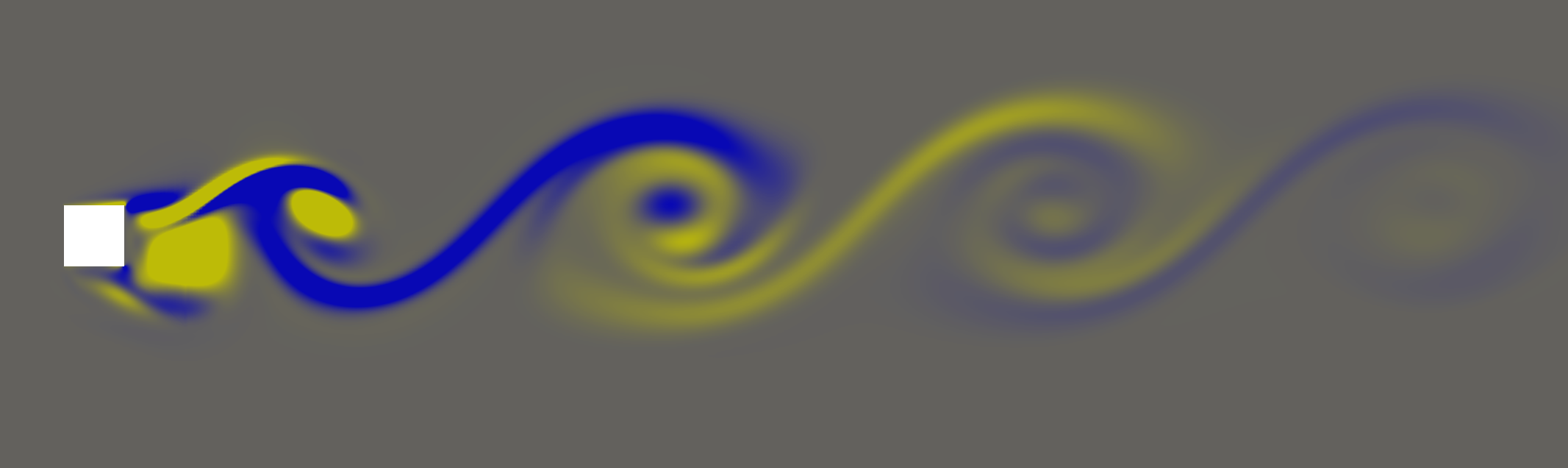} &
    \includegraphics[width=0.31\linewidth]{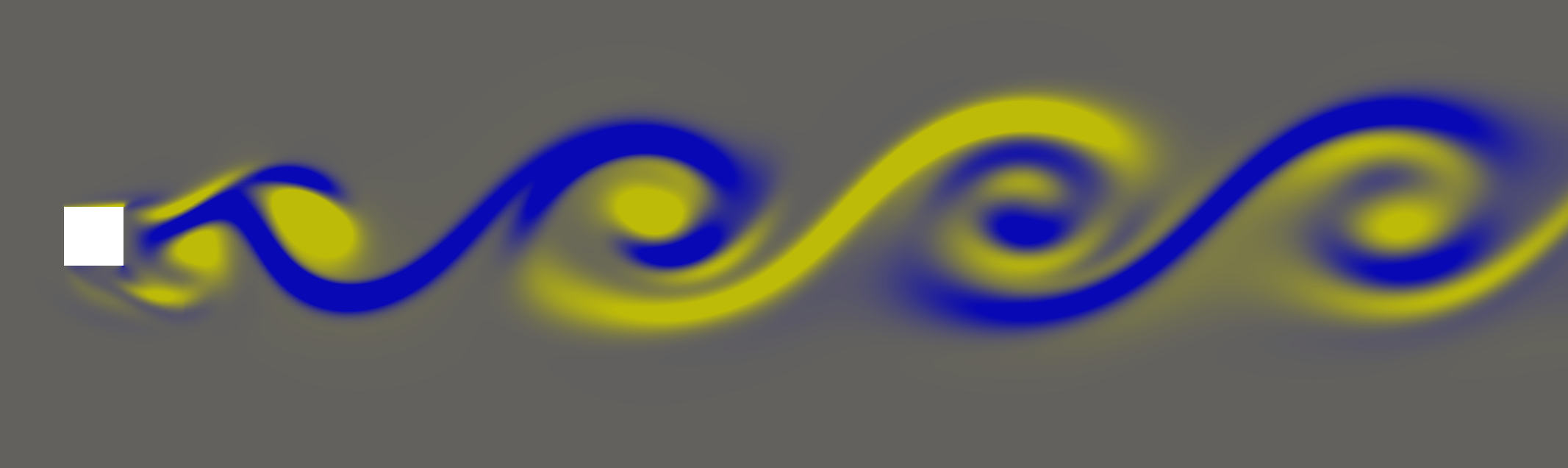}\\
    \raisebox{0.06\linewidth}{1} &
    \includegraphics[width=0.31\linewidth]{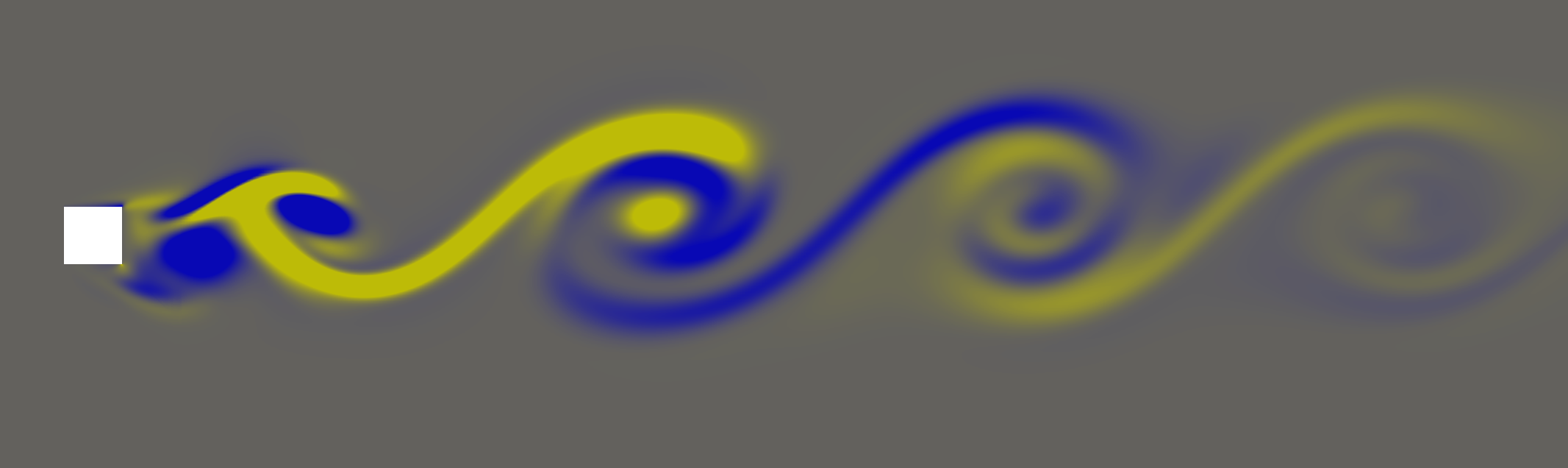} &
    \includegraphics[width=0.31\linewidth]{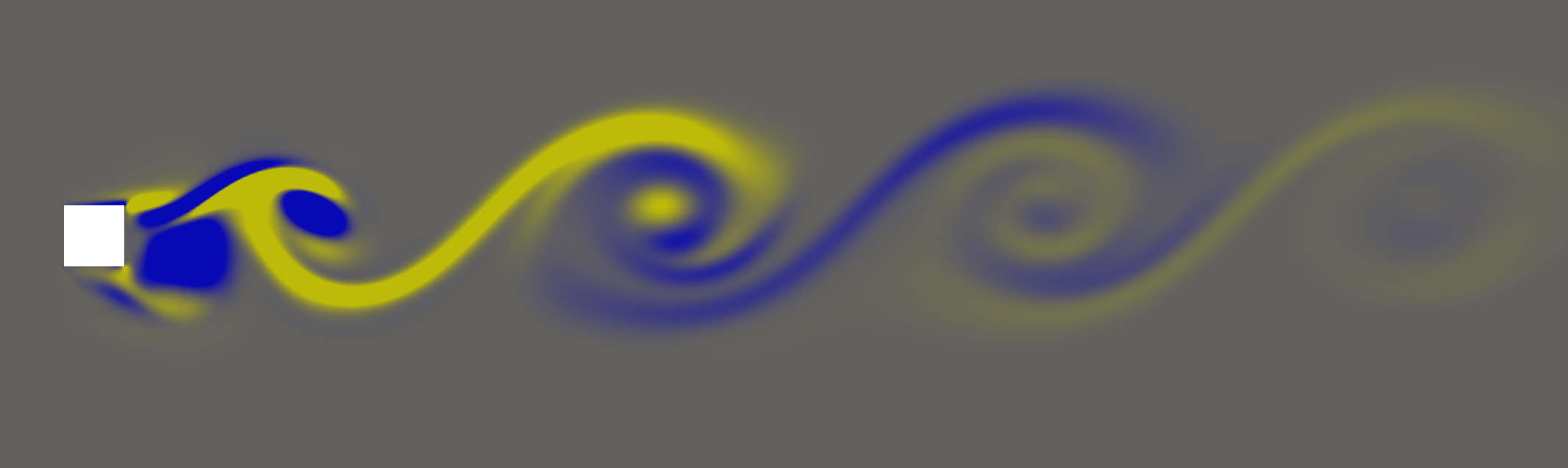} &
    \includegraphics[width=0.31\linewidth]{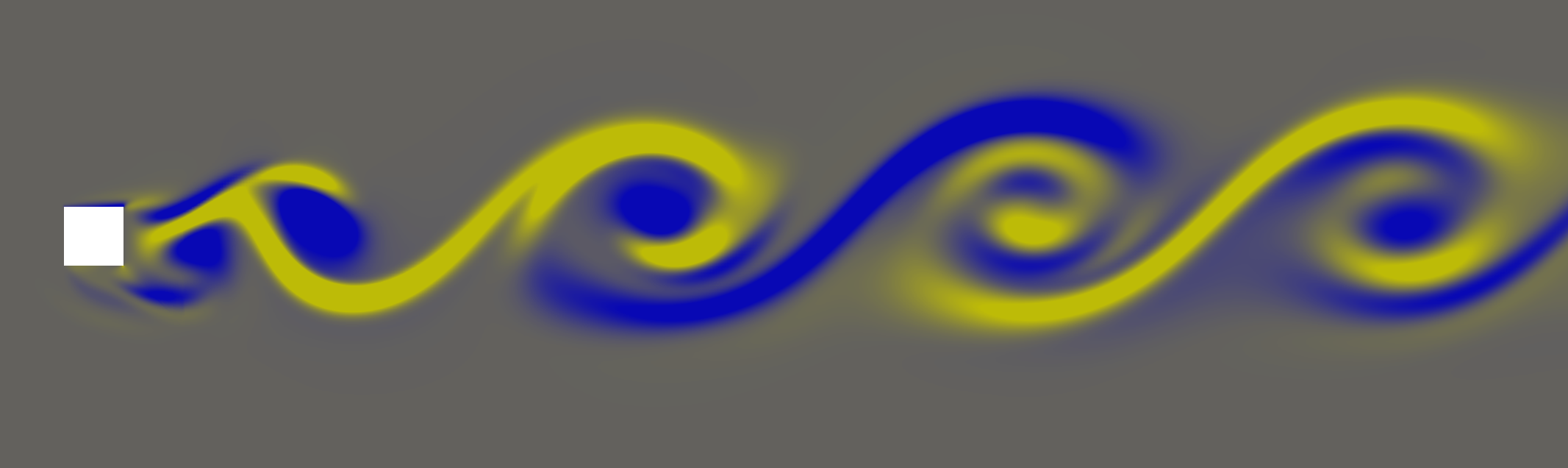}\\
    \raisebox{0.06\linewidth}{2} &
    \includegraphics[width=0.31\linewidth]{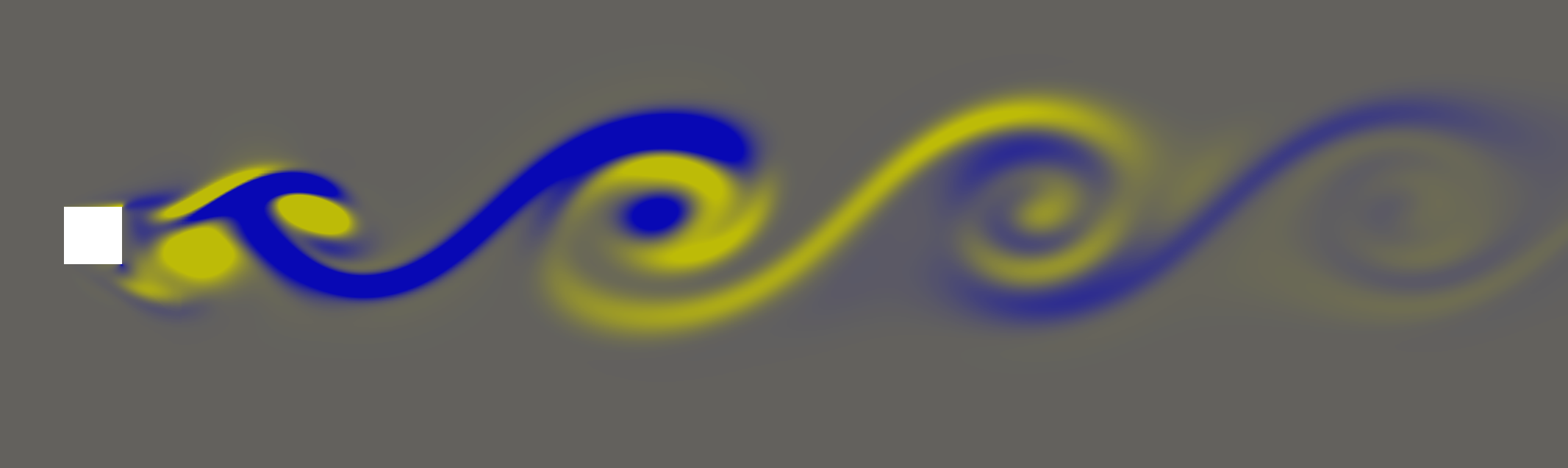} &
    \includegraphics[width=0.31\linewidth]{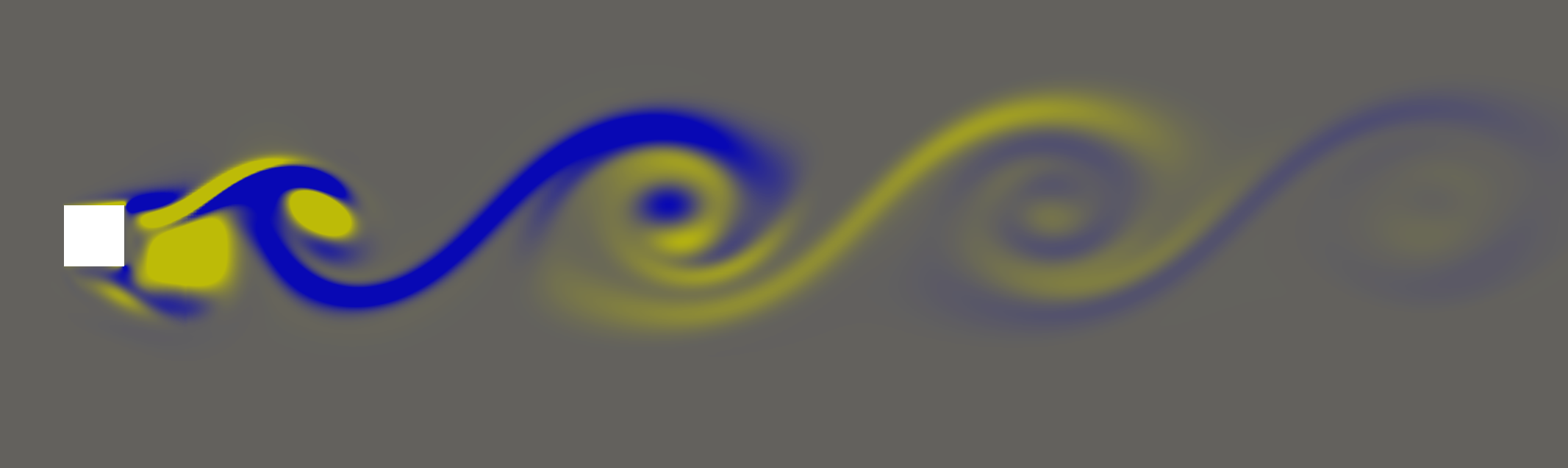} &
    \includegraphics[width=0.31\linewidth]{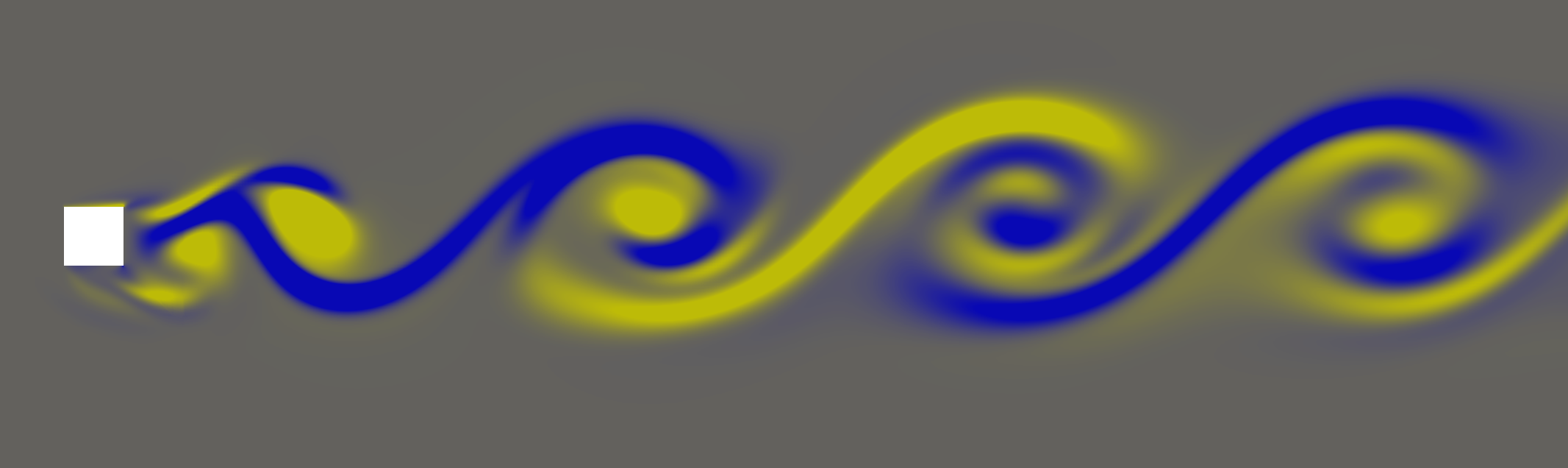}\\
    \raisebox{0.06\linewidth}{3} &
    \includegraphics[width=0.31\linewidth]{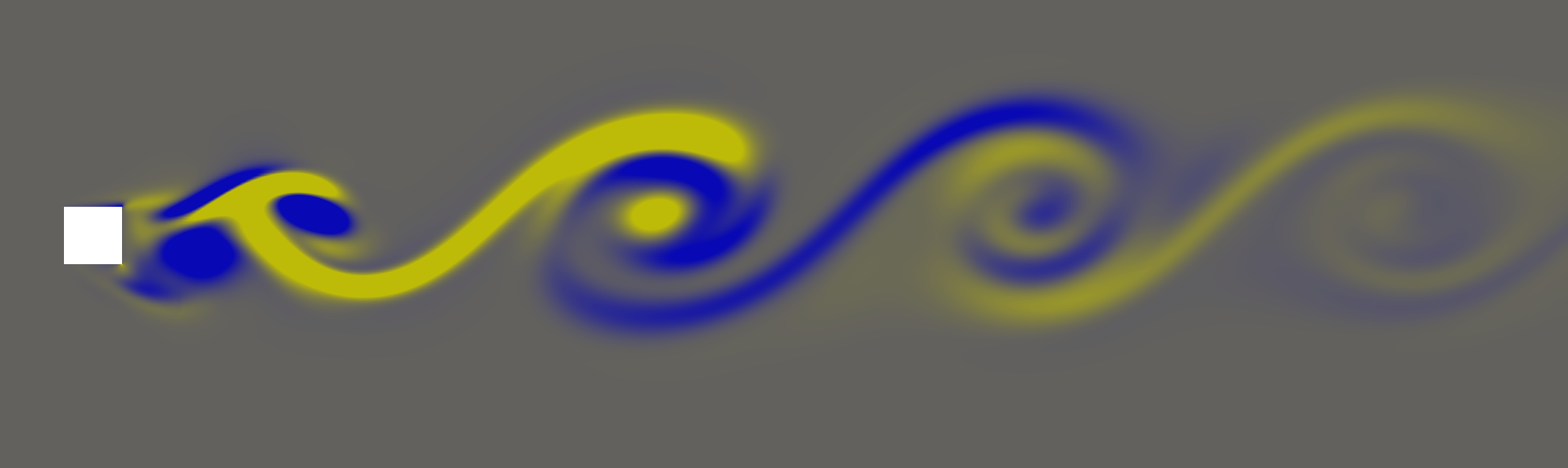} &
    \includegraphics[width=0.31\linewidth]{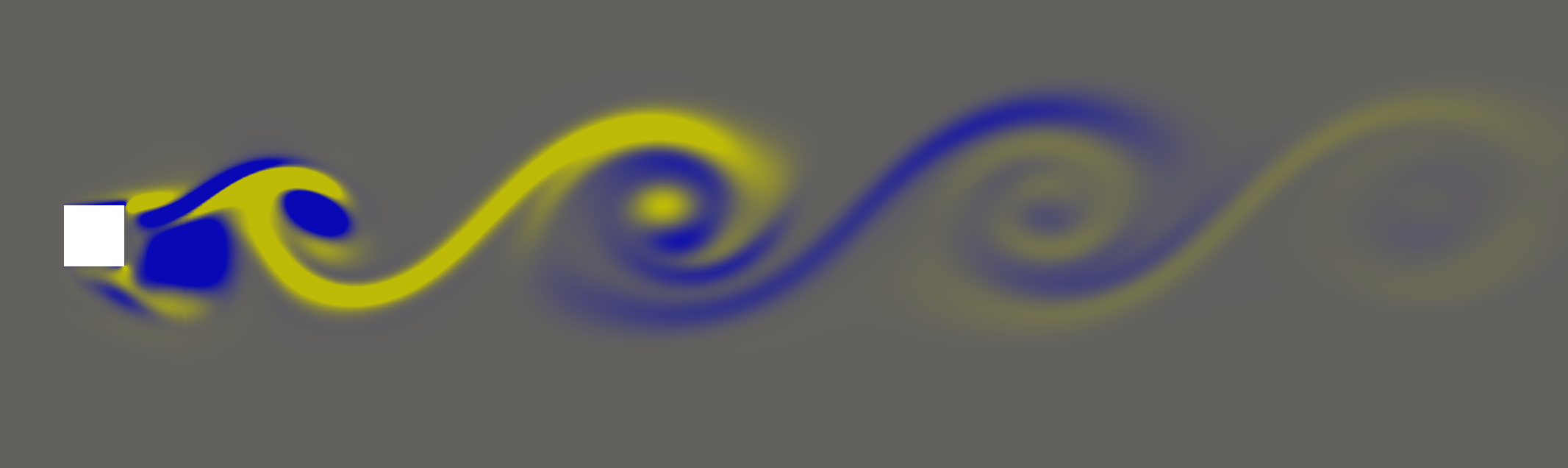} &
    \includegraphics[width=0.31\linewidth]{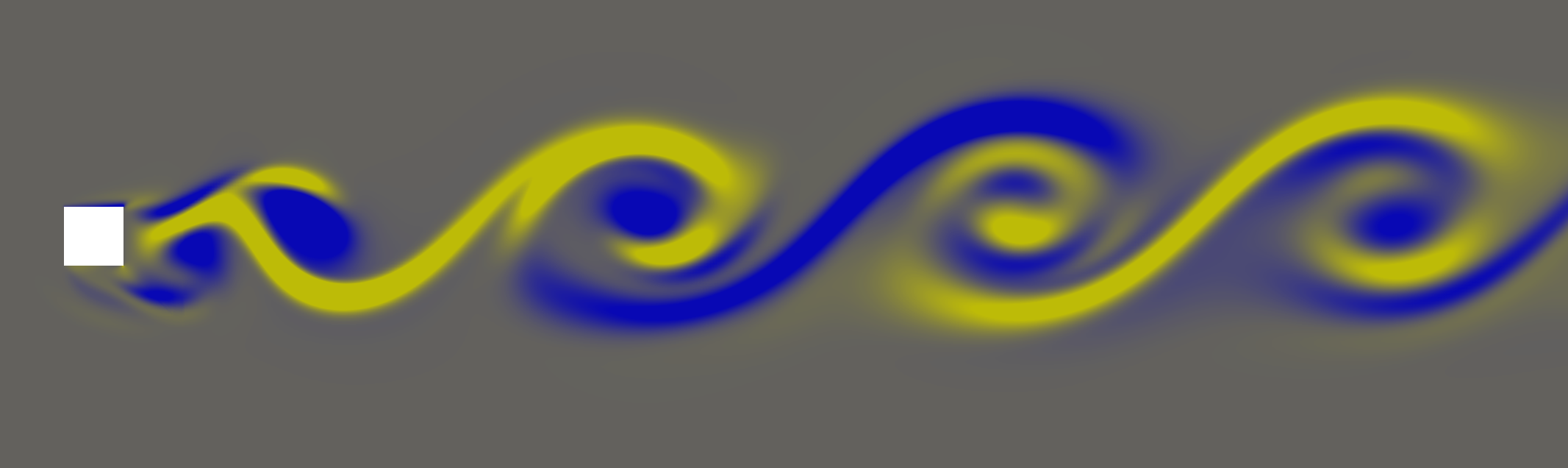}\\
  \end{tabularx}
  \caption{Most unstable mode for $Re_B=56$ and (a) $(R,\lambda_z)=(3.4,2.5)$ (Movie 6), (b) $(R,\lambda_z)=(3.8,2.5)$ (Movie 7), and (c) $(R,\lambda_z)=(3.8,5)$ (Movie 8). Normalised streamwise vorticity ($\omega_x\in[-1,1]$) colourmaps at plane $z=z_0+\lambda_z/4$ at four consecutive Poincar\'e crossings (marked and labeled in figure~\ref{fig:floquet}a for case (a)).}
  \label{fig:mode1}
\end{figure*}

Although the dominant eigenmode curves in the Floquet spectrum of figure~\ref{fig:floquet} are smooth in the range of wavelengths observed in nonlinear solutions, evidence that they represent the same actual instability requires a close inspection of the eigenfields. Four snapshots taken at consecutive crossings within the linear regime of a purposefuly designed Poincar\'e section, well before three-dimensional modal energy reaches nonlinear saturation, illustrate the flow topology of the instability for $(R,\lambda_z)=(3.4,2.5)$, $(3.8,2.5)$ and $(3.8,5)$ in figure~\ref{fig:mode1}.
Shown are normalised streamwise vorticity $\omega_x$ colourmaps on a streamwise-cross-stream plane halfway between two consecutive symmetry planes of the eigenmode. Comparison of all three cases points at a common origin of all three instabilities, which in fact are one and the same mode just deformed by a continuous change of the parameters. The only apparent effect, besides the stretching/compressing associated to changes in $\lambda_z$ which was already obvious from figures~\ref{fig:3DQR38}b and c, is that the instability dissipates faster along the wake when $R$ is increased or the wavelength shortened towards criticality.

The instability grows preferentially in the very near wake, where the perturbation $|\omega_x|$ is maximum, and the period-doubling nature of the mode is obvious from the coincident normalised $\omega_x$ fields of any two snapshots taken exactly two vortex-shedding cycles apart. The appropriate choice of the spanwise plane used for the representation also substantiates the evident fulfilment of a space-time symmetry through a mere change of sign of $\omega_x$ between any two consecutive Poincar\'e crossings. The topology and symmetries of the eigenmode consumately foreshadow the properties of the nonlinear three-dimensional solution observed for $R=3.4$ and act as harbingers of the main features of the flow arrangement for $R=3.8$, but cannot by themselves explain the actual spanwise periodicity or the period-doubled nature of the latter. We surmise that the $\lambda_z=5$ instance of the eigenmode might be interfering with the dominant $\lambda_z=2.5$ eigenmode at $R=3.8$, such that the nondimensional features arise from an Eckhaus- or Benjamin-Feir-type instability. Simulations in much longer spanwise domains would be required to sufficiently approximate the continuous band of unstable wavelengths that would be required to allow for mode competition and thus analyse the bifurcation in the light of modulational instabilities.

\section{Conclusions}\label{sec:conc}

We have investigated the incompressibe Newtonian flow past an infinite square cylinder immersed in the wake of an upstream flat plate separating two streams of different velocities. The cylinder is therefore subject to an incoming shear flow consisting of a (nearly) step velocity profile instead of the more classic case of homogeneous shear, which presents serious limitations in its realisation both experimentally and numerically, particularly so when three-dimensionality is to be expected and Couette flow has become prone to subcritical transition. The splitter plate length and upstream distance from the cylinder being fixed, the problem is governed by two parameters, namely the bulk Reynolds number $Re$ and the velocity ratio $R$. Here we have chosen to keep the bottom stream Reynolds number $Re_T=56$ fixed and vary only $R$, such that the closely related shear parameter $K\equiv2(R-1)/(R+1)$ and $Re\equiv(R+1)Re_B/2=28(R+1)$ are linked.

Several aerodynamic performance parameters have been assessed along the path from the steady symmetric solution towards the highly asymmetric and spatio-temporally chaotic wake dynamics. Pressure forces, particularly on the upper half of the front wall, have been shown to be the main contributor to the drag coefficient $C_d$, which increases with $R$, particularly after the onset of vortex shedding.
%The stagnation point climbs fast on the front surface as $R$ is increased, an effect that can be ascribed solely to the associated increase in $K$, and settles in the vicinity of the top front corner with an increasing value of the pressure coefficient $C_p^s$. At the rear, the base point first descends and then climbs, on average, as the wake becomes three-dimensional. The base point location wanders wildly at high values of $R$, but the base pressure coefficient $C_p^b$ is quite insensitive and describes a slowly descending trend with barely perceptible time-fluctuations.
The lift coefficient $C_l$ is dominated by downward friction on the front wall at low values of $R$, thus producing a net downforce. Time-dependence sets the pressure difference between top and bottom on an increasingly growing trend that ends up yielding positive lift. Force coefficient fluctuations $C_l'$ and $C_d'$ increase while the solution remains two-dimensional but stagnate to rather constant values once three-dimensionality sets in. The Strouhal umber $St$ is rather stable, with a slightly increasing trend mainly pulled by the increase in $Re$, while the independent effect of $K$ is unknown but probably decreasing.
%The recirculation bubble length $l_r$ grows with $R$ for the steady wake and recedes thereafter, while the vortex formation region $l_f$ shrinks all along, in analogy to the circular cylinder case. It would seem that the $Re$ increase rather than that of $K$ dictates the evolution of both. 

Much of the behaviour exhibited by the flow configuration analysed here bears compelling analogy to the somewhat related case of bluff bodies immersed in homogeneous upstream shear. It is therefore probable that the velocity jump across the bluff body cross-stream length-scale is in fact to be held responsible for most of the observed phenomena, the actual shape of the velocity profile (linear or step-like) playing only a subsidiary role.

The symmetric case $R=1$ ($Re=56$, $K=0$) is steady, implying that the presence of the splitter plate has a stabilising effect, through the viscous blocakge effect, on the flow past a square cylinder, which is known to undergo a Hopf bifurcation somewhat earlier at $Re^H\simeq45$. The first bifurcation upon increasing $R$ remains the two-dimensional onset of time-periodicity, but this only happens for $R=2.1\pm1$ ($Re=86.8\pm2.8$, $K=0.71\pm0.04$). It cannot be concluded from present results whether the delay in the onset of time-dynamics is due to the presence of the splitter plate or whether the upstream velocity ratio also has a stabilising effect.

Once vortex-shedding is in place, the K\'arm\'an vortex street in the cylinder wake follows the same trends observed in the literature. The vortices on the high velocity side become stronger and rounder with shear, while those on the low velocity side are weak, elongated and dissipate fast in the wake.

Stability analysis of the time-periodic two-dimensional vortex-shedding solution to three-dimensional perturbations reveals three distinct modes. A long wavelength mode, possibly related to the mode A that is ubiquitous in the wake transition regime of the classic circular and square cylinders, is dominant but stable for velocity ratios $R\leq3$. At the high $R$ end, a very short wavelength mode arises and bifurcates for $R\gtrsim6$, but is never even close to dominant. The last mode, of an intermediate wavelegth, prevails for $R>3$ and is the one responsible for wake three-dimensionalisation along the increasing $R$ path followed here. This mode bifurcates at $R^{\mathrm{C}}\simeq3.1$ ($Re^{\mathrm{C}}\simeq115$, $K^{\mathrm{C}}\simeq1.02$) with critical wavelength $\lambda_z^{\mathrm{C}}\simeq2.4$, exhibits elongated streamwise-cross-stream vortical structures along the braids connecting consecutive clockwise K\'arm\'an vortices, and is subharmonic (period-doubling). While breaking the spanwise translational invariance SO(2) of the two-dimensional time-periodic solution, which belongs to the O(2)=Z$_2\times$SO(2) symmetry group class, it still preserves a collection of symmetries. Besides repeating every two vortex-shedding cycles, the eigenfields are mirror symmetric with respect to a discrete number of spanwise planes every $\lambda_z/2$, and also invariant under the evolution by half a period (one vortex-shedding cycle) followed by either reflection about spanwise planes located halfway between contiguous reflection-symmetry planes or shift by half a wavelength $\lambda_z/2$. The wavelength, topology and symmetries of the mode are typical of the mode C that is characterisitc of the wake transition regime of open rings or square cylinders at incidence, and known to evolve from the quasiperiodic mode QP previously reported for circular and square cylinders. Converged three-dimensional nonlinear solutions slightly beyond the bifurcation point exhibit time dynamics and wake structures sharing all properties that are characteristic of mode C.

At somewhat higher $R\in(3.4,3.8)$, the flow undergoes a tertiary bifurcation that doubles the period yet again (nonlinear solutions repeat every 4 vortex-shedding cycles) and also doubles the spanwise wavelength. The instability is therefore of the modulational type and amplifies perturbations that have double the wavelength of the original spanwise-periodic pattern. The resulting nonlinear solution has been observed in a computational domain with $L_z=5$, allowing only for a discrete number of periodic patterns with wavelengths $\lambda_z=L_z/j=\{5,2.5,1.667,1.25,...\}$. The spanwise mirror symmetry has finally been broken and the only remaining invariance besides the four-fold time-periodicity is a space-time symmetry that recovers the solution by aplying a reflection after the evolution by half a period (two vortex-sheedding cycles). There is no evidence of a secondary bifurcation of the two-dimensional solution exhibiting a $\lambda_z=5$ other than a stretched version of the same mode C, which has also become unstable. As a matter of fact, the range of unstable wavelengths for which mode C is unstable increases with $R$ but the dominant wavelength remains quite fixed at $\lambda_z\simeq2.5$ for $R\leq4$ and only migrates to increasingly wider values beyond this point. It is probable that the observed period-doubling bifurcation must be understood in the light of an Eckhaus instability, whereby a periodic pattern becomes modulationally unstable following the bifurcation of a widening spectrum of competing wavelengths.

Very shortly after the period-doubled three-dimensional pattern, a further increase in $R$ brings spatio-temporal chaos with it. Simulations in much longer domains, allowing for a larger set of unstable wavelengths, would be required in the wake transition regime so as to elucidate the actual nature of the route to chaotic dynamics and turbulence, but this is beyond the scope of the present investigation.

\section{Acknowledgements}
This work was supported by the Spanish Government grants FIS2016-77849-R and PID2020-114043GB-I00, and by the Catalan Government grant 2017-SGR-00785, respectively. Part of the computations were done in the Red Espa\~{n}ola de Supercomputaci\'{o}n (RES), the Spanish supercomputer network, under grants FI-2019-1-0023 and FI-2018-3-0030.

% \bibliographystyle{elsarticle-harv} 
% \bibliography{mybibfile}

\appendix

\section{Resolution study}\label{app:validation}
\setcounter{table}{0}
\setcounter{figure}{0}

The time-stepping scheme and space discretisation used here for the flow past a square cylinder immersed in the wake of an upstream splitter plate separating two streams with different velocities have been validated against literature results using as benchmark the closely related problem of the flow past a stand-alone square cylinder in homogeneous flow at zero angle of attack and at comparable Reynolds numbers.

In the present study, the bottom stream Reynolds number has been kept fixed to $Re_B=56$ throughout, while the top-to-bottom velocity ratio has been varied in the range $R\in[1,5.357]$. The corresponding top stream Reynolds number is then $Re_T=R  Re_B\leq300$, and it makes therefore sense to use benchmark results for the classical square-cylinder-flow configuration at $Re=300$ for mesh validation. The flow remains nevertheless two-dimensional for $R\lesssim3.1$, so that the in-plane mesh for two-dimensional simulations might be safely based on comparison against square cylinder results anywhere beyond $Re\geq115$.

The stand-alone square cylinder at $Re=150$, just short of entering the wake transition regime, has been selected as the benchmark for the design of the in-plane two-dimensional mesh for all two-dimensional cases. Three different levels of refinement have been tested as shown in table~\ref{tab:gridres2d}.

\begin{table}
  \begin{center}
    \begin{tabular}{lrrrrrrrrr}
      & \multicolumn{1}{c}{$d_w$} & \multicolumn{1}{c}{$A_c$} & \multicolumn{1}{c}{$N_{xy}$} & $St$ & [\%] %$\varepsilon_{St} (\%)$ 
      & $\langle C_{d}\rangle$ & [\%] %$\varepsilon_{C_{d}} (\%)$ 
      & $C_{l}'$ & [\%] %$\varepsilon_{C_{l}^{\mathrm{rms}}} (\%)$
      \\   
      \hline
      Mesh A & $1.0\times10^{-2}$ & $1.0\times10^{-4}$ & 84678 & 0.162 & [+1.25] & 1.543 & [+3.10] & 0.302 & [+5.59]\\
      Mesh B & $2.4\times10^{-3}$ & $2.4\times10^{-5}$ & 95294 & 0.160 & [+0.00] & 1.500 & [+0.33] & 0.291 & [+1.57]\\
      Mesh C & $1.2\times10^{-3}$ & $1.2\times10^{-5}$ & 125204 & 0.160 &      & 1.495 &       & 0.286 &       \\
      \cite{WangTTU2016} [Case 1] & & $1.7\times10^{-4}$ & & 0.160 & & 1.474 & & 0.285 & \\
      \cite{Ali2009} [Case E] & $1.0\times10^{-2}$ & $ 1.0\times10^{-4}$ & 228800 & 0.160 & & 1.470 & & 0.285 & \\
      \cite{DoolanAIAAJ2009} & & & & 0.156 & & 1.440 & & 0.293 & \\
      \cite{FrankeJWEIA1990} & $3.8\times10^{-3}$ & & 6688 & 0.165 & &  1.560 & & & \\
      %	\hline
    \end{tabular}
  \end{center}
  \caption{Resolution study for the two-dimensional flow past a stand-alone square cylinder at $Re=150$. $d_w$ denotes the maximum thickness of wall-adjacent elements, $A_c$ is the minimum cell area, and $N_{xy}$ is the total cell count. Reported are the vortex-shedding non-dimensional frequency $St$, the mean drag coefficient $\langle C_d\rangle$ and the rms of the lift coefficient $C_l'$, alongside the relative error with respect to mesh C results in brackets.}\label{tab:gridres2d}
\end{table}

In-plane cell counts of $N_{xy}=84678$ (Mesh A), 95294 (Mesh B) and 125204 (Mesh C) have been deployed, with maximum thickness of wall elements $d_w=10^{-2}$, $2.4\times10^{-3}$ and $1.2\times10^{-3}$, respectively, and minimum cell areas of $A_c=10^{-4}$, $2.4\times10^{-5}$ and $1.2\times10^{-5}$. The vortex-shedding frequency (Strouhal number $St$) compares favourably with literature results, particularly so with the well-resolved computations of case 1 by \citet{WangTTU2016} and case E by \citet{Ali2009}. The same goes for the r.m.s. of the lift coefficient $C_l'$. The mean drag coefficient $\langle C_d\rangle$ is a little off by about 1.5\%, still within acceptable deviation. Futhermore, while the coarsest mesh A produces still considerable error with respect to the finest Mesh C, the intermediate Mesh B falls within 1.5\% accuracy for all three monitored parameters. The maximum wall distance of the first layer of cells, measured in wall units ($y^+\equiv y u_{\tau}/\nu$, with $u_{\tau}\equiv\sqrt{\tau_w/\rho}$ the friction velocity) resulted in 1.4, 0.3 and 0.14 for meshes A, B and C, respectively, always well inside the viscous sublayer of the boundary layer. In combination with a very slow progression of cell thickness away from the wall, this is a good indication that boundary layers are well resolved. Accordingly, Mesh B has been taken as sufficiently accurate for two-dimensional simulations with $R\leq3.1$.

Mesh B has been selected also as the base for three-dimensional simulations at $R>3.1$ and two different three-dimensional resolutions have been tested for the stand-alone cylinder at $Re=200$. Results and comparison with published data are summarised in table~\ref{grids3d}.
\begin{table}
  \begin{center}
    \begin{tabular}{ l c c c c  c c p{1.35cm} c c c}
      & \\ % put some space after the caption
      %	\hline
      & $L_{z}$ & $L_x^u$ & $L_x^d$ & $H$ & $N_{z}$ &$N_{xy}$ &$N (\times10^{6})$  & $St$ & $C_l'$ & $\langle C_{d}\rangle$ \\
      \hline
      %		Mesh B 2D& 0.156&0.338&1.530&---&9D&25.5D&16D&0&95294&---\\
      Mesh B$_{1}$
      &18 &9 &25.5 &16 &32 & 95294 &3.05
      & 0.156 &0.298  &1.497
      \\
      Mesh B$_{2}$
      &9&9&25.5&16&32&95294&3.05
      &0.154&0.285&1.493 \\
      \cite{MahirIJTS2017}
      &6&7.5&19.5&15&25&22479&0.56
      &0.154&---&1.518
      \\
      \cite{LuoJFS2007}&---  &---&---&---&---&---&---&0.160&---&---\\
      \cite{SahaIJHFF2003}
      &6&6&17.5&10&22&14240&0.31
      &0.166&---&1.590
      \\
      \cite{LuoPoF2003}&--&---  &---&---&---&---&---&0.159&---&---\\
      \cite{SohankarPoF1999}
      &6&D&12.5&18&25&20449&0.51
      &0.157&---&1.390
      \\	
      \cite{OkajimaJSMEIJF1990}
      &---   &---&---&---&---&---&---&0.136-0.154&---&---\\
      \cite{OkajimaJFM1982}
      &---   &---&---&---&---&---&---&0.139-0.145&---&---\\
      %	\hline
    \end{tabular}
  \end{center}
  \caption{Mesh assessment for three-dimensional computations at $Re=200$.}\label{grids3d}	
  %	\hline
\end{table}
The three-dimensional meshes result from the extrusion of the in-plane mesh in the spanwise direction, this coordinate being discretised using Fourier spectral expansions of order $N_z=32$ over a periodic length $L_z=18$ and 9 for meshes B$_1$ and B$_2$, respectively. Agreement with published results is fair in both cases.

As in the present study some of the simulations are performed at a velocity ratio of $R=5.357$, corresponding to a top Reynolds numbers about $Re_T=300$, the appropriate mesh must be able to capture spatio-temporal chaos at sufficiently large values of $Re$.

A final grid refinement study has been undertaken at $Re=1000$ and the results listed in table~\ref{gridsRe1000}.
\begin{table}
  \begin{center}
    \begin{tabular}{ l c c c c  c c c c c c}
      %	\hline
      &
      $ L_{z}$& $L_z^u$&$L_z^d$ & $H$ &$N_{z}$ &$N_{xy}$ &$N (\times10^{6})$
      &$St$ & $C_l'$ & $\langle C_{d}\rangle$\\
      \hline 
      Mesh B$_{3}$&2&9&25.5&16&32&95294&3.04&0.125&1.529&2.240\\
      
      Mesh C$_{1}$&2&9&25.5&16&32&125204&4.01&0.123&1.562&2.225\\
      %	\hline
      Mesh C$_{2}$&2&9&25.5&16&50&125204&6.26&0.120&1.603&2.214\\
      Mesh C$_{3}$&4&9&25.5&16&100&125204&12.52&0.119&1.609&2.212\\
      \cite{BaiPoF2018} LES&4&8&16&16&48&62000&2.98&0.125&---&2.210\\
      \cite{NorbergJWEIA1993}&---&---&---&---&---&---&---&0.127&---&2.227\\
      \cite{OkajimaJSMEIJF1990}&---&---&---&---&---&---&---&0.118:0.136&---&---\\
      \cite{DavisPoF1984}&---&---&---&---&---&---&---&0.137&---&2.046\\
      \cite{OkajimaJFM1982}&---&---&---&---&---&---&---&0.119:0.123&---&---\\
      %	\hline
    \end{tabular}
  \end{center}
  \caption{Three-dimensional mesh assessment at $Re=1000$ for the flow past a square cylinder in homogeneous upstream flow.}\label{gridsRe1000}	
\end{table}

In-plane meshes B and C have been extruded in the spanwise direction to $L_z=2$ and $4$, and the spanwise resolution increased to $N_z=50$ and 100 for the $C$ mesh. Results using in-plane mesh C are in very good agreement with literature results, although mesh B also yields fairly acceptable errors.

\begin{figure}
  \begin{center}
    {\includegraphics[width=0.85\textwidth]{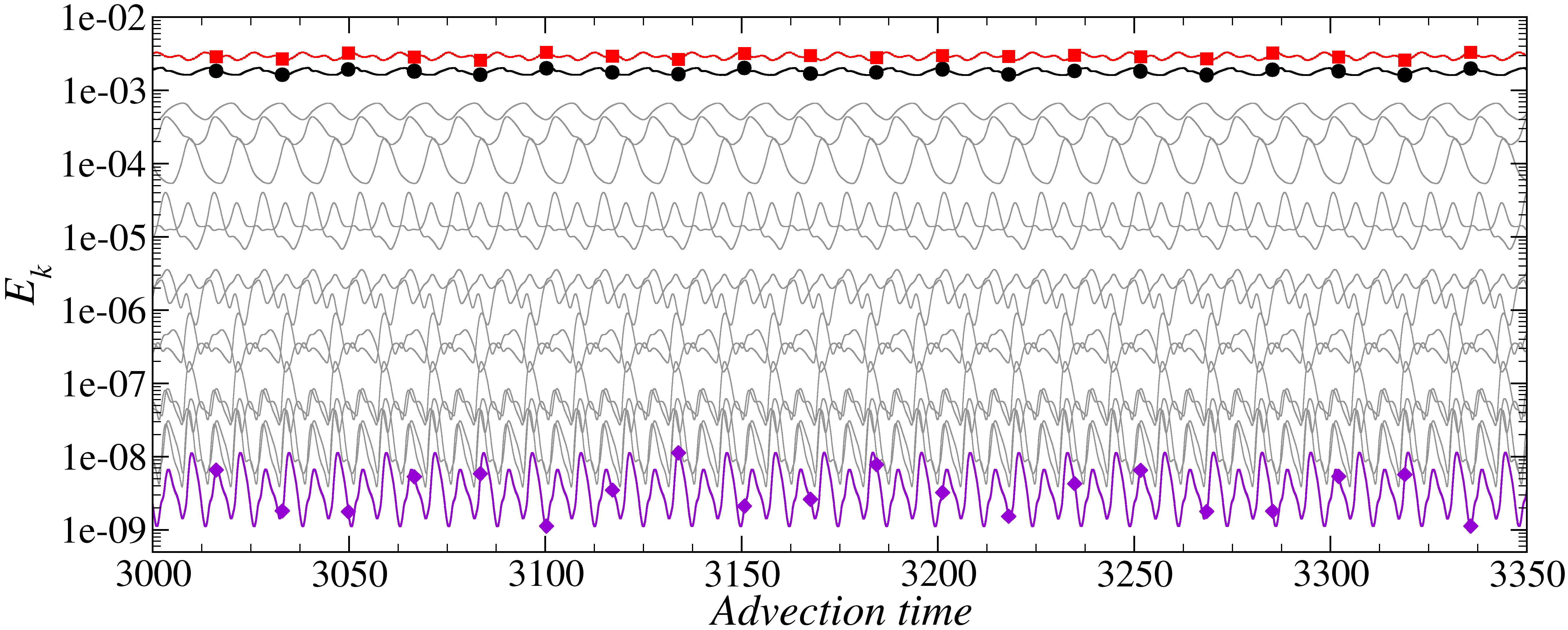}}
  \end{center}
  \caption{Time evolution of the modal energies $E_{k}$ contained in the 16 spanwise-dependent modes of the converged solution at $R=3.8$ in a domain $L_z=5$. Shown in colour are the spanwise wavenumbers 1 (black line), 2 (red line), and 16 (violet line).}
  \label{fig:R38_Modal_energy}
\end{figure}

In the present paper, the spanwise extent of the domain for three-dimensional simulations has been adjusted to fit an integer number of times (commonly 2) the most unstable wavelength as obtained from stability analysis of the two-dimensional time-periodic underlying solution. The spanwise resolution has been chosen in each case to ensure a modal energy decay of about 6 orders of magnitude from the largest non-zero spanwise mode. An example is provided by the modal energy evolution for the converged solution at $R=3.8$ shown in figure~\ref{fig:R38_Modal_energy}.
The decay from spanwise mode 1 to the last mode 16 is good enough to claim sufficient spanwise resolution.

\section{Stability analysis of periodic solutions}\label{app:StAn}

The spanwise translational invariance of the two-dimensional vortex-shedding flow past an infinite square cylinder breaks as the Reynolds number is increased beyond a critical threshold. Floquet theory provides the framework for analysing the stability of time-periodic base flows and has been employed successfully to shed light on the three-dimensionalisation of the wake behind cylinders \citep{NoackJFM1994,HendersonPoF1996,BarkleyJFM1996}. The instantaneous flow is decomposed into the additive superposition of the periodic solution and a perturbation following
\begin{equation}\label{eq:soldecomp}
  ({\bf u},p) = ({\bf U},P) + (\tilde{\bf u},\tilde{p}).
\end{equation}
Formal substitution into the Navier-Stokes equations (\ref{eq:NS}) and linearisation yields the governing equations for the perturbation field
\begin{equation}\label{eq:linNS}
  \begin{array}{rcl}
    \dfrac{\partial \tilde{\mathbf{u}}}{\partial t}+(\tilde{\mathbf{u}} \cdot \nabla) \mathbf{U} + (\mathbf{U} \cdot \nabla) \tilde{\mathbf{u}} & = & -\nabla \tilde{p}+\dfrac{1}{Re} \nabla^{2} \tilde{\mathbf{u}} \\
    \nabla \cdot \tilde{\mathbf{u}} & = & 0,
  \end{array}
\end{equation}
with homogeneous conditions on all boundaries of the same typology as for the nonlinear problem.

The two-dimensionality of the base flow and the translational invariance in the spanwise direction allow for the following modal ansatz:
\begin{equation}
  \tilde{\bf u}({\bf r};t)=\int_{-\infty}^{\infty}\hat{\bf u}(x,y,\beta;t)\mathrm{e}^{\mathrm{i}\beta_z z}d\beta_z,
\end{equation}
corresponding to a Fourier decomposition, such that spanwise modes $\hat{\bf u}(x,y,\beta;t)$ of different spanwise wavenumber $\beta_z$ decouple exactly.

Since we are not interested in computing the full eigenspectrum nor even a subset of it, we follow here a simple time-stepping approach that provides the most unstable (or least stable) multiplier for any desired Reynolds number $Re$ and spanwise wavenumber $\beta_z$. We perturb the two-dimensional periodic vortex-shedding solution, duly computed through time stepping in the two-dimensional domain where it remains stable, with a tiny spanwise-dependent random disturbance. The initial condition thus obtained is then evolved in time integrating the full nonlinear Navier-Stokes equations but using a single spanwise Fourier component, besides the homogenous component, in a domain of spanwise size $L_z=2\pi/\beta_z$. After the initial transients, the energy contents on the stable manifold vanish and the perturbation aligns with the dominant eigenmode. At this stage, the dominant eigenmode can be extracted directly from the only non-homogeneous Fourier component of the simulation as long as it does not approach zero-machine (in the stable case) or saturation unto the nonlinear regime (for the unstable case), where it starts feeding energy back into the homogeneous Fourier component via the nonlinear advection term and the simulation becomes utterly under-resolved.

While still in the linear regime, but well past the initial transients, the already modal perturbation follows an oscillating behaviour with the periodicity of the underlying two-dimensional solution on top of an exponential decay/growth. The decay/growth rate of the dominant Floquet mode (as represented by the Floquet multiplier $\mu$) can be derived from the energy time-series of the $\beta_z$-mode by fitting a power-law to the sequence of crossings ($E_{\beta_z}^k=E_{\beta_z}(t_k)$) of a purposely defined Poincar\'e section (in our case, picking $C_l=0$, and numbering the consecutive crossings at times $t_k$ by an index $k$)
\begin{equation}\label{eq:energygrowth}
  E_{\beta_z}^k=E_{\beta_z}^0\lvert\mu\rvert^{2k}.
\end{equation}
Here $E_{\beta_z}^0$ is an irrelevant fitting parameter, while the $2$ in the exponent deals with the square relation that exists between the velocity field and its corresponding kinetic energy. A more sophisticated fit of the form
\begin{equation}\label{eq:energygrowth2}
  a_{\beta_z}^k=a_{\beta_z}^0 r^{k}\cos{(k\theta+\psi)},
\end{equation}
where $a$ is any primitive degree of freedom of the perturbation field (a point velocity component, for example), $\psi$ is another fitting parameter representing an initial phase and $\mu\equiv r\mathrm{e}^{\pm\mathrm{i}\theta}$ a complex multiplier, allows discrimination between synchronous and quasiperiodic modes \citep{BoJFM2019}. We have not used such a fit here, as the only mode present is clearly subharmonic and appears as synchronous when inspecting the modal energy evolution.

\begin{figure}
  \centering
  \includegraphics[width=0.55\textwidth]{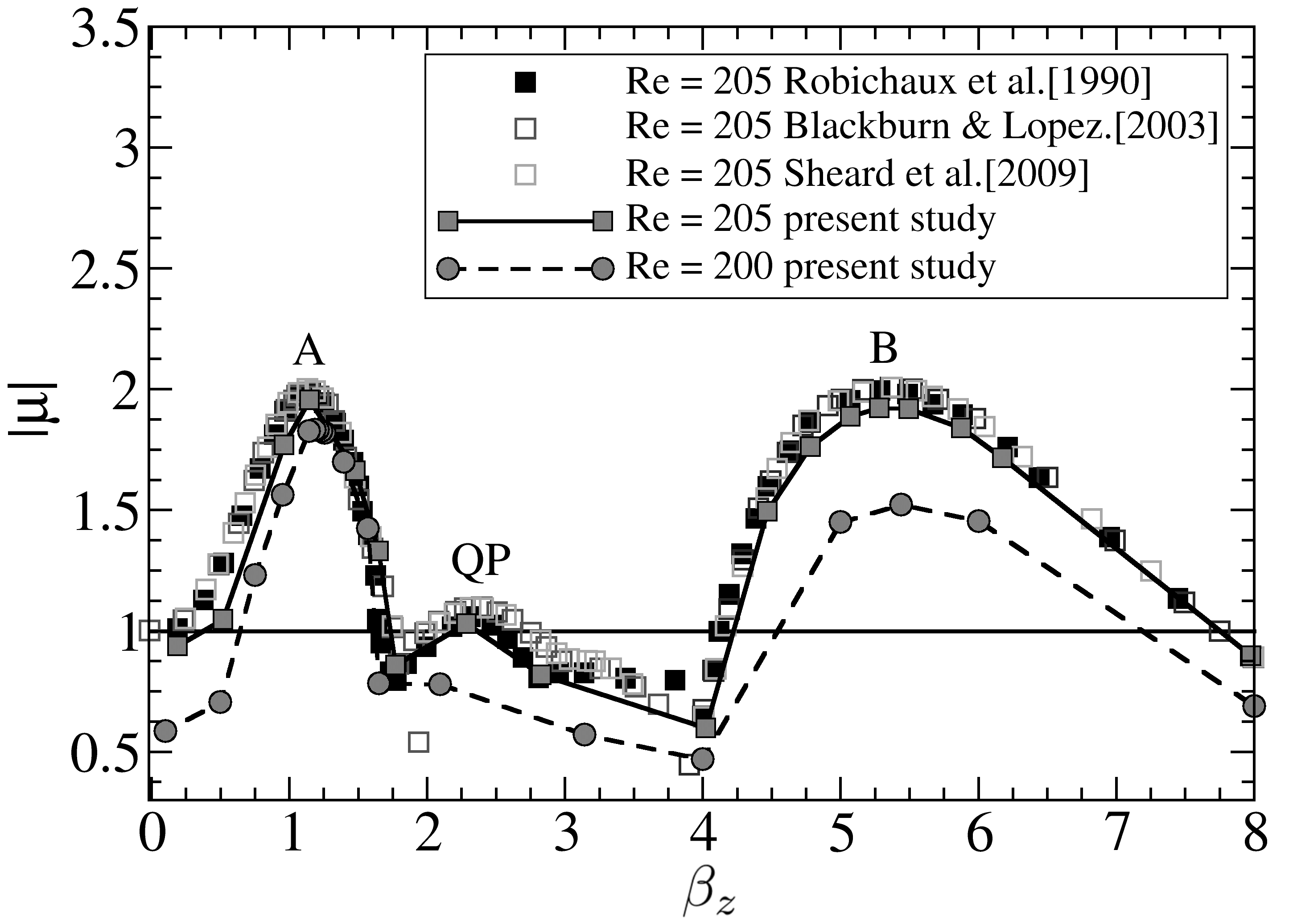}
  \caption{Dominant Floquet multipliers at Reynolds numbers $Re=200$ (circles) and $Re=205$ (squares) for the square cylinder without splitter plate. Shown are our numerical results (grey) alongside those by \citet{RobichauxPoF1999} (black). Labels indicate modes A, B and QP.}
  \label{fig:Floquet200}
\end{figure}

The envelope of the spectrum $\lvert\mu\rvert(\beta)$ for $Re=200$ and 205 has been computed as proof of concept of the aforementioned method, and compared to published results \citep{RobichauxPoF1999,BlackburnPoF2003,SheardJFM2009} in figure~\ref{fig:Floquet200}.
The agreement is fair at $Re=205$, for which all three modes, A, B and QP, are duly observed and exhibit the right growth rate dependence on the wavenumber. The most unstable modes at $Re=200$ have wavenumbers $\beta_z^{\mathrm{A}}\simeq1.21$ (wavelength $\lambda_z^{\mathrm{A}}=5.2$) and $\beta_z^{\mathrm{B}}\simeq5.5$ ($\lambda_z^{\mathrm{B}}=1.15$), in line with experimental \citep{LuoPoF2003} and experimental \citep{LuoJFS2007} results. A third unstable mode arises at $Re^{\mathrm{QP}}\simeq205$ with wavenumber $\beta_z^{\mathrm{QP}}\simeq2.2$ ($\lambda_z^{\mathrm{B}}\simeq2.8$), that was not clearly discernible at $Re=200$. This mode, which was initially thought subharmonic and called mode S \citep{RobichauxPoF1999}, is in fact quasiperiodic \citep{BlackburnPoF2003,SheardJFM2009}, hence the name QP, and has already been detected, albeit at much higher $Re$, for the flow past circular cylinders \citep{BarkleyJFM1996,BlackburnPoF2003}.

In the light of these results, non-linear three-dimensional simulations at these values of the Reynolds number would need to fit several times the longest mode (A) so that a decent range of the known unstable linear modes are allowed to play their part in selecting/driving the fully non-linear solution.

\FloatBarrier

\bibliographystyle{jfm}

\bibliography{local}

\end{document}